\documentclass[
aip, 
graphicx,
reprint
]{revtex4-1}

\draft % marks overfull lines with a black rule on the right

\usepackage{graphicx}% Include figure files, %Right column sets wrong without it
\graphicspath{{./images/}}
\usepackage{dcolumn}% Align table columns on decimal point
\usepackage[utf8]{inputenc}
\usepackage[T1]{fontenc}
\usepackage{etoolbox}
\usepackage{subcaption}
\usepackage{amsmath}
\usepackage{amssymb}
\usepackage{wasysym}
\usepackage{nccmath}
\usepackage{MnSymbol}
\usepackage{diagrams_48}
\usepackage{placeins}
\usepackage{algorithm2e}
\usepackage{graphicx}% Include figure files
\usepackage{dcolumn}% Align table columns on decimal point
\usepackage[utf8]{inputenc}
\usepackage[T1]{fontenc}
\usepackage{etoolbox}

\makeatletter
\def\@email#1#2{%
 \endgroup
 \patchcmd{\titleblock@produce}
  {\frontmatter@RRAPformat}
  {\frontmatter@RRAPformat{\produce@RRAP{*#1\href{mailto:#2}{#2}}}\frontmatter@RRAPformat}
  {}{}
}%
\makeatother

\makeatletter
\def\foo#1\endgraf\unskip#2\foo{\def\row@to@buffer{#1\endgraf\unskip\unskip#2}}
\expandafter\foo\row@to@buffer\foo
\makeatother
\begin{document}

%\preprint{}

\title{Advances in the Simulation and Modeling of Complex Systems using Dynamical Graph Grammars} %Title of paper

%\preprint{AIP/123-QED}

%\title[Sample title]{Sample Title:\\with Forced Linebreak}
% Force line breaks with \\
\author{E. Medwedeff}
\affiliation{ 
Computational Science Research Center, San Diego State University, 5500 Campanile Drive, San Diego, CA 92182, United States of America
}
\affiliation{ 
Department of Computer Science, University California Irvine, Irvine, CA 92697-3435, United States of America
}

\author{E. Mjolsness}%
 \email{emj@uci.edu}
 \affiliation{ 
Department of Computer Science, University California Irvine, Irvine, CA 92697-3435, United States of America
}
\affiliation{Department of Mathematics, University California Irvine, Irvine, CA 92697-3875, United States of America
}

\date{\today}
\begin{abstract}
%250 exactly 
The Dynamical Graph Grammar (DGG) formalism can describe complex system dynamics with graphs that are mapped into a master equation. An exact stochastic simulation algorithm may be used, but it is slow for large systems. To overcome this problem, an approximate spatial stochastic/deterministic simulation algorithm, which uses spatial decomposition of the system’s time-evolution operator through an expanded cell complex (ECC), was previously developed and implemented for a cortical microtubule array (CMA) model. Here, computational efficiency is improved at the cost of introducing errors confined to interactions between adjacent subdomains of different dimensions, realized as some events occurring out of order. A rule instances to domains mapping function \(\varphi\), ensures the errors are local. This approach has been further refined and generalized in this work. Additional efficiency is achieved by maintaining an incrementally updated match data structure for all possible rule matches. The API has been redesigned to support DGG rules in general, rather than for one specific model. To demonstrate these improvements in the algorithm, we have developed the Dynamical Graph Grammar Modeling Library (DGGML) and a DGG model for the periclinal face of the plant cell CMA. This model explores the effects of face shape and boundary conditions on local and global alignment. For a rectangular face, different boundary conditions reorient the array between the long and short axes. The periclinal CMA DGG demonstrates the flexibility and utility of DGGML, and these new methods highlight DGGs' potential for testing, screening, or generating hypotheses to explain emergent phenomena.
\end{abstract}


\maketitle %\maketitle must follow title, authors, abstract and \pacs

\section{Introduction} \label{intro}
\subsection{Overview}
Graphs are a fundamental concept in mathematics and computer science \cite{graph_theory}, and intuitively allow us to represent relationships between objects and concepts. However, graphs are static and independent of time. Dynamic graphs, on the other hand, are graphs that can change over continuous time and allow us to intuitively represent changing relationships. Theories\cite{Mjolsness_2013, Mjolsness2022, graph_grammars_rozen} of graph grammars provide a comprehensive mathematical framework to understand how dynamic graphs become dynamic using expressive rewriting systems. A dynamic graph can be further equipped with a high-level language that maps graphs to a master equation in an operator algebra framework, effectively enabling the dynamic graph to become dynamic through a stochastic rewriting process. Dynamical Graph Grammars (DGGs) \cite{Mjolsness2019} allow for an expressive and powerful way to declare a set of local rules, both stochastic and differential-equation bearing, to model a complex dynamic system with graphs. 

DGGs can be simulated using an exact algorithm, which is derived using operator algebra \cite{Mjolsness_2013}. The exact algorithm works for parameter-bearing objects, and in a special case reduces to Gillespie's Stochastic Simulation Algorithm (SSA) \cite{GILLESPIE1977_OG} which has a stochastic Markov chain formulation\cite{Mjolsness_2013}. The idea behind the SSA is similar to the kinetic Monte Carlo algorithms of statistical physics \cite{Kinetic_WMYoung1966}. What makes DGGs so powerful is that they allow more complicated objects and processes to be described using a formalism that generalizes chemical kinetics to spatially extended structured objects. DGGs can also be simulated using an approximate algorithm that is faster, but at the cost of some reactions occurring out of order.

Preliminary work on an approximation \cite{Medwedeff_2023} of the exact algorithm demonstrated the potential for modeling and simulation of more complex systems. In this work, we build beyond the initial formulation of the approximate algorithm and provide an expanded description and improved version of the approximate algorithm, along with demonstrating its viability to be used as a framework for computational testing, screening, or inventing hypotheses to explain emergent phenomena by providing an example plant periclinal cortical microtubule array (PCMA) model. Every ``with'' rule in a DGG is stochastic, and in the PCMA DGG, twenty out of the twenty-two have this property. The algorithm, model, and framework are realized by the Dynamical Graph Grammar Modeling Library (DGGML). Together, they represent the far-reaching impact of Monte Carlo methods in contemporary research \cite{metropolis}.

\subsection{Stochastic Chemical Kinetics and Beyond}
When modeling chemical systems, there are two broad approaches \cite{modeling}: \textit{stochastic} and \textit{deterministic}. In the \textit{deterministic} approach, the reaction-rate equations used \cite{reaction_rate} comprise a system of coupled ordinary differential equations (ODEs), and the process is predictable. Alternatively, a form of \textit{stochastic} modeling arises whenever a continuous time Markov process is completely described by a \textit{master equation} and sample trajectories are simulated using the SSA. 

The master equation (ME) itself is a high-dimensional linear differential equation, $P^{'}(t) = W \cdot P(t)$. A key difference between it and the reaction rate equations is that it governs the rate at which probability $p$ flows through different states in the modeled system. However, the actual system to model can become very large due to an exponential state-space explosion concerning the number of parameters involved, and the ME may have an infinite-dimensional state space. An analytical solution to the master equation is often computationally intractable or impossible. Typically only smaller, simpler models have analytical solutions. Therefore, algorithms that can allow for simulation of a sample trajectory are important and are used to attempt to recover the probability density functions. 

Methods like Kinetic Monte Carlo have been applied to Ising spin systems \cite{kinetic_monte_carlo} and used in Gillespie's work \cite{GILLESPIE1992_RIG}, where he derives an exact stochastic simulation algorithm (SSA) for chemical kinetics using Monte Carlo and kinetic theory. This derivation relies on assumptions such as a large number of well-mixed molecules at thermal equilibrium. Reaction rates, which are essential, can be determined through lab measurements, large \textit{ab initio} quantum mechanical calculations, machine learning generalizations, or parameter optimizations with system-level observations and known reaction rates. An event is sampled from a conditional density function (CDF) representing a reaction. While the Monte Carlo method doesn't solve the master equation analytically, it provides an unbiased sample trajectory of a system through numerical simulation, which is important for making computationally taxing problems more tractable. 

Despite its power the exact SSA is slow, as it computes each reaction event sequentially. Various methods have been proposed to speed up the SSA. $\tau$-Leap \cite{GILLESPIE2001_TAU} fires all reactions within a time window $\tau$ before updating propensity functions, saving computation time at the cost of some errors, and was later made more efficient \cite{TauLeapEfficient}. $R$-Leaping \cite{RLeap} allows a preselected number of reactions per simulation step, also sacrificing some accuracy. The Exact $R$-Leap (``$ER$-Leap'') \cite{ER_Leap} refines $R$-Leaping to be exact, significantly speeding up the SSA, and was later improved and parallelized in $HiER$-Leap \cite{HiER_Leap}. More recently, $S$-Leap \cite{SLeap} has been introduced as an adaptive, accelerated method combining $\tau$-Leaping and $R$-Leaping. 

Numerous other methods have also been developed to expedite the original SSA. However, what all of these methods lack is generality. This is one of the features the DGG formalism \cite{Mjolsness_2013, Mjolsness2006, Mjolsness2019} provides, and the approximate algorithm and methodology presented in this work aims to provide an alternative point of view on the acceleration of this class of algorithms and beyond, providing a valuable degree of generality by representing the dynamics of spatially extended objects using graphs. By virtue of the framework, chemical systems can be simulated, but so can spatially extended and structured biological systems as well.

\subsection{Generalizing Beyond Chemical Reactions using DGGs}
Dynamical Graph Grammars comprise of a set of declarative modeling rules. Pure chemical reaction notation contextualizes what we mean by declarative modeling. Pure reaction rules can themselves be represented in a production rule notation. For example, if we have a pure reaction system, the standard chemical reaction notation \cite{Mjolsness2010, Mjolsness2019} would be of the form:

\begin{equation}
    \label{eq:chem_mass}
    \sum^{A_{max}}_{\alpha = 1} m_\alpha^{(r)} A_{\alpha} \overset{k_{(r)}}{\longrightarrow} \sum^{A_{max}}_{\beta = 1} n_\beta^{(r)} A_{\beta}
\end{equation}

Here we use $r$ to represent the $r$-th reaction channel. We have $m_\alpha^{(r)}$ and $n_\beta^{(r)}$ as the left and right-hand side nonnegative integer-valued stoichiometries, since we cannot physically have negative chemical reactants. Finally, we let $k_{(r)}$ be the forward reaction rate function. 

We can write equation \ref{eq:chem_mass} in a more compact form \cite{Mjolsness2019} using multi-set notation denoted as $\{\cdot\}_*$, where $\cdot$ is a mathematical placeholder and $*$ symbolically indicates multi-set. So:

\begin{equation}
    \label{eq:chem_mass_compact}
     \{m_\alpha^{(r)} A_{\alpha}\}_* \overset{k_{(r)}}{\longrightarrow}  \{n_\beta^{(r)} A_{\beta}\}_*
\end{equation}

A simple example of the notation in action is the reaction rule for the formation of water:

\begin{equation}
    \label{eq:chem_example}
    2H_2 + O_2 \overset{k_r}{\longrightarrow} 2H_2 O \quad \text{or} \quad \{2H_2,O_2\}_* \overset{k_r}{\longrightarrow} \{2H_2, O\}_*
\end{equation}

In equation \ref{eq:chem_example} the left-hand side of the equation represents the reactants, consisting of two molecules of hydrogen gas $H_2$ and one molecule of oxygen gas $O_2$. The right-hand side represents the products, consisting of two molecules of water $H_2 O$. The coefficients in the equation indicate the stoichiometric ratios of reactants and products involved in the reaction. The reaction occurs with a rate of $k_r > 0$. We can make this more generic and summarize the resulting process as ${LHS}_r \longrightarrow {RHS}_r$ for rule $r$. Consequently, we can encode this formal description into the master equation and derive Gillespie's SSA \cite{GILLESPIE1977_OG} for a well-mixed system using operator algebra \cite{Mjolsness_2013}.

Dynamical grammar rules \cite{Mjolsness2009} subsume these stoichiometric production rules, but are instead written using labeled graphs, which allows for more expressiveness by adding parameters to the object nodes and graph edges between them. Using a formal graph notation \cite{Mjolsness2019}, DGGs can be represented using a simple form:  
    \begin{flalign} \label{def:general} \nonumber
       & G\llangle\lambda\rrangle \longrightarrow G^\prime\llangle \lambda^\prime\rrangle \\  
       & \quad \textbf{with } \rho_r(\lambda, \lambda^\prime) \textit{ or } \textbf{solving } d{\lambda}_i/dt = v(\lambda)\text{.} 
    \end{flalign}

Here $G\llangle\lambda\rrangle$ is the left-hand side labeled graph with label vector $\lambda$ and $G^\prime\llangle\lambda^\prime\rrangle$ is the right-hand side labeled graph with label vector $\lambda^\prime$. $G$ and $G^\prime$ without their label vectors $\lambda$ and $\lambda^\prime$ are numbered graphs so that the assignment of label component $\lambda_i$ to graph node member $i$ is unambiguously specified. We also have the \textbf{solving} and \textbf{with} clauses. A rule that has a \textit{with} clause is the stochastic rule for instantaneous events that occur at a rate determined by the function $\rho$, whereas a rule that has a \textbf{solving} clause is one in which the parameters are updated continuously in time according to a differential equation.

A simple way to read the rule notation in equation \ref{def:general} is graph $G$ transforms into graph $G^\prime$ either with a rate and newly sampled parameters from $\rho_r(\lambda, \lambda^\prime) \equiv \rho_r(\lambda) * P(\lambda^\prime \: \mid \: \lambda)$ or by solving an ODE(s) associated with the label parameters.

Since DGG rules can extend beyond reactants, we can use them to describe the structural dynamics of larger polymers of proteins such as microtubules. For example:

\begin{flalign}\label{rule:ex_disc_retract}
\nonumber & \textbf{Stochastic Retraction:} \\ 
\nonumber & 
\left(
\begin{diagram}[size=1em, textflow, loose]
\;
\blacksquare_1 & \rLine & \Circle_2 & \rLine & \Circle_3
\end{diagram}
\right)
\llangle
(\text{${ \boldsymbol x}$}_{1}, \text{${ \boldsymbol u}$}_{1}), 
(\text{${ \boldsymbol x}$}_2, \text{${ \boldsymbol u}$}_2), 
(\text{${ \boldsymbol x}$}_3, \text{${ \boldsymbol u}$}_3)
\rrangle
& \\ 
\nonumber & \longrightarrow
\left(
\begin{diagram}[size=1em, textflow, loose]
\;
\blacksquare_1 & \rLine & \Circle_3
\end{diagram}
\right)
\llangle
(\text{${ \boldsymbol x}$}_{1}, \text{${ \boldsymbol u}$}_{1}), 
\emptyset, 
(\text{${ \boldsymbol x}$}_3, \text{${ \boldsymbol u}$}_3)
\rrangle
& \\ 
& \quad \quad \text{\boldmath $\mathbf{with}$} \ \
H(\cdot) &
\end{flalign}
is a rule where the structural dynamics of a depolymerizing microtubule are encoded as a graph rewrite rule. The rule in equation \ref{rule:ex_disc_retract} has only a single component on the left-hand side (LHS). Other rules, such as those of interacting molecules could have multiple components. Here $H(\cdot)$ is some nonegative function of some or all of the $x$ and $u$ parameters, the probability per unit time of rule firing.

\subsection{Related Work}
The dynamical graph grammar modeling library (DGGML) draws direct inspiration from Plenum \cite{yosiphon_2009}, which implements dynamical graph grammars within the Mathematica \cite{Mathematica} environment by leveraging its symbolic programming capabilities. Despite its expressive and mathematically friendly interface, Plenum faces scalability challenges due to its reliance on the exact algorithm, making it less suitable for larger, more complex systems. Additionally, Plenum's use of unique object identifiers (OIDs) to indirectly handle graphs, rather than treating them as native data structures, further contributes to performance issues.

The exact algorithm in Plenum supports the expressive nature of dynamical graph grammars (DGGs) but suffers from diminished performance when applied to large systems, further exacerbated by its symbolic implementation in a computer algebra system. Plenum’s design choice to use Mathematica trades performance for expressiveness. Mathematica's pattern matching capabilities are used to find matches for left-hand side (LHS) grammar rules by treating LHS patterns as tuples of object combinations and eliminating invalid matches through constraints and constraint-solving \cite{bucket_elimination, constraint_processing}. Plenum integrates discrete events and continuous rules, solving differential equations using Mathematica's numerical ODE solver. In contrast, DGGML is less expressive but implements the exact algorithm with significantly less overhead and supports native graph structures directly.

Related modeling libraries and tools similar to Plenum and DGGML include MGS \cite{MGS_long}, MCell \cite{mcell}, PyCellerator \cite{pycellerator}, BionetGen \cite{bionetgen2, bionetgen}, and Kappa \cite{kappa}. MGS is a declarative spatial computing programming language for cell complexes with applications in biology such as neurulation \cite{MGS_short}, which is the process by which the neural tube is formed, in vertebrate development. Other works such as by Lane\cite{Lane2015CellCT} used cell complexes for the developmental modeling of plants and simulating cell division. MCell is used for the simulation of cellular signaling and focuses on the complex 3D subcellular microenvironment in and around living cells. It is designed around the assumption that at small subcellular scales, stochastic behavior dominates. Therefore, MCell uses Monte Carlo algorithms to track the stochastic behavior of discrete molecules in space and time as they diffuse and interact with other molecules. PyCellerator, on the other hand, is a computational framework for modeling biochemical reaction networks using a reaction-like arrow-based input language (a subset of Cellerator \cite{cellerator}). BioNetGen is a software tool used for the rule-based modeling and simulation of biochemical systems, allowing researchers to create and analyze complex biological networks. Kappa is another rule-based modeling language for molecular biology, and has been applied to protein interaction networks \cite{deeds}. Although these tools share similarities with DGGML and are worth independent exploration, this work concerns the methodology and simulation of spatially embedded dynamic graphs, modeled by dynamical graph grammars including both stochastic and differential equation rules and applications. 

\section{Methods}
\subsection{Organization}
First, we introduce the previous version of the approximate algorithm (Algorithm \ref{alg:original_approx}) and its improved version (Algorithm \ref{alg:approx}). Both algorithms build upon the exact simulation algorithm \cite{Mjolsness_2013} and speed it up the by splitting the simulation into small sub-simulations of a decomposed domain where the cost is some reactions being processed out of order on a short time scale, potentially trading speed for accuracy. After presenting the approximate algorithm and its improvements, we highlight other unique aspects of the algorithm and clarify some of the details of its requirements. We briefly describe Yet Another Graph Library (YAGL), discuss the choices for the rule match to geocell mapping function $\varphi$, and propose an incremental match data structure that includes all rule instances and all possible left-hand side connected component instances. We detail the process for updating this data structure, which significantly reduces the need to recompute matches after each rule firing event. Additionally, we discuss graph transformation in terms of operator algebra, category theory and the Dynamical Graph Grammar Modeling Library's (DGGML)'s implementation of the theory. Finally, we briefly touch on grammar analysis and pattern matching, along with including insight for how its time complexity can be upper-bounded.

A key challenge in enabling the reproducibility of complex algorithms such as Algorithm \ref{alg:approx} is ensuring that the concepts and implementation choices are clearly conveyed and delineated whenever possible. This involves not only providing comprehensive descriptions of the algorithms themselves but also elucidating the underlying theoretical frameworks and practical considerations that inform their design and operation. Moreover, it is useful to illustrate how these concepts are instantiated in the code and to explain the rationale behind specific implementation decisions. The details provided below are meant to enable others to understand, reproduce, and potentially extend the work. 

\subsection{Improving the Approximate Algorithm}
While the DGG formalism may be expressively powerful, just as for the SSA the exact algorithm \cite{Mjolsness_2013} becomes prohibitively slow for large systems, making it computationally intractable to sample a single trajectory. In line with other Monte Carlo methods, many simulations are needed to compute meaningful statistics or recover outcome density functions. The exact algorithm quickly becomes impractical since it must compute the next event based on every potential rule firing in a large system. To address this, we introduce the basis of the approximate \cite{Medwedeff_2023} Algorithm \ref{alg:original_approx} and the significantly improved Algorithm \ref{alg:approx}, which is the algorithm used by DGGML.

\begin{figure}[!ht]
  \centering
  \includegraphics[width=0.95\linewidth]{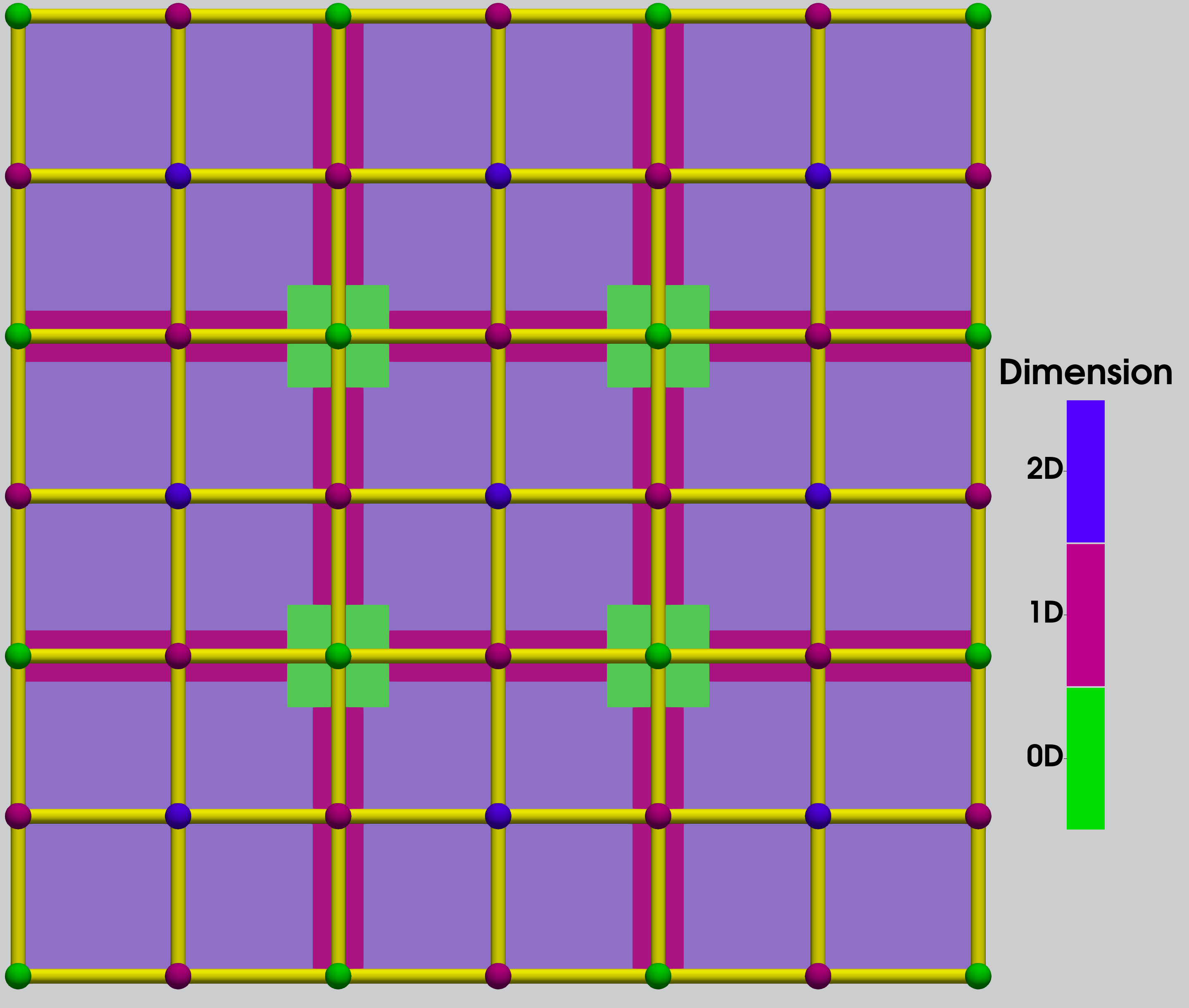}
  \caption{The expanded cell complex for a $3\times3$ decomposition. Dimensions of the same color are well separated from one another. The $0D$ regions are expanded points, the $1D$ are expanded edges. Exterior cells in the cell complex do not need to be expanded.}
  \label{fig:post-expand}
\end{figure}

Two key assumptions are made in approximating the exact algorithm: spatial locality of the rules and well-separateness of cells of the same original dimensionality within the cell complex used to decompose the simulation space. Spatial locality, allows us to decompose the simulation space into smaller, well-separated geometric cells, or ``geocells.'' In the simulation algorithm, a ``cell'' refers to a computational spatial domain, not a biological cell. A geocell is a cell of an expanded cell complex (ECC), labeled by the dimension of the corresponding cell in the unexpanded complex. Lower-dimensional cells are expanded perpendicularly, so as to prevent rule instances from spanning multiple same-dimensional geocells. By setting these geocells wide enough (several times larger than the exponential ``fall off distance'' of local rule propensity functions), we can map rule instances using the to-be-discussed $\varphi$ function to well-separated geocells. An example of an ECC can be found in figure \ref{fig:post-expand}.

The system state itself is a labeled graph with nodes labeled by vector-valued position parameters, also allowing additional parameters without spatial constraints. Our graph grammar rules are used to find matching graph patterns in the system graph and are made spatially local by their propensity functions, which define local neighborhoods of rule firings. Rule instantiations outside this neighborhood requires a fast enough e.g. exponential falloff with distance in the ``with'' rule propensity functions, and in turn they will have zero or near-zero propensity outside the neighborhood. Thus, interactions between distant objects are unlikely and can be ignored with minimal error e.g. in a chemical reaction, reactants that are very far away from one another do not interact.

The operator $W$ in $P^{'}(t) = W \cdot P(t)$ is defined in the state space for all extended objects in the system. Let $W = \sum{W_r}$ be the sum of the DGG rules mapped to operators. We can approximate the solution, $e^{tW}$, to the master equation by using operator splitting that imposes a domain decomposition by means of an expanded cell complex that corresponds to summing operators, $W = \sum_{(d)} W_{(d)} = \sum_{(d,c)} W_{(c,d)}$, over pre-expansion dimensions $d$, and cells $c$ of each dimension:
\begin{subequations}
	\begin{alignat}{3}
	\label{eq:op_split1} e^{tW} & 
	\approx \Bigg( \prod_{d \downarrow} e^{\frac{t}{n} W_{(d)}} \Bigg)^{n \rightarrow \infty} &
	\\
	\nonumber e^{t^\prime W_{(d)}} & 
	= \prod_{c \; \subset \; d} e^{t^\prime W_{(c,d)}} & \\ \label{eq:op_split2} &
	\text{where} \quad [W_{(c,d)}, W_{(c^\prime,d)}] \approx 0
	\quad \text{and} \quad t^\prime \equiv \frac{t}{n} &\\
	\label{eq:op_split3} W_{(c,d)} &
	= \sum_r W_{r,c} 
	\equiv \sum_r \sum_{\Bigl\{\substack{R \; | \; \varphi(R) =c, \\ R \text{ instantiates } r}\Bigl\}} W_r (\; R \; | \; c,d) &
	\end{alignat}
\end{subequations}

Sub-equation (\ref{eq:op_split1}) is a first-order operator splitting, by solution phases of fixed cell dimension, where $d \downarrow$ means we multiply from right to left in order of highest dimension to lowest. It incurs an approximate error of $\mathcal{O}((t/n)^2)$. Sub-equation (\ref{eq:op_split2}) is an even stronger refinement of sub-equation (\ref{eq:op_split1}) because it uses the fact that the resulting cells $c$ of fixed dimension $d$ are all well-separated geometrically with enough margin (due to the expanded regions\cite{collars} of dimension $d^\prime \neq d$) so that rule (reaction) instances $R$, and $R^\prime$ commute to high accuracy if they are assigned to different cells $c,c^\prime$ of the same dimensionality, by some rule (reaction) instance allocation function $\varphi$. The commutators of equation (\ref{eq:op_split2}) can be calculated as derived in \cite{Mjolsness2022}, but they will inherit the product of two exponential falloffs with separation that we assumed for the rule propensities (c.f. \cite{Mjolsness2022}, equation 12 therein), which is an even faster exponential falloff. Hence the dynamics $e^{tW_{(c,d)}}$ of different cells $c,c^\prime$ of the same original dimension $d$ and can be simulated in any order, or in parallel, at little cost in accuracy. The logic is expressed in Algorithm \ref{alg:original_approx}.

\begin{algorithm}[!ht] 
\SetKw{kwParFor}{ParFor}
\caption{Original Approximate Spatially Embedded Hybrid Parameterized SSA/ODE Algorithm}
\While{$t \leq t_{max}$}{
\ForEach{dimension $d \in \{D_{max}, D_{max}-1,\dots, 0\}$}{
\textit{using function $\varphi$ map rule instances to the geocells of the expanded cell complex}\;
\kwParFor expanded geocell $c_i \in ExpandedCellComplex(d)$ \textbf{do}\\
\quad\textbf{run} Exact Hybrid Parameterized SSA/ODE algorithm for $\Delta t$ in $c_i$\;
}
$t \: += \: \Delta t$\;
}
\label{alg:original_approx}
\end{algorithm}

Algorithm \ref{alg:original_approx} is the original version of the approximate algorithm and was run in serial for the model discussed in previous work \cite{Medwedeff_2023}. On the highest level, it specifies that $\varphi$ must be used to map reaction instances to geocells and those have the potential to be processed in parallel or serial asynchronously. It fully encapsulates sub-equations (\ref{eq:op_split2}) and (\ref{eq:op_split3}). However, it leaves unspecified the details of how to keep the state of the rule matches in the simulated system consistent across geocells after a rule fires and a rewrite occurs. 

\begin{figure*}[!ht]
    \centering
    \includegraphics[width=0.9\linewidth]{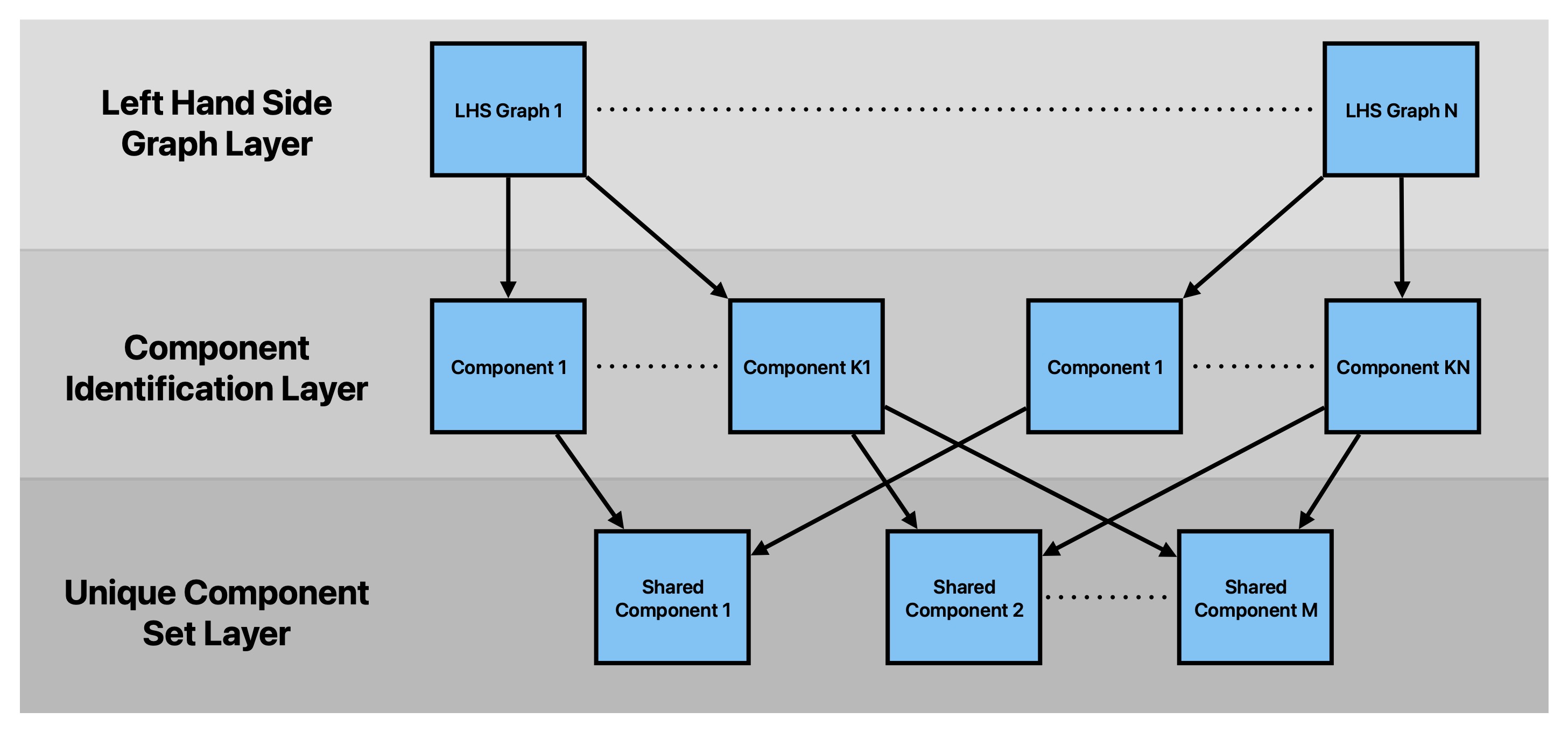}
    \caption{An overview of how the rule match data structure is organized. In this hierarchy, squares represent data, with arrows leaving a square indicating its composition with the pointed to elements. Dotted lines denote the possible omission of additional instances at that particular layer.}
    \label{fig:grammar_comp_top}
\end{figure*}

Algorithm \ref{alg:approx}, which is the algorithm used in DGGML, improves upon the original by adding in explicit requirements to efficiently keep the state of rule matches in the simulated system consistent across geocells after a rule fires and a rewrite occurs. To keep a consistent state of rule matches, we specify how rule matches are represented. The \textit{match data structure} (visualized in figure \ref{fig:grammar_comp_top}), which is initialized at the beginning of Algorithm \ref{alg:approx}, is a data structure that stores all possible rule instances matching the grammar rules to all possible corresponding left-hand side (LHS) connected component instances. The match data structure is then the representation of all current rule instances and must be updated every time a rule fires and the system graph is rewritten. To make this work we take advantage of an incremental updating procedure (section \ref{sec:incremental}) within a geocell, which allows us not to have to recompute all new matches after distant events occur in the system.

\begin{algorithm}[!ht] 
\SetKw{kwParFor}{ParFor}
\SetKw{kwPar}{Parallel}
\caption{Improved Approximate Spatially Embedded Hybrid Parameterized SSA/ODE Algorithm}
\textit{initialize the match data structure with all rule instances}\;
\While{$t_{global} \leq t_{max}$}{
\ForEach{dimension $d \in \{D_{max}, D_{max}-1,\dots, 0\}$}{
\textit{using function $\varphi$ map rule instances to the geocells of the expanded cell complex}\;
\kwParFor expanded geocell $c_i \in ExpandedCellComplex(d)$ \textbf{do}\\
\quad$t_{local} = t_{global}$\;
\quad\textit{factor} $\rho_r([x_p], [y_q]) = \rho_r([x_p]) * P([y_q] \: | \: [x_p])$\;
\quad\While{$t_{local} \leq t_{global} + \Delta t_{local}$}{
    \textit{initialize} SSA propensities as $\rho_r([x_p])$\;
    \textit{initialize} $\rho^{(total)} := \sum_r \rho_r([x_p])$\;
    \textit{initialize} $\tau := 0$\;
    \textbf{draw} effective waiting time $\tau_{max}$ from $\exp{(-\tau_{max})}$\;
    \While{$\tau < \tau_{max}$ \text{and} $t_{local}  \leq t_{global} + \Delta t_{local}$}{
        \textbf{solve} ODE system, plus an extra ODE updating $\tau$\;
        \quad$d\tau/dt_{local} = \rho^{(total)}(t_{local})$\;
    }
    \textbf{draw} rule instance $r$ from distribution $\rho_r([x_p])/\rho^{(total)}$\;
    \textbf{draw} $[y_q]$ from $P([y_q] \: | \: [x_p])$ and \textit{execute} rule instance $r$\;
    \textit{incrementally update match data structure}\;
}
\textit{synchronize and remove invalid rule instances from the data structure of matches}\; 
}

\textit{recompute rule level matches (as needed)}\; 
$t \: += \: \Delta t_{global}$\;
}
\label{alg:approx}
\end{algorithm}

Algorithm \ref{alg:approx} also includes the point of synchronization and rule invalidation that comes after processing all expanded cells of a given dimension. After $\varphi$ maps every rule to a cell $c_i$ of dimension $d$, the exact algorithm running within $c_i$ is responsible for incrementally updating its state and keeping track of any rule invalidation that may share objects with rules belonging to neighboring cells of differing dimensions. Hence, the need for the \textit{synchronization} point in Algorithm \ref{alg:approx}. Synchronizing for consistency is a consequence of the commutator approximation in equation (\ref{eq:op_split2}). When running in serial the synchronization point is still present and functions as a logical point in the algorithm for inserting any custom code for correcting and creating a consistent state of rule matches and potentially measuring any errors. Algorithm \ref{alg:approx} also has the potential to be made parallel in a lock-free way thanks to the synchronization point, because it leaves the resolution of the global state until after the operation of all geocells of dimension $d$, making it a substantial improvement over Algorithm \ref{alg:original_approx}. 

Algorithm \ref{alg:approx} also substantially improves upon Algorithm \ref{alg:original_approx} by introducing the aforementioned match data structure with an incremental update process to ensure that distant rule firings do not require the matches to be recomputed and that full system matches are only ever recomputed when needed. The primary need for recomputation even after synchronization occurs because spatially embedded and spatially local graph grammar rules are composed of one or more distinct connected components. By virtue of \textbf{solving} rules, these components may move in or out of the propensity ``fall-off'' distance. While these connected components themselves can always be incrementally updated i.e. the set of connected component matches is \textit{fully online}, the rule instances that comprise them can only be incrementally updated for a short period of time. As a result, they need to be recomputed periodically, making the match data structure \textit{semi-online}. 

In the absence of domain decomposition Algorithm \ref{alg:original_approx} reduces to the exact algorithm, and Algorithm \ref{alg:approx} reduces to a version of the exact algorithm but with the incremental update of the rule match data structure. A more complete mathematical treatment of the approximate algorithm using DGG commutators computed \cite{Mjolsness2022} to bound operator splitting errors is a topic for future work. Algorithms \ref{alg:original_approx} and \ref{alg:approx} are specifically designed for spatially embedded graphs, but there are situations where they could be used in other non-spatial instances if we had a way to reasonably measure locality in the parameter space and that is another topic for future work. For DGGML Algorithm \ref{alg:approx} is used, but we present Algorithm \ref{alg:original_approx} as well to highlight the difference. 

\subsection{Yet Another Graph Libary (YAGL)}
The DGG simulator relies on graphs as its core data structure but does not specify how to implement graph rewrites or represent graphs computationally. To address this, the Yet Another Graph Library (YAGL) was introduced\cite{Medwedeff_2023} and further developed as a versatile, header-only library designed for compatibility with modern C++17 features. Unlike the monolithic Boost Graph Library\cite{BGL} (BGL), YAGL emphasizes flexibility and future extensibility without external dependencies. YAGL defines nodes, edges, and adjacency lists using static polymorphism and template meta-programming, with nodes stored in an unordered map for efficient lookup and edges in an unordered multi-map to support multiple edges between nodes. The adjacency list, realized as an unordered map, allows constant-time lookup of incoming and outgoing edges, facilitating dynamic graph operations.

YAGL includes operations such as adding/removing nodes and edges, identifying connected components, various graph searches, generating spanning trees, determining graph isomorphism, and solving the injective homomorphism problem for subgraph pattern recognition. For DGGML, which uses spatial graphs, YAGL implements a specialized spatial variadic templated variant node type, enabling dynamic type changes during graph rewrites while maintaining spatial data and user-defined values.

\subsection{Phi Function}
The process by which rules get mapped to a particular dimensional cell in Algorithm \ref{alg:approx} hinges on the choice of the $\varphi$ function in Equation (\ref{eq:op_split3}). This function plays a pivotal role in establishing a suitable mapping from reactions (rules) to geocells, ensuring that each reaction instance maps to exactly one geocell. This approach guarantees complete accountability for all rules. Because $\varphi$ assigns a geometric cell to every match, $\varphi$ serves to partition the set of matches. Various implementation choices exist for defining $\varphi$, each presenting distinct trade-offs. We will focus on two primary approaches to demonstrate how they work.

Assume that the following is a match of the left-hand side of a graph grammar rule, and the associated parameters and unique IDs are as follows:

\begin{flalign}\label{rule:ex_match}
& 
\left(
\begin{diagram}[size=1em, textflow, tight]
\;
\blacksquare_{10} & \rLine & \Circle_{11} & \rLine & \Circle_{12}
\end{diagram}
\right)
\begin{array}{l}
\llangle
(\text{${ \boldsymbol x}$}_{10}), 
(\text{${ \boldsymbol x}$}_{11}), 
(\text{${ \boldsymbol x}$}_{12})
\rrangle
\end{array}
\end{flalign}

Here, the match has a closed square type with integer ID $10$ and two open circle types with integer IDs $11$ and $12$. Each of the graph grammar rules has an associated position vector $x$. Let $\text{c}_{\text{id}}(x_i)$ be a function that returns the unique cell id to which an object at position $x_i$ belongs and let the function $\text{dim}(\text{c}_{\text{id}}(x_{i}))$ be a function that returns the dimension of that cell. For example, assume that $\text{dim}(\text{c}_{\text{id}}(x_{10})) = 2$, $\text{dim}(\text{c}_{\text{id}}(x_{11})) = 1$, and $\text{dim}(\text{c}_{\text{id}}(x_{12})) = 1$. Also, assume $\text{c}_{\text{id}}(x_{10}) = 4$, $\text{c}_{\text{id}}(x_{11}) = 5$, and $\text{c}_{\text{id}}(x_{12}) = 5$. Therefore, $x_{10}$ belongs to a cell $4$ of dimension $2$, and the other nodes belong to cell $5$ of dimension $1$. Also, keep in mind that the spatial-locality and well-separated constraints ensure that all matched objects in the rule will not belong to more than one cell of the same dimension. Thus, restricted to a given match, the $\text{dim}(\cdot)$ function has an inverse $\text{dim}^{-1}(\cdot)$.

The first approach we discuss uses the single-point anchored $\varphi$ function, which involves selecting an ``anchor node'' from the left-hand side and determining the geocell containing its spatial coordinates. The selection of another node may be arbitrary or may be guided by a user-defined heuristic such as a root of a spanning tree. Using the section example, we could pick the root to be node $10$. In this way, we have $\varphi(x_{10}) = \text{c}_{\text{id}}(x_{10}) = 4$, which is a cell of dimension $2$. Hence, a single-point anchored mapping function maps node $10$ to cell $4$. This method boosts computational efficiency by necessitating only a single point for lookup. However, a trade-off arises wherein automorphisms of the same rule match may be assigned to different dimensions. When an invalidation occurs, each geocell containing the invalidate rule must be updated. 

Alternatively, the minimum dimensions function $\varphi$ can be defined to map the rule match to the minimum dimension of the geocells to which each node maps. Using the section example, it would be:
\begin{flalign}
\nonumber \varphi(x_{10}, x_{11}, x_{12}) = & 
\begin{array}{l}
\text{dim}^{-1}(\text{min}[\text{dim}(c_{id}(x_{10})), \\ \quad \text{dim}(c_{id}(x_{11})),\text{dim}(c_{id}(x_{12}))]) 
\end{array} \\ 
= & \text{ } 5\text{,} 
\end{flalign}
which is a cell of dimension $\text{min}[\text{dim}(\dots)] = 1$ where $\text{dim}^{-1}$ exists by well-separatedness. This function can be further extended to consider all rules in which any node of the current rule participates. In either scenario, each instance of the $\varphi$ function calculates the minimum dimension for any match and assigns the rule instance to the corresponding geocell. The well-separated property ensures that no node in the rule or its associated rules can be assigned to multiple geocells of identical dimensions. The minimum dimension $\varphi$ ensures that all automorphisms are consistently assigned to the same dimension and cell because the $\min$ function returns the same result, regardless of the ordering of the input. Computationally, the cost of using the minimum dimension $\varphi$ means checking the spatial coordinates of each node on the left-hand side.

\subsection{Incremental Update} \label{sec:incremental}
The incremental update in Algorithm \ref{alg:approx} is a critical improvement and results in a more efficient algorithmic realization of the DGG formalism. Once a rule is selected to fire and fires, the system's state must be updated. The most straightforward approach would involve recomputing all matches in the system\cite{Medwedeff_2023}, which is clearly resource intensive. The following discussion presents an alternative solution by incrementally updating the matches of the LHS of DGG rules found in the system.

Before proceeding, it is useful to clarify some terminology. In the context of DGGML, matches of the LHS graph for a rule $r$ are stored hierarchically. We store them in this way because the LHS graph of a grammar rule may be a graph with more than one connected component. This is what we mean by a \textit{multicomponent} rule/graph. For example a collision-induced catastrophe rule from the PCMA grammar in Appendix \ref{pcma_dgg}:

\begin{flalign}\label{ex:multi}
 & 
\left(
\begin{diagram}[size=1em]
\nonumber & \Circle_1 & \rLine & \CIRCLE_2 & \text{, } & \Circle_3 & \rLine & \Circle_4 & & 
\end{diagram}
\right)
\begin{array}{l}
\llangle
(\text{${ \boldsymbol x}$}_1, \text{${ \boldsymbol u}$}_1), 
(\text{${ \boldsymbol x}$}_2, \text{${ \boldsymbol u}$}_2),  \\
(\text{${ \boldsymbol x}$}_3, \text{${ \boldsymbol u}$}_3),
(\text{${ \boldsymbol x}$}_4, \text{${ \boldsymbol u}$}_4)
\rrangle
\end{array}
& \\ 
\nonumber & \longrightarrow
\left(
\begin{diagram}[size=1em]
\;
\Circle_1 & \rLine & \blacksquare_2 & \text{, } & \Circle_3 & \rLine & \Circle_4 
\end{diagram}
\right)
\begin{array}{l}
\llangle
(\text{${ \boldsymbol x}$}_{1}, \text{${ \boldsymbol u}$}_{1}), 
(\text{${ \boldsymbol x}$}_2, \text{${ \boldsymbol u}$}_2),  \\
(\text{${ \boldsymbol x}$}_3, \text{${ \boldsymbol u}$}_3),
(\text{${ \boldsymbol x}$}_4, \text{${ \boldsymbol u}$}_4)
\rrangle
\end{array}
& \\ 
\end{flalign} is a rule with multiple components on the left-hand side. From the grammar, there may be several rules that have connected components that when taken as subgraphs are isomorphic to one another. This common set of connected component patterns is what we mean by \textit{motifs} for a grammar.

The \textit{component match set} (represented as a map data structure) is the set of all matches (injective homomorphisms) of these motifs. Since LHS grammar rules can be multicomponent, combinations of the matches in the component set are what form the \textit{rule match set} (also represented as a map data structure). Together, the \textit{component match set} and the \textit{rule match set} form what we mean by the \textit{match data structure}. To accelerate the search for new rule matches, components matches are also sorted into a \textit{cell list}\cite{cell_list} i$.$e$.$ a data structure allowing for fast queries of components matches nearby in space.

The goal of the incremental update is to keep the match data structure up to date following a graph rewrite. By using the incremental update, the connected component match set can be kept fully online, since connected component matches are invariant to motion. The rule match set, on the other hand, is incrementally updated within a geocell, but is periodically recomputed after all dimensional geocells have run. The recomputation occurs because the connected components of the rule matches move with respect to one another and may eventually go out of or come into the ``fall-off'' distance for the propensity function. So, we can say that the match data structure is \textit{semi-online}.

Next, it is important to clarify the impact of certain rewrite operations on the system graph. Adding a node or an edge inevitably requires new matches to be found, as it opens up new search paths and the possibility for more patterns to be found and added to the component and rule match sets. Conversely, removing a node or an edge solely invalidates, without initiating a search, owing to our graph rewrite semantics which never require the \textit{absence} of an LHS object. In the context of DGGML, altering the node type is identical to adding a new node in terms of match invalidation, whereas modifying parameters such as position results in a passive rewrite. These passive rewrites affect multi-component spatial pattern matches and are addressed periodically by recomputation of all multicomponent matches globally. As rewrites always involve a combination of these operations, there exists a systematic approach to processing them.

Following a rewrite event, the incremental update process involves the following steps: \textit{component invalidation}, \textit{rule match invalidation}, \textit{component matching}, \textit{component validation}, \textit{rule matching}, and \textit{rule match validation}. The ordering is similar to reading the semantics of equation \ref{eq:op_graph_trans} below from right to left, where all relevant nodes/edges are destroyed and then created. However, the incremental update differs from the operator equation in being an implementation rather than formal mathematical semantics, because it is cast in the context of the data structures and internal implementation of DGGML.

Initially, \textit{component invalidation} requires searching the component set for all connected component match instances of any components containing the nodes and edges marked for removal from the graph rewrite discussed in the previous section, and removing those components. Due to the well-separated nature of geocells and the localized execution of rewrites on the matches homomorphic to a subgraph of the spatially embedded system graph, concurrent processing of geocells of the same dimensions is possible. By using a suitable data structure for the set of component instances and optimizing the function $\varphi$, component matches can be safely removed in parallel or in a serialized manner by locating them in the component match data structure.

\begin{figure}[!ht]
    \centering
    \includegraphics[width=0.95\linewidth]{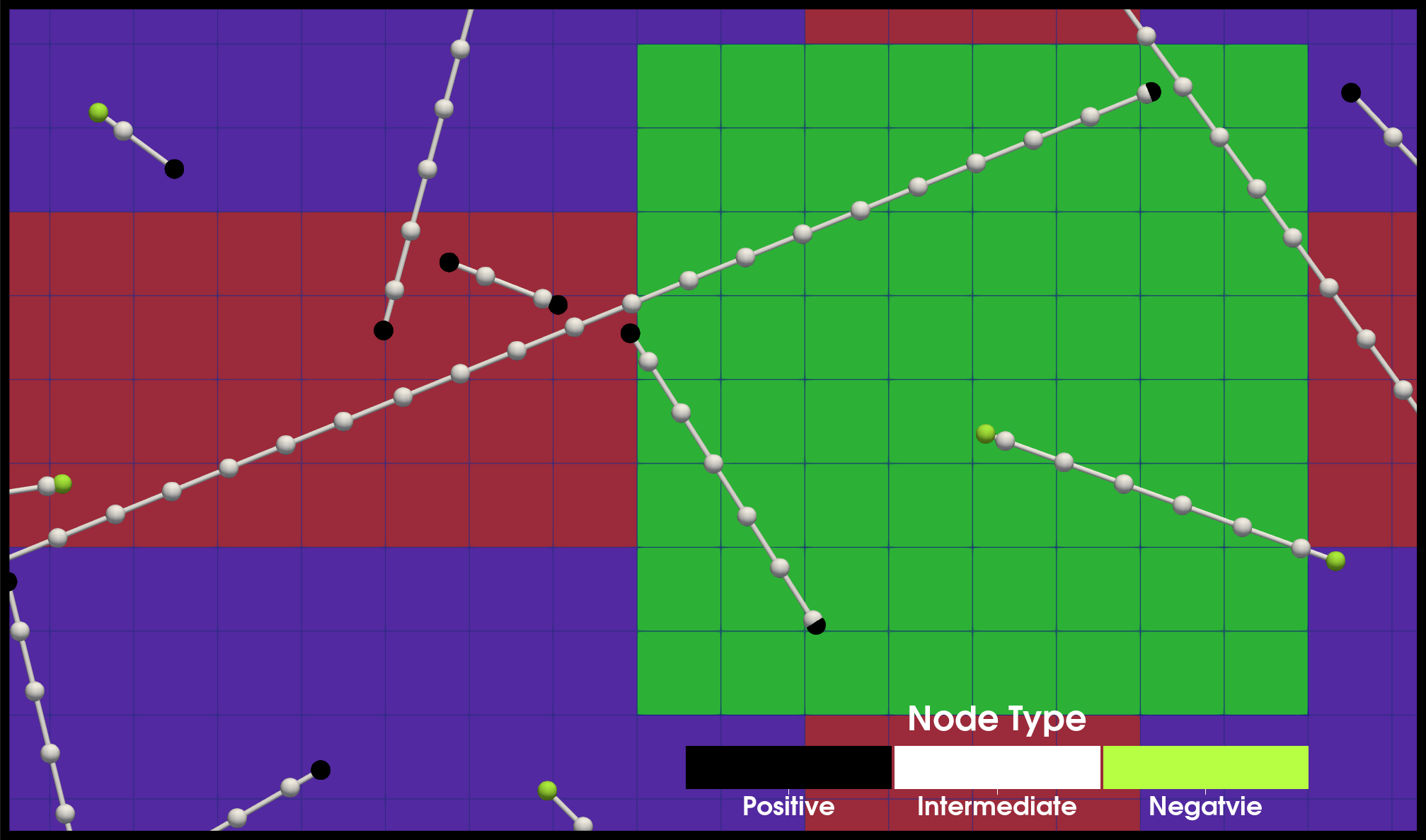}
    \caption{Zoomed in version of expanded cell complex (ECC) with well-separated lower dimensions and spatially embedded graph objects. Regions of the same pre-expansion dimension are separated from each other. Note how only interior lower dimensional cells are expanded. A reaction grid is aligned with the geocells, and reaction grid cells are smaller than the geocells.}
    \label{fig:ecc}
\end{figure}

\textit{Rule match invalidation} within a geocell, e.g. in a parallel computation, presents a more complicated challenge. Because rules are directly mapped to the geocell, only those rules mapped to the current geocell via $\varphi$ can be invalidated. Given that the current geocell lacks the state of rules mapped to other geocells, it must temporarily store the list of invalidated components for future potential invalidation of rules owned by nearby geocells, which are of different dimension and therefore not being simulated at the same time. An optimization would be to track only the invalidated components near the boundary of the geocell. The future invalidation of rules occurs during the \textit{synchronization} point in Algorithm \ref{alg:approx}. Also, for both component and rule match invalidation, spatial locality can be leveraged to improve the search efficiency for components containing an invalid node/edge or rules containing an invalid component. Using a grid of sub-cells i.e. a ``reaction grid'' and a list of components found in each sub-cell, neighboring sub-cells and their components can be quickly queried. To illustrate these concepts, a zoomed in region of an example system and a reaction grid can be found in figure \ref{fig:ecc}. So, instead of searching everywhere for a potentially invalidated component, the search is only performed within a local region. 

In \textit{component matching}, once all the invalidations of components are processed, new component matches must be found. A simple and slightly greedy way to accomplish this is to precompute the height of the tallest rooted spanning tree generated from all left-hand side motifs in the common set of components for the grammar. For each of the newly created nodes, we perform a depth-first search to the depth of the precomputed height and add newly visited nodes to a visited set. This does not need to be done for any removed node, since removal will only generate invalidations. Using the visited set, a subgraph $S_{search} \in G_{sys}$ is induced. Subgraph specific pattern recognition (SSPR) is then run on subgraph $S_{search}$ to find all new component matches. The search is specifically designed to accept matches and add them into the component match data structure only if they contain a new node or newly added edge from the creation set generated by graph rewrite - this is what is meant by \textit{component validation}. Once the components are validated, a spatial hash function is applied to map them to their respective cells in the cell list. 

For \textit{rule matching}, the set of cells in the cell list that new components were inserted into, and their collective neighbors, constitute a new search space $S_{nbr}$. This time the search space is not an induced subgraph to find component matches, but instead comprises all component matches in nearby cells to be combined to find new rule matches. To accomplish this, we search for all combinations of component matches that match the component(s) in the LHS of a rule. To ensure that only new rule matches are identified, matches involving newly validated components are exclusively considered and the others are rejected - this is what is meant by \textit{rule match validation},. Additionally, $\varphi$ is applied to determine the geocell to which the new rule matches are mapped. If the geocell is the current one, the rule matches are inserted into the rule list. Otherwise, they are set aside and inserted into the correct geocell during the \textit{synchronization} phase of Algorithm \ref{alg:approx}, provided they still exist.

While alternative approaches exist for this incremental update procedure, the algorithmic idea proposed squarely addresses the problem as a graph algorithm. Moreover, given the assumption of a system significantly larger than the local search space, new matches can be identified and updated with considerably less overhead when compared to having to search the space globally. An alternative approach involves storing an entire copy of the full search space in the form of completed and partial matches, resulting in a complex hierarchy of search trees. While potentially faster than the approach employed in DGGML, this method would necessitate more storage.

\subsection{Graph Transformation}
Graph transformations are a fundamental part of Algorithm \ref{alg:approx} and can be described and given formal semantics using the operator algebra in the mathematical framework for DGGs. The following equation from \cite{Mjolsness2019} demonstrates this fact:

\begin{equation} \label{eq:op_graph_trans}
    \begin{aligned}
    \hat{W}_r & \propto \rho_r(\lambda, \lambda^\prime) \times \\ 
    & \sum_{\langle i_1 \dots i_k \rangle_\neq} 
    \biggr[ \Bigr( \prod_{p \in B_r} \prod_{i \neq i_q | \forall q \in \overline{B}_rp} E_{i_{p}\;i} \Bigl)
    \Bigr( \prod_{p \in C_r} \prod_{i \neq i_q | \forall q \in \overline{C}_rp} E_{i\;i_{p}} \Bigl) \biggl] \\
    & \times \biggr[ \Bigr( \prod_{p^\prime,q^\prime \in \text{rhs}(r)}  \Bigr( \hat{a}_{i_{p^\prime} i_{q^\prime}}\Bigl)^{g^{\prime}_{p^\prime q^\prime}}\biggl] 
    \biggr[ \Bigr( \prod_{p^\prime \in \text{rhs}(r)}  \Bigr( \hat{a}_{i_{p^\prime} \lambda^{\prime}_{p^\prime}}\Bigl)^{h_{p^\prime}}\biggl] \\
    & \times \biggr[ \Bigr( \prod_{p,q \in \text{lhs}(r)}  \Bigr( a_{i_p i_q}\Bigl)^{g_{p q}}\biggl] 
    \biggr[ \Bigr( \prod_{p \in \text{lhs}(r)}  \Bigr( a_{i_p \lambda_p}\Bigl)^{h_{p}}\biggl] \text{.}
    \end{aligned}
\end{equation}

Equation \ref{eq:op_graph_trans} is a generalized form of graph rule semantics that allows for the elimination of hanging edges as part of the grammar's operator algebra. When reading from right to left, all vertices and vertex labels of the LHS graph are annihilated (destroyed) in an arbitrary order and then all edges are annihilated in any order, then all vertices and vertex labels of the RHS graph are created in any order, then all edges of the RHS are created in any order. The final operators on the first line are the dangling edge erasure operators that enforce that edges must connect active vertices. In this way, the erasure operators presented in this context ensure that hanging edges are removed. For a more detailed explanation please see \cite{Mjolsness2019}.

 Within a category-theory approach there are two algebraic category-theoretic constructions for graph transformation \cite{graph_grammars_rozen}: single-pushout (SPO) \cite{single_pushout} and double-pushout (DPO) universal diagrams \cite{double_pushout}. SPO focuses on local transformations, directly replacing a subgraph in the original graph with another subgraph. In contrast, DPO has more flexibility, allowing for the replacement subgraph to be embedded in a larger context. To accomplish this, DPO requires two additional constraints to ensure uniqueness and keep transformations well-defined. The dangling condition, the first constraint in DPO, requires all edges incident to a deleted node to be deleted as well. The second constraint, the identification condition, ensures that each element set for deletion has only one copy in the left-hand side subgraph. The combination of the two conditions is the gluing condition, which allows for a consistent and unambiguous transformation.

In DGGML, the methodology used to perform graph rewrites follows the semantics presented in the operator algebra and works similarly to the DPO approach. The logic takes advantage of sets for the structural aspect of rewrites in the library. To clarify this, first let $r: G_{{LHS}} \longrightarrow G_{{RHS}}$ be the production of a grammar rule $r$. For shorthand $G_{LHS} = L$ and $G_{RHS} = R$. The vertices of the $L$ are then $V(L)$ and the vertices of $R$ are then $V(R)$. The edges of $L$ are $E(L)$ and the edges of $R$ are $E(R)$. 

To determine how to structurally transform the $L$ into $R$ following equation \ref{eq:op_graph_trans}, we need four sets, arising from the creation of new vertices and edges and the annihilation with the restoration of no longer valid vertices and edges. Let $K$ be the creation graph, and $D$ be the destruction graph. The set of vertices to create consists of the vertices found in the right-hand side vertex set, but not in the left. Hence, $V(K) = V(R) \setminus  V(L)$, i.e. the set difference. The edges to be created are then, $E(K) = E(R) \setminus V(R)$. For the creation graph, we simply have $K = R \setminus  L$. The set of vertices and edges to destroy is the opposite of the vertex and edge sets of the creation graph. So, vertices and edges in the left-hand side that are not found in the right-hand side must be removed. Hence, the destruction graph is $D = L \setminus  R$. Once we have the creation and destruction sets for a particular rule type $r$, we can take the node/edge numberings of any injective homomorphism of $L$ denoted as $L_i$ found in the system graph, $G_{SYS}$, and build corresponding sets $K_i$ and $D_i$. Here the new vertices and edges in $K_i$ are created using newly generated keys.

To get the new state of vertices in the system graph, $G_{SYS}^\prime$ we can use the following equation:

\begin{eqnarray} \label{eq:rewrite_set}
    V(G_{SYS}^\prime) & = & (V(G_{SYS}) \setminus  V(D_i)) \cup V(K_i) \nonumber \\
    & = & (V(G_{SYS}) \setminus (V(L_i) \setminus V(R_i))) \nonumber \\
    & & \cup (V(R_i) \setminus  V(L_i))
\end{eqnarray}

When removing nodes, YAGL in DGGML cleans up hanging edges automatically. Finally, it should be noted that $L$ and $R$ are technically labeled graphs i.e. $G = (V, E, \alpha)$ where $\alpha$ is the labeling function. So, the labels of $L$ are $\alpha(V(L))$ and the labels of $R$ are $\alpha(V(R))$. However, in the context of the library, YAGL and in turn DGGML store the label information as data within the graph node itself, meaning the label is a type in the context of C++. So, the final step after building the creation and destruction sets for the vertex and edges sets and structurally transforming the graph is to set the new label data for any newly created nodes by copying the types from $K$ to $K_i$, effectively equating to a relabelling. In DGGML, this is realized by switching all the newly created nodes to the correct type in the specialized spatial variadic templated variant node in YAGL. In line with the DGG formalism, after the structural rewrite and type change of the rewrite occurs, the label data is updated based on the newly sampled parameters.

\subsection{Grammar Analysis and Pattern Matching}
In any generalized implementation of the DGG formalism, grammar analysis is a fundamental. Both stochastic and deterministic grammar rules should be converted into a semantically analyzable data format when created. In DGGML, these rules are efficiently mapped into a structure with unique rule names, enabling direct function binding and resolution at runtime. The analysis begins by precomputing transformations from the left-hand side (LHS) to the right-hand side (RHS) to ensure predefined steps during rewrites. Each rule's LHS graphs are analyzed to identify their connected components, known as \textit{motifs}, which are then consolidated into a set of unique motifs using graph isomorphism.

The motifs are matched within the system graph using a ``subgraph-specific pattern recognizer'' (SSPR). A component match map data structure manages the relationship between motifs and their instances, enabling operations involving components and rule matches. Each rule match consists of a list of unique keys representing the associated components, stored in a structure for efficient querying during simulation.

Previously \cite{Medwedeff_2023}, graph pattern recognition was hard-coded and did not account for multi-component LHS grammar rule analysis. Algorithm \ref{alg:approx} in DGGML includes a generalized version \cite{Valiente2021} of subgraph-specific pattern recognition. Within DGGML, grammar rules are analyzed and parsed into motifs, and when searched for the subgraph algorithm generates a rooted spanning tree for each connected component to aid in the search for the component match.

For optimization, certain LHS graph rules might share isomorphic connected components, which is why all motif matches should be found first. DGGML's focus on spatially embedded graphs then means we can treat connected component matches of an LHS as separate nearby graphs to be searched for, and can find complete pattern matches by finding nearby combinations of matching objects. This eliminates the need for extensive redundant searches for each component in each rule. A ``cell list'' \cite{cell_list} data structure expedites this search process. Further optimization could be found by combining searches for rules with patterns in common or by merging their rooted spanning trees and setting additional ways to reach acceptance states.

To address the scaling of pattern matching in large graphs, it is important to clarify reasonable bounds on computational complexity through a brief upper bound analysis. For instance, let $N$ represent the number of vertices in the large system graph, and $ 0 \leq c \leq 1$ be the fraction of system nodes that match an LHS component spanning tree root node based solely on type. Let $n \ll N$ denote the maximum number of system graph nodes within a reaction radius (or diameter) of any given node, and $s$ be the number of nodes in the LHS component.

To incorporate the effect of node connectivity, let $f$ represent the maximum fan-out of a node, which is the number of edges leaving that node. The number of nodes within the reaction radius to be searched can be influenced by the local connectivity. So, the number of nodes in $n$ that are searched can be further reduced and put in terms of $f$:
\begin{equation}
    \text{min}(f,n)
\end{equation}
The cost of matching an LHS component is then upper bounded by:
\begin{equation}
    \mathcal{O}((cN)\;\text{min}(f,n)^s)
\end{equation} This expression is linear in $N$, polynomial in $f$ and $n$, and exponential only in the fixed, small integer $s$. It demonstrates how the structure of the graph, particularly the connectivity of nodes, affects the cost of pattern matching.

In case of multi-component rules, i.e. we need to find a combination of items. Let $p$ be the number of components required in the LHS of a rule, and $M$ be the total number of components found in the system graph.  If 
$m$ is the maximum number of components within a reaction radius of any given component, then an upper bound cost estimate is:
\begin{equation} \label{eq:combination}
    \mathcal{O}(M \cdot m^{p-1})
\end{equation} In equation \ref{eq:combination} the upper bound remains the product of the cost to find each item.

Several factors can further reduce the component and multi-component search cost:
\begin{enumerate}
    \item The choice of rooted spanning tree and its topological sorting, typically prioritizing rare node types so that search loops can be exited early.
    \item The specificity of parameter checks and non-tree graph link checks, which can result in early termination of the match search (higher specificity is preferable).
    \item Decomposing large LHS rules into multiple smaller rules, thereby reducing \( s \).
\end{enumerate}

The pattern matching method used in DGGML is then a heuristic tree search with backtracking, constraints, and spatial locality, designed for matching rules with LHS graphs that have multiple connected components. This approach effectively addresses the NP-completeness of the general (sub-)graph isomorphism problem by casting it into a special case when small user-chosen graphs and labels enable the algorithm to run in low-degree polynomial time as a function of system size. In the PCMA grammar in Appendix \ref{pcma_dgg} the LHS graph labels and structure enable finding component matches in polynomial-time as a function of system size. To fully complete the rules, combinations of components must be found. A reasonable LHS is constant in size relative to the system graph. This makes DGGs suitable for modeling complex systems and capable of integrating other modeling paradigms based on the complexity scale of the grammar rule graph objects.

\section{Results}
\subsection{Highlights}
To demonstrate what a practical implementation of the improvements looks like we have developed the Dynamical Graph Grammar Modeling Library (DGGML) in C++, the first open-source implementation of a generalized DGG framework. The other generalized DGG software Plenum \cite{yosiphon_2009} was written in the proprietary language Mathematica\cite{Mathematica}, which limits its availability to a broader community and limits the extent to which users can modify and reuse it for their purposes. 

To emphasize the utility and actual practicality of using the DGG modeling paradigm, we have included a grammar (Appendix \ref{pcma_dgg}) with DGGML for the periclinal cortical microtubule array (PCMA) in plant cells. In these results, we have included a unique finding from the PCMA DGG model, that demonstrates the viability of using DGGML and in general the DGG formalism for hypothesis testing and exploring what mechanisms can lead to observed emergent behavior. We also compare the performance of the previous prototype simulator CajeteCMA \cite{Medwedeff_2023} to DGGML on a simple microtubule growth model. Many underlying factors can lead to one performing better than the other, but the key improvement is that DGGML provides far more expressiveness, combined with reasonable performance, when compared to CajeteCMA which is limited to only a specific model.

\subsection{The Dynamical Graph Grammar Modeling Library}
The Dynamical Graph Grammar Modeling Library (DGGML) realizes the approximate algorithm and implements all of the forgoeing improvements. In the current state, the library itself functions as a tool to be used for $2$D spatially embedded DGG simulations, and the design serves as a template for future implementations and improvements such as extending to $3$D or leveraging scalability offered by the algorithm when applied to large scale systems and offloading work to accelerators. 

\begin{figure*}
    \centering
    \includegraphics[width=0.95\linewidth]{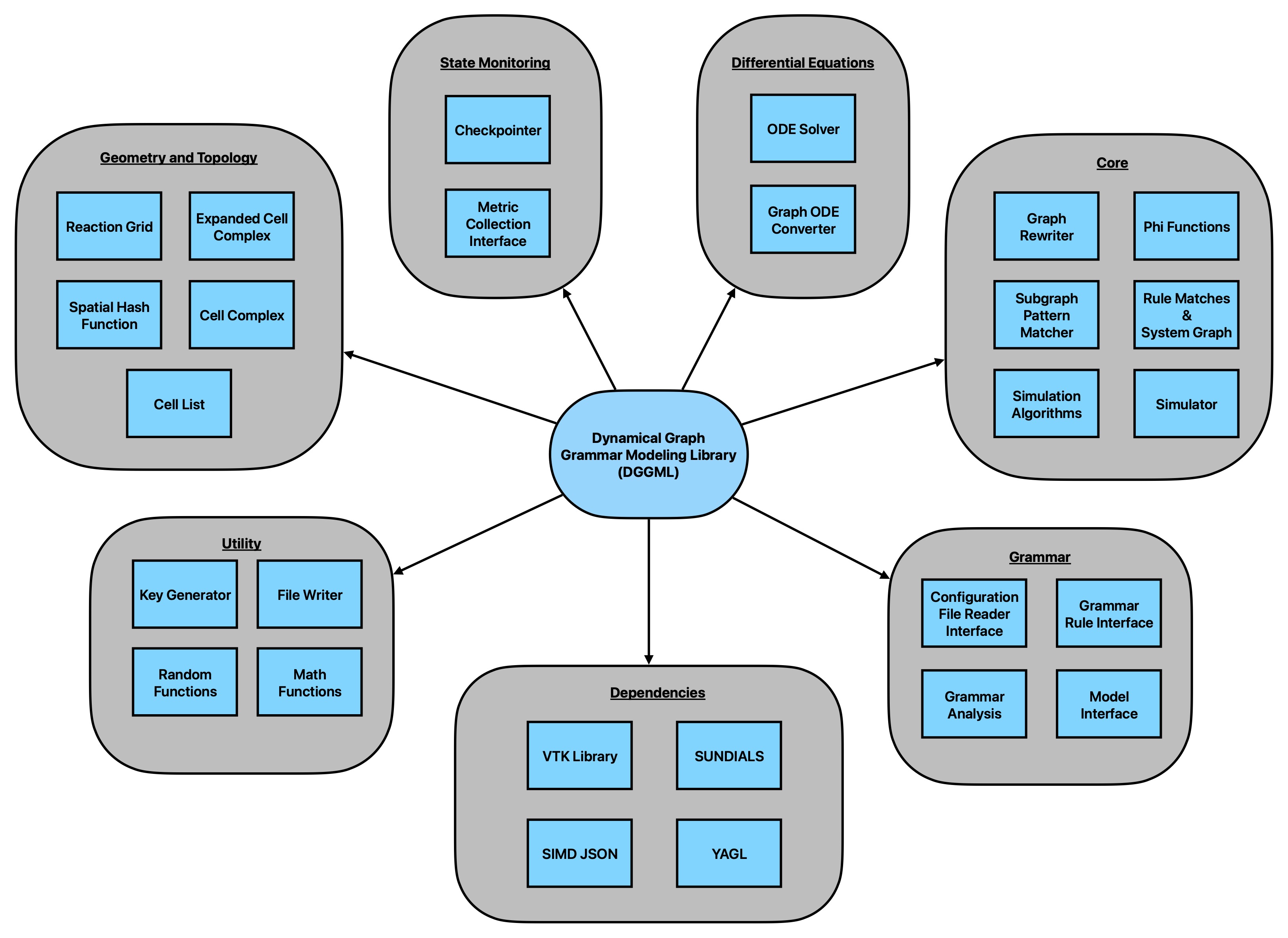}
    \caption{Conceptual overview of library design and organization. The library organization is grouped by components falling into the core, grammar, geometry and topology, state monitoring, differential equations, utility, and dependency categories. The hierarchy of the grouping is visualized as a central blue squircle (square with rounded corners) designating the library itself surrounded by exterior grey squircles, each representing a group of components, where the black directional arrows depict compositional relationships. Within these exterior squircles, blue rectangular boxes represent individual components. We have acronyms ODE (ordinary differential equations), VTK (visualization toolkit), SUNDIALS (SUite of Nonlinear and Differential/ALgebraic equation Solvers), SIMD JSON (single input, multiple data JavaScript object notation), and YAGL (yet another graph library).}
    \label{fig:dggml_library}
\end{figure*}

The initial implementation of the approximate simulation algorithm was specific to a plant cell cortical microtubule array model \cite{Medwedeff_2023} and not generalized to work with different model types. Since then it has been reworked into the modeling library DGGML, to enable reuse and generic construction of simulations for different grammars. Further, the modeling library is written in C++17 with minimal dependencies, making it portable and accessible. A conceptual overview of the library is shown in figure \ref{fig:dggml_library}. 

In figure \ref{fig:dggml_library}, the library is structured into seven distinct categories: core, grammar, geometry and topology, state monitoring, differential equations, utility, and dependency. The core group contains essential functionality required for simulating a grammar.  Within the grammar category are components necessary to define a grammar as a model and to initialize data structures for interaction with the core components. The geometry and topology components facilitate spatial organization, partitioning, and search queries. State monitoring components track and collect information on simulation progress. While differential equations could be grouped within the core, their potential for extensibility to accommodate more complex solvers favors a separate categorization. The utility grouping contains components that are used by the core and other categories. Finally, the dependencies are used throughout the library, but are the components that could be replaced by alternatives or for which a custom version of the type of component could be specifically written and included as part of the library itself. Figure \ref{fig:dggml_library} serves as a conceptual point of reference, but with any design, there are unavoidable relationships among components within and between groups.

\subsection{The PCMA DGG Example}
\subsubsection{Biological Motivation}
Using DGGML, an example of a model for the Periclinal Cortical Microtubule Array (PCMA) has developed and been tested. It builds upon a previous CMA model \cite{Medwedeff_2023} and has been further refined to be more biologically realistic. This refinement includes the addition of new rules, which effectively demonstrates the capabilities of the DGG formalism. A key change in the update to how microtubule zippering works. On the modeling side, these new rules are used to investigate the mechanisms underlying the formation of specific array patterns of cortical microtubules without other cell information by incorporating dynamics observed in the outer periclinal face of plant cells.

The mechanics of plant morphogenesis are complex and play out on many different scales of space \cite{plant_mechanics}, and these mechanics are well suited to be used to develop DGG models \cite{Mjolsness2019}. The reorganization of the cortical microtubule array (CMA) \cite{shaw_reorg} is one part of the scale hierarchy. The CMA is significantly complex and there are still many outstanding questions \cite{update_cma}, such as what general principles govern the organization of cortical microtubules into functional patterns that govern the direction of cell division and cellular responses to mechanical stress?

The Periclinal CMA (PCMA) model approximates the periclinal face of a cell, which is the face parallel to the organ's surface. The anticlinal faces are perpendicular to it \cite{cell_div}. A common approximation of the shape of a plant cell for CMA simulations is a cube \cite{allard_sim} or polygonal prism, or a curved surface such as a cylinder \cite{cylinder_sim, cylinder_sim2}. In our case, the cell is approximated to be a rectangular prism, which includes the cube and could be extended to other polyhedral prisms. Therefore, the approximated cell has six faces, 12 edges, and 8 vertices. Using this notion of prisms, we choose the base faces of the prism to be the top and bottom periclinal faces of the cell. During \textit{in vivo} experiments, the top periclinal face is easier to observe, so we chose this location as a reasonable representation of the simulation space for our model. 

CLASP (Cytoplasmic linker-associated protein) is a well-studied microtubule-associated protein (MAP) that is found in different types of plant, fungal, and animal cells. In the plant cells of Arabidopsis it has been observed to facilitate transitions between CMA patterns \cite{CLASP_facilitates}. The localization of the Arabidopsis thaliana CLASP (AtCLASP) protein to specific cell edges was shown experimentally and computationally to mediate microtubule (MT) polymerization when encountering cell corners of high curvature, and the selective localization biased the alignment of the CMA \cite{Ambrose2011}. In our model, we use this observed behavior to make broad idealized assumptions about the effects it has on the boundaries.

\begin{figure}[!ht]
\centering
  \includegraphics[width=.65\linewidth]{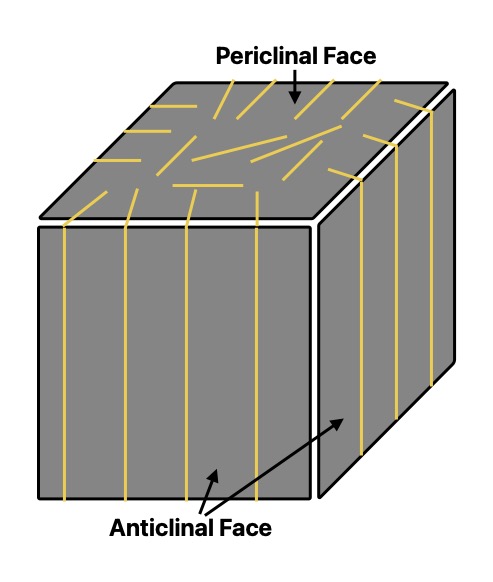}
  \caption{Sketch of our approximation of the cell as a polyhedral prism (top: periclinal, sides: anticlinal) with our approximation of the ``picket fence'' phenomena, where microtubules (MTs) in the anticlinal faces are aligned in a perpendicular array pattern \cite{shaw_reorg}. Newly created MTs forming in the anticlinal face or those aligned to existing MT orientation in the anticlinal faces are likely oriented to enter the periclinal faces at angles perpendicular to the edge with some variance \cite{shaw_reorg}.}
  \label{fig:picket_fence}
\end{figure}

In terms of CMA orientation, a common hypothesis for MT array pattern formation is ``survival of the aligned'' \cite{aligned_survival}, where lower rates of crossover and higher rates of zippering lead to an organized array, and MTs that are created but do not align do not survive. A common pattern in the CMA is observed band formation \cite{band_fromation} where several bands form along the cell face's horizontal or vertical axis, indicating the network has fully wrapped around the surface in that alignment configuration. There is also a ``picket fence'' phenomenon, whereby MTs forming and aligned to existing MT orientation in the anticlinal faces are oriented to enter the periclinal faces at angles perpendicular to the edge, with some variance \cite{picket_fence}. The entering MTs are, to an extent, aligned in parallel arrays in the anticlinal plane \cite{shaw_reorg}. A sketch of the approximation we make can be seen in figure \ref{fig:picket_fence}. There is also evidence showing that light can have an effect on the reordering of the array and can result in an influx of MTs from the lateral anticlinal walls in light-sensing mutants and hormone-stimulated plants \cite{light_reorg}.

%\FloatBarrier
\subsubsection{Model}
%\FloatBarrier
The Periclinal CMA (PCMA) DGG used extends and refines the previous CMA grammar\cite{Medwedeff_2023}. A full listing of all the grammar rules for the PCMA along with their corresponding parameters can be found in Appendix \ref{pcma_dgg}. We also note that the CMA model used artificial parameters and this new PCMA model uses more bio-inspired parameters, so the parameter regimes for their results are not clearly comparable. 

In this section, we highlight several of the grammatical distinctions. For example, we have added an alternate zippering rule that does not zipper directly into the extant microtubule (MT) graph segment (i.e. have the snap on action) but rather functions as an entrainment rule with a stand-off separation distance \cite{sep_dist}. By introducing this alternative zippering rule, MTs are now capable of ``piling up'' and forming high alignment on the local scale for the array. A zippering guard rule has been added to act as an approximation to bundling proteins \cite{update_cma}, to prevent a funneling effect that can otherwise occur. Essentially, another MT can enter the gap between zippered MTs unless there is a rule to prevent it. 

The previous simulation of the CMA DGG started with a fixed number of MTs that did not change over time. The PCMA now includes nucleation and destruction of MTs to simulate full depolymerization and removal from the system. For nucleation, we have included a grammar rule for uniformly random nucleation of MTs in the background using a nucleation rate \cite{allard_sim} functioning independently from the existing array of MTs. We have not, however, included a rule for MT-dependent nucleation, where new MTs are nucleated on existing MTs \cite{NucleateMurata2005, NucleateNakamure2009} or MT creation by means of severing at crossover site \cite{crossover_severing}. In our previous CMA model \cite{Medwedeff_2023} the closest comparison to nucleation was what occurred when an MT depolymerized. In that case, it would enter a dormant state until a rescue occurred to effectively change the state of one of the ends back to a growing node.

\begin{figure}[!ht]
  \centering
  \includegraphics[width=.65\linewidth]{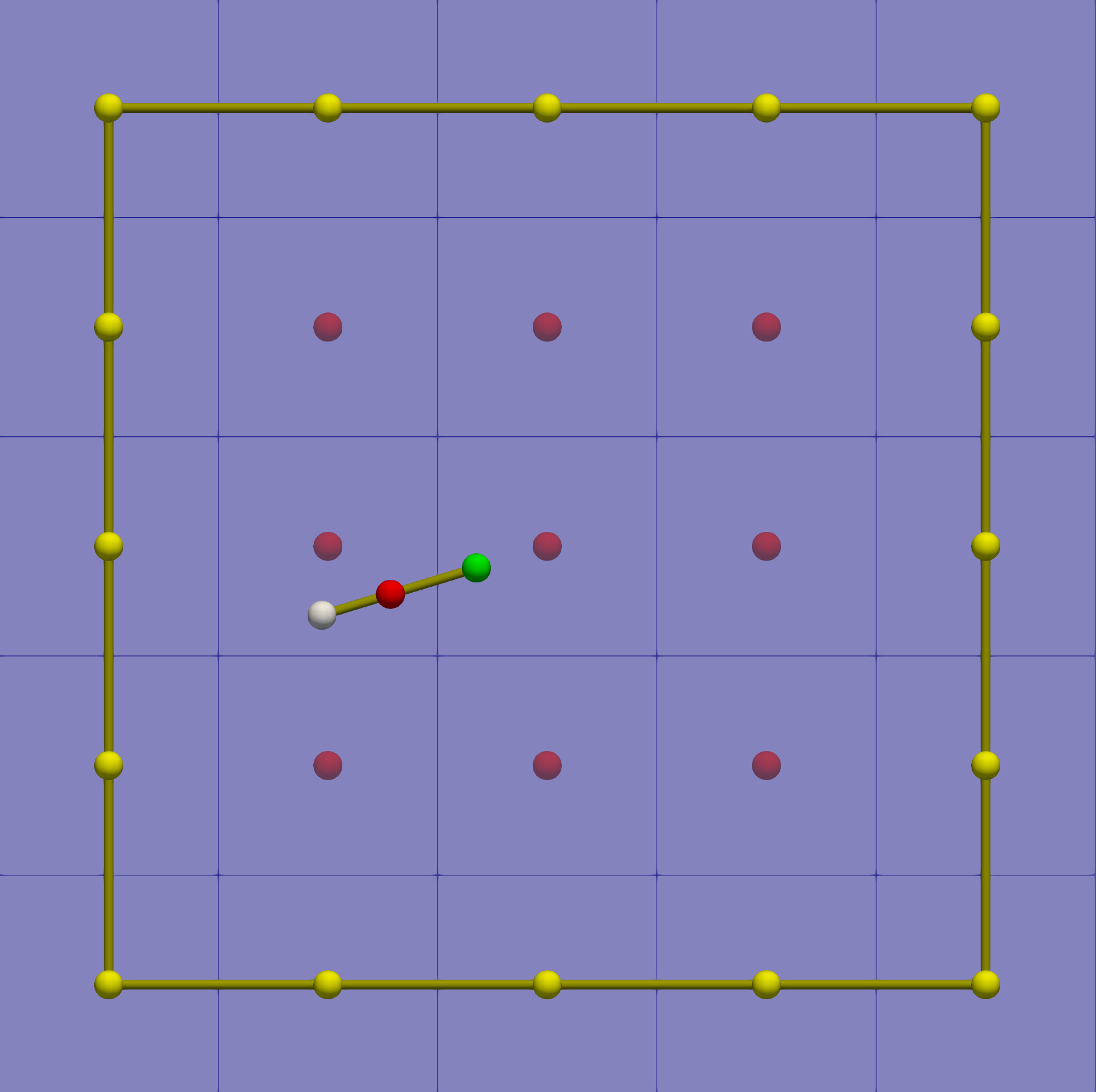}
  \caption{System after a short time of simulation. Nucleation points are shown in the interior, and the exterior graph is the boundary graph. A single microtubule exists within the simulation box.}
  \label{fig:boundary_node_on}
\end{figure}

We have also modified how the boundary works. In the previous model, the boundary worked by implicitly exploiting the functionality of the simulator in ignoring any MTs that moved outside the simulation space, effectively functioning to capture the MTs. In the new model, the boundary is now explicitly defined as a graph. An example of this can be seen in figure \ref{fig:boundary_node_on}. To prevent any MTs from leaving the interior of the space bounded by the boundary graph, an induced catastrophe rule has been added. Alternatively, the ODE in the deterministic growth rule could have an additional constraint that decays the velocity to zero when a boundary is reached, which would work as another route for using the DGG to specify boundary conditions.  

The Periclinal CMA (PCMA) also includes new special boundary rules, to approximate the effects of CLASP in the case of high curvature on the boundary between face transitions. Biologically, there are different types of CLASP, but for the example simulation study, we use a generic CLASP as a means to explore one of its functions, the ability to mediate MT polymerization from the periclinal into the nearest anticlinal face in the context of an idealized plant cell. In reality, CLASP likely can localize to only parts of the edge \cite{allard_sim}, but we make the assumption that CLASP has localized uniformly on the edge. When an MT reaches the boundary, two outcomes may occur: the MT will cross over to the periclinal plane mediated by CLASP or it will collide and induce catastrophe. 

MTs can also fully depolymerize from the periclinal plane into the anticlinal, and this is modeled as a rewrite rule that removes the depolymerizing MT attached to the boundary from the system. There is also a rule to allow the MT to unstick from the boundary which models the collision of an MT within the anticlinal face, and encapsulates the unknown dynamics that play out in this ``hidden'' plane. One of these dynamics is the ``picket fence'' phenomena mentioned in the background section and illustrated in figure \ref{fig:picket_fence}. 
\FloatBarrier

\subsubsection{Design and Measurements}
The results are divided into three experiments. In the first experiment, we have a collision-induced catastrophe (CIC) boundary, where any microtubule (MT) that collides with the boundary will begin to rapidly depolymerize. The CIC boundary is used as a baseline for the default behavior for predicting the alignment axis and the remaining experiments investigate perturbations of the system through means of rule changes and minor parameter variations. In the second experiment, we enable the CLASP-mediated crossover boundary conditions to stabilize the array at the boundaries and explore the effects of high MT crossover rate on the respective array. By high we mean we change the initial rate found in Appendix \ref{pcma_dgg} from $200$ to $8,000$ (a fortyfold increase), which makes crossover nearly as likely as CIC (which has a default rate of $12,000$). In the third experiment, we introduce an influx of MTs from the boundary and increase the initial rate of $0.001$ to $0.008$, meaning new MTs enter the system $8$ times as often.

Each of the individual experiments is run as ensembles of sixteen simulations using DGGML. Unless otherwise specified, the experiment will use the default parameters. For the grammar rules and the default parameters, please see Appendix \ref{pcma_dgg}. Each of the simulations is run for two hours of biological time, and checkpoints are taken every 24 seconds of that time. At each step, the angles of all intermediate, retracting, and growing MT nodes are collected and projected to the right half of the plane with $0^\circ$ being the positive x-axis ($-90^{\circ}$ (negative y-axis) to $90^{\circ}$ (positive y-axis)). These projected angles are binned by increments of $15^{\circ}$, creating a histogram \cite{orientation_angle_measure}. The correlation length vs. distance histogram is also computed at each checkpoint. From the correlation length vs. distance histogram, we take the average over the first third of the correlation vs distance function to get the local correlation average, and over its entirety to get the global average. While there are other measures of alignment \cite{angular_cost}, the local and global mean correlations work as a sufficient measure to quantify alignment when combined with the histogram. 

We compare how aligned the two ending states are by by computing an MT orientation correlation function defined as the squared cosine between the orientation of the MT segments, and average within bins of roughly constant distance. (This measure can be derived \cite{Medwedeff_2023} as the trace of the product of the two rank-one projection matrices defined by the two unit vectors; it is invariant to sign reversals of these unit vectors. We then have: \begin{equation}
    \text{Tr}((u_i \otimes u_i) \cdot (u_j \otimes u_j)) =  (u_i \cdot u_j)^2 = \text{cos}^2(\theta_{ij})\text{.}
\end{equation} The function measures on average how ``aligned'' MT segments a distance away are from one another. The square is needed to remove anti-symmetry, since nearby MTs may be aligned but in anti-parallel directions, but anti-parallel alignment is not visibly distinguishable from parallel alignment in typical MT imagery. Values close to $0.5$ using this measure indicates orthogonality and therefore no alignment, whereas values close to $1$ indicate complete parallel or anti-parallel alignment. 

\subsubsection{Boundary conditions may reorient arrays in rectangular domains}
\begin{figure*}[!ht]
    %\centering
    \includegraphics[width=0.75\linewidth]{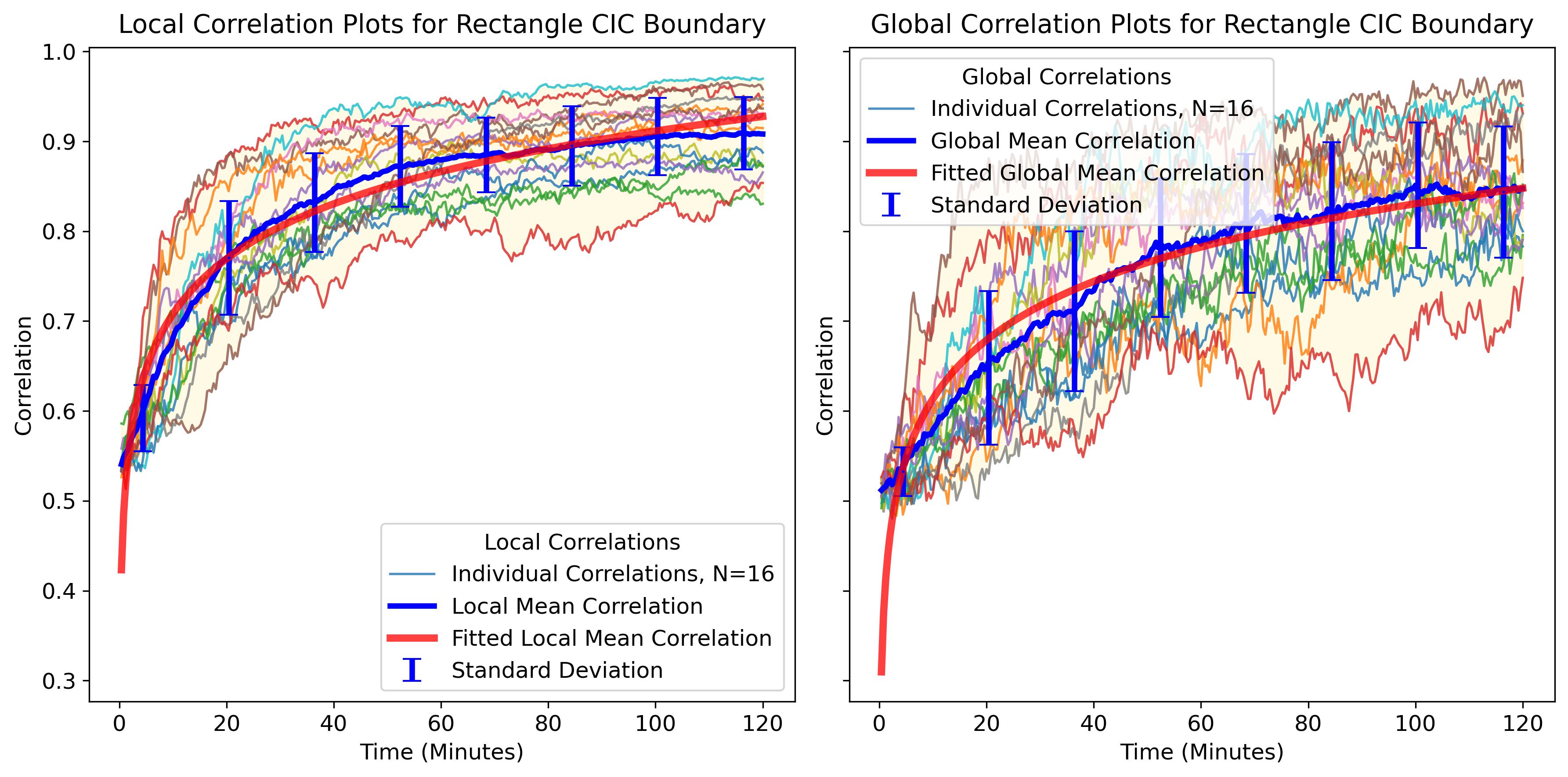}
   \caption{For the rectangular experiment with CIC Boundary, the plot contains the following: all $N=16$ samples of local correlations (left) and global correlations (right), the corresponding mean of all the samples with the standard deviations, and the best-fit logarithmic curves ($f(t) = a\log(bt)+c$) for the mean on the $120$ minute range of interest.}
    \label{fig:curve_fit_rect}
\end{figure*}

The area of the simulated rectangular domain is initially empty and has dimensions of $8.33\mu\text{m} \times 3\mu\text{m}$. For each of the experiments, we have $N = 16$ samples of the local and global correlations. In figure \ref{fig:curve_fit_rect} we present the rectangular CIC boundary experiment data used to calculate the mean local and global correlations.

\begin{figure}[!ht]
  \centering
  \includegraphics[width=.65\linewidth]{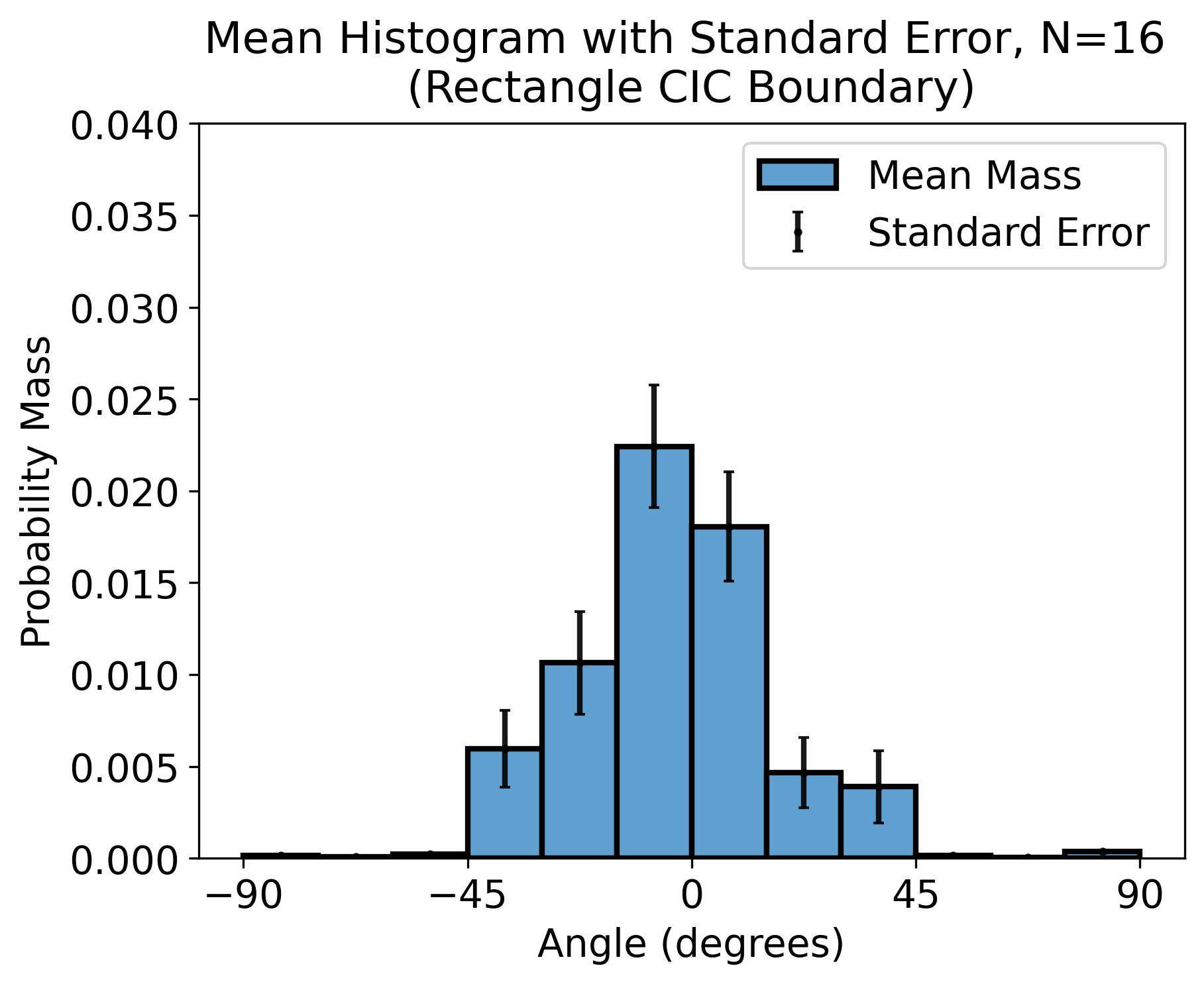}
  \caption{Histogram estimating the probability mass function for array orientation of rectangular domain with CIC boundary, including standard error bars and averaged over $16$ runs.}
  \label{fig:rect1_hist1}
\end{figure}

\begin{figure}[!ht]
  \centering
  \includegraphics[width=.65\linewidth]{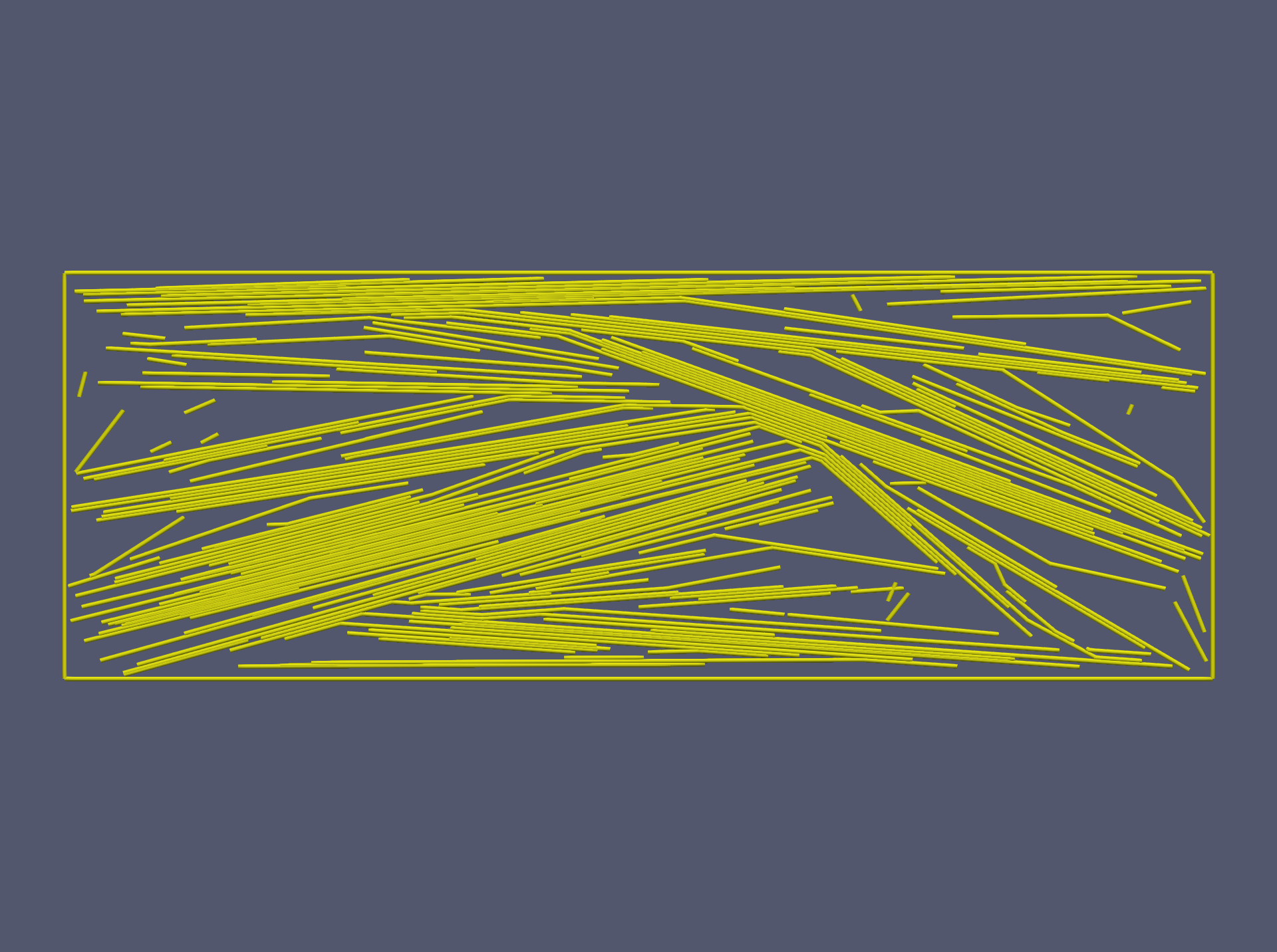} 
  \caption{Example of one of the samples used to calculate the array orientation. The array is biased towards being oriented horizontally, with some variation. Local correlation $= 0.93$ and global correlation  $= 0.83$.}
  \label{fig:rect1_sim1}
\end{figure}

The defining alignment behavior for the CIC boundary experiments is illustrated in figures \ref{fig:rect1_hist1} and \ref{fig:rect1_sim1} by a mean histogram and a sample from the data used to compute it. The histogram in figure \ref{fig:rect1_hist1} prominently centers around $0^{\circ}$, which indicates a high occurrence of angles oriented in the horizontal (long axis) direction. The sample array in figure \ref{fig:rect1_sim1} is mostly aligned horizontally, with some deviation towards the diagonals, which can be seen in the histogram as well. The local alignment is $0.93$ and the global is $0.83$. We found that given enough time, under these conditions, the array orientation may be biased towards uni-modal behavior, and it has a tendency to be oriented horizontally i.e. the long direction of the rectangle. Given a CIC at the boundary, the array likely orients horizontally because the longest axis allows the MTs to survive for longer indicating potential verification of a key property in CMA experiments, ``survival of the aligned''\cite{aligned_survival}.   

\begin{figure}[!ht]
  \centering
  \includegraphics[width=.65\linewidth]{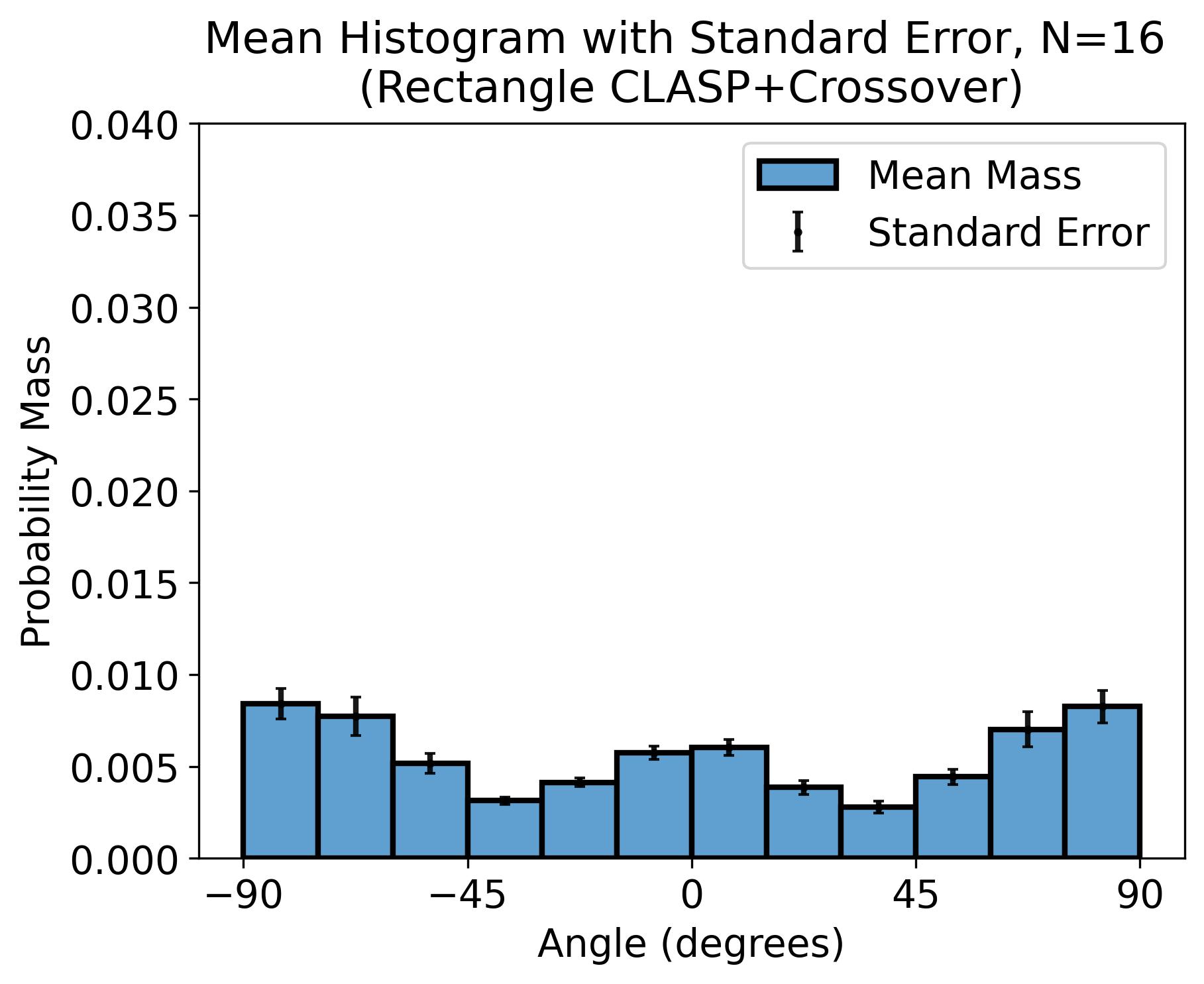}
  \caption{Histogram estimating the probability mass function for array orientation of rectangular domain with a CLASP boundary and a crossover rate of $8,000$, including standard error bars and averaged over $16$ runs.}
  \label{fig:rect1_hist2}
\end{figure}

\begin{figure}[!ht]
  \centering
  \includegraphics[width=.65\linewidth]{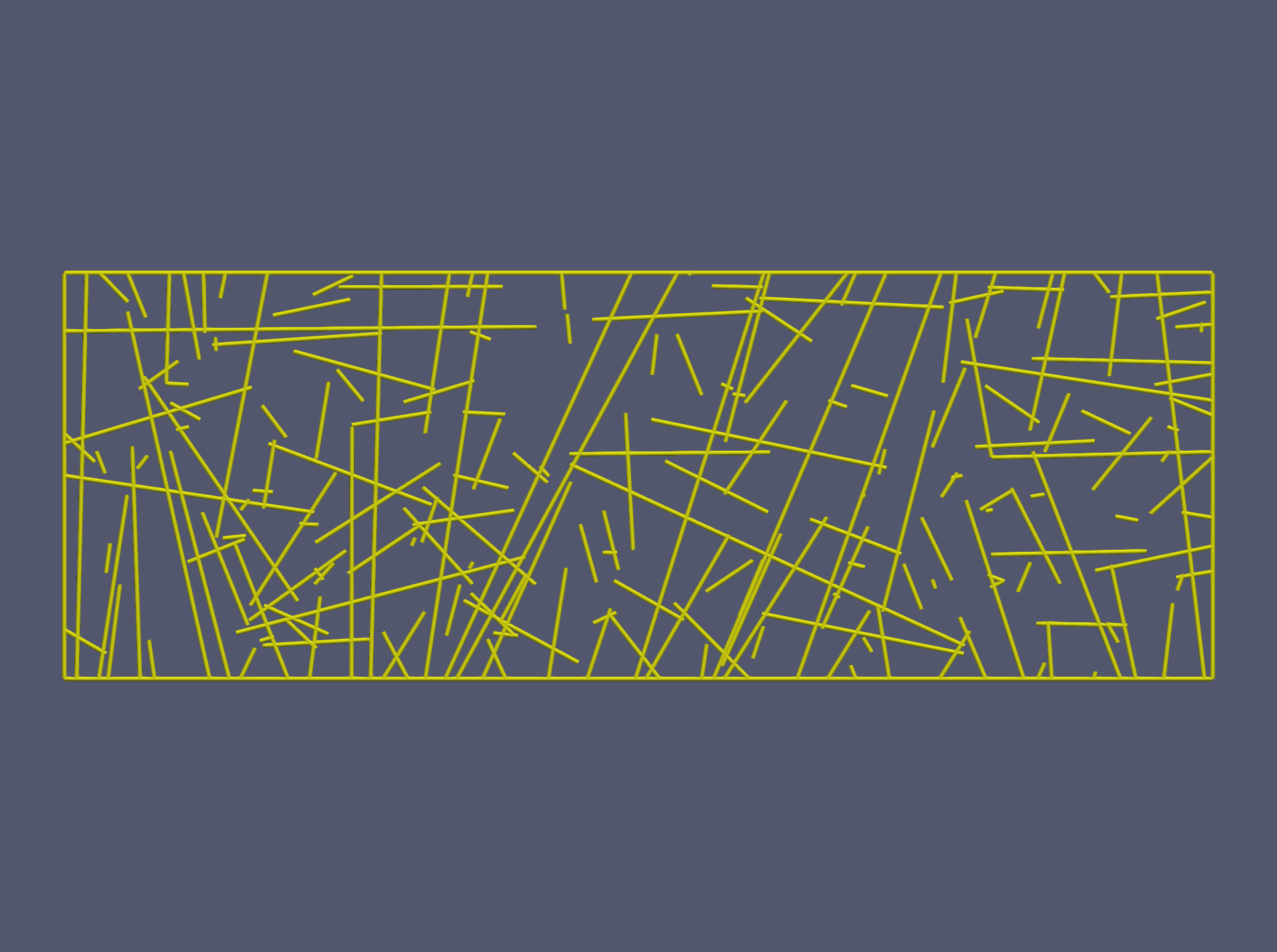} 
  \caption{Example of one of the samples used to calculate the array orientation. The array has no preferred orientation as has network-like behavior. Local correlation $= 0.5$ and global correlation $= 0.49$}
  \label{fig:rect1_sim2}
\end{figure}

Given the uni-modal behavior of this baseline case, we attempt to see if we can modulate (switch) it to another mode. We have increased the rate of crossover from $200$ (default value in Appendix \ref{pcma_dgg}) to $8,000$ and enabled the CLASP boundary conditions to stabilize the edges of the array. The resulting histogram in figure \ref{fig:rect1_hist2} is markedly flatter than the histogram for the CIC case. In figure \ref{fig:rect1_sim2} we have a sample ending state of the simulation used to calculate the histogram. It has a local correlation $= 0.5$ and a global correlation $= 0.49$. The MTs in the array for the sample are completely uncorrelated and a network-like behavior can be observed.

\begin{figure}[h]
  \centering
  \includegraphics[width=.65\linewidth]{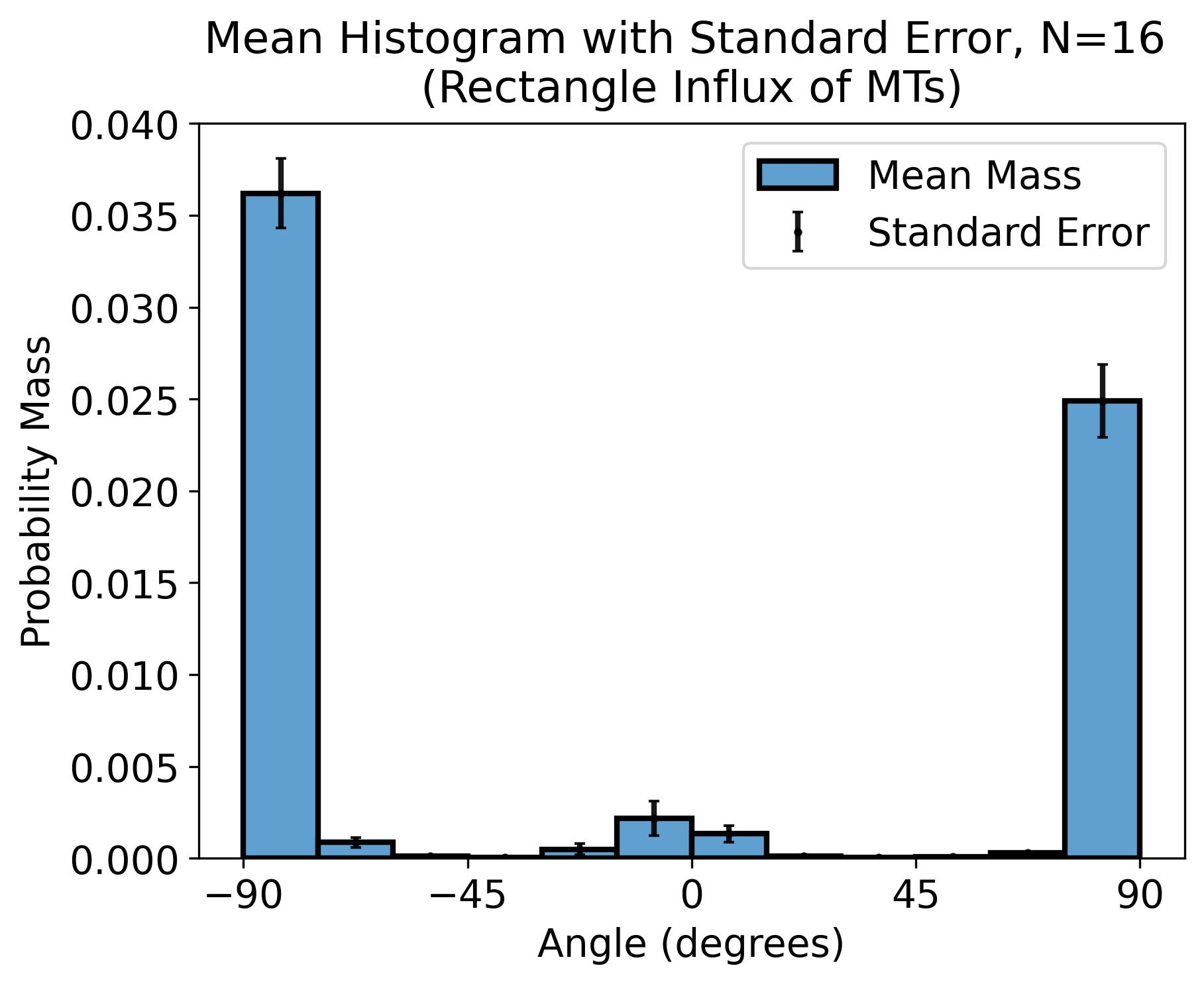}
  \caption{Histogram estimating the probability mass function for array orientation of rectangular domain with a CLASP boundary and an influx of microtubules entering, including standard error bars and averaged over $16$ runs.}
  \label{fig:rect1_hist3}
\end{figure}

\begin{figure}[h]
  \centering
  \includegraphics[width=.65\linewidth]{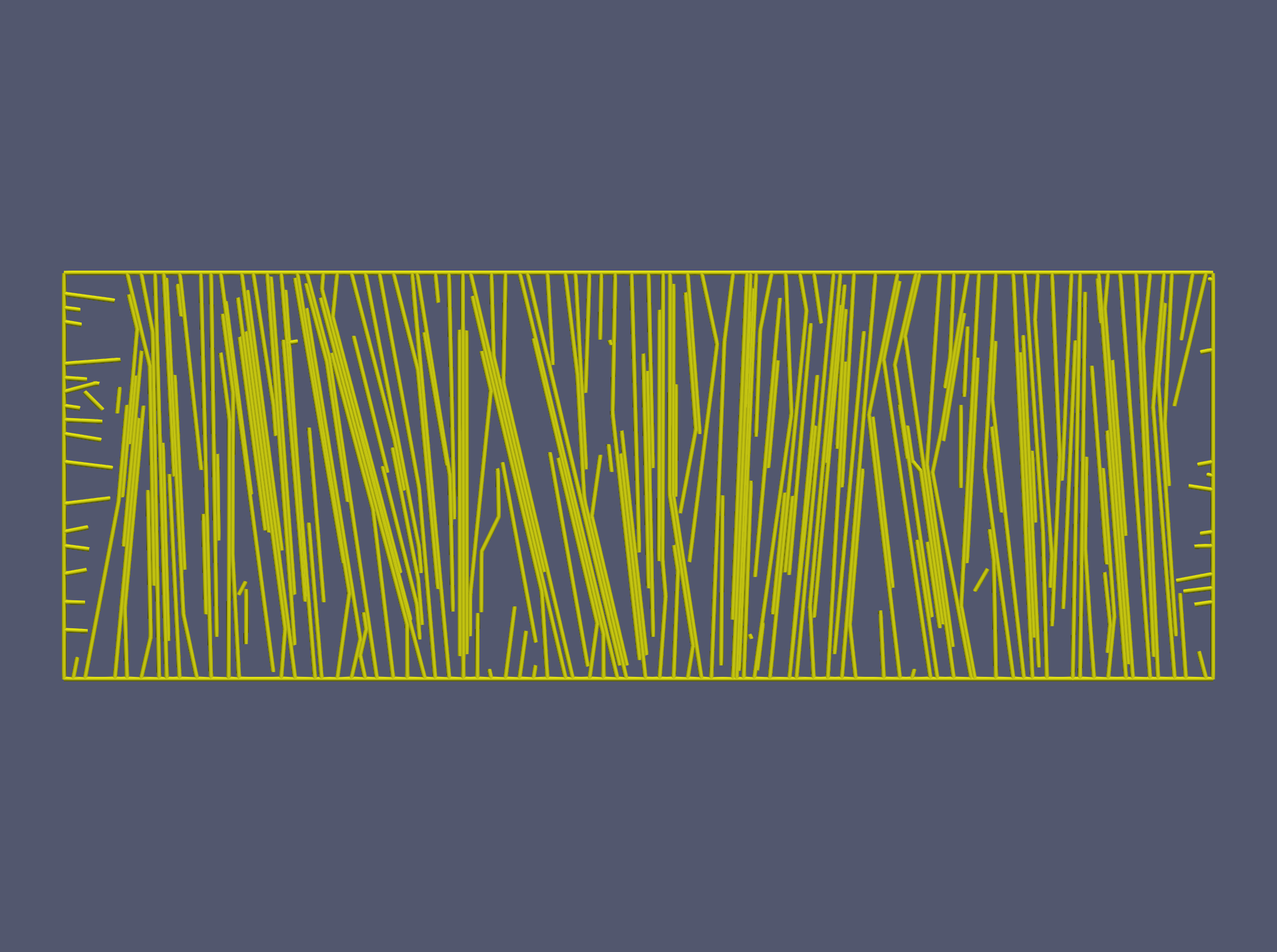} 
  \caption{Example of one of the samples used to calculate the array orientation. The array is oriented vertically. Local correlation $= 0.94$, global correlation $= 0.86$.}
  \label{fig:rect1_sim3}
\end{figure}

In the third experiment, we activate an influx of microtubules (by changing the influx rate from $0.001$ to $0.008$) on the boundaries and lower crossover back to the default rate of $200$. Microtubules (MT)s can still exit the simulation through the CLASP exit rule as well. We have another comparison of the mean histogram for the ending state of the realizations of the experiment, along with a sample. In the histogram in figure \ref{fig:rect1_hist3}, we observe the angle orientations have switched to be clustered around $-90^{\circ}$ and $90^{\circ}$. In this case, this indicates a vertical alignment i.e. aligned with the shortest side of the rectangle (projecting the angles to the right half of the plane causes vertical alignment to appear this way in the histogram). The sample shown in figure \ref{fig:rect1_sim3} verifies the behavior. The sample has a local correlation $= 0.94$ and a global correlation $= 0.86$. In the rectangular domain, the influx of MTs is enough to allow for the majority of MTs on the longer axis to reach the other side first and have a chance to cross over into the anticlinal face. Once they do, they will stay attached until periodically released from the boundary to approximate collision in the hidden plane and any other potential severing effects at the boundary not included in the model. Any newly nucleated MTs will then be forced to align with the dominant axis and entering MTs in the non-dominant axis will be unable to overcome the effects of alignment. 

As a future path forward for the experimental setup outlined, we could instead simulate the top periclinal face of the cell using different domain shapes. For example, we could use a hexagonal face for the simulation space. As a further variation, different stretching and skewing transformations could be applied the the cell face. More experiments can also be added with different variations of parameters and rules as well. For example, we could introduce the severing of MTs at crossover sites with a Katanin rule \cite{katanin_mulder} or add in a stochastic wobble to growth direction, to model thermal fluctuations \cite{mt_thermal}. 

\subsection{Performance Comparison: Trading Expressive Power}
The previous implementation, CajeteCMA \cite{Medwedeff_2023} used Algorithm \ref{alg:original_approx} and was hard coded for a specific cortical microtubule array (CMA) DGG. DGGML, which uses Algorithm \ref{alg:approx}, on the other hand, is more expressive and can be run with different grammars. However, expressivness does not always come without out a computational cost. The cost is evident the original DGG formalism implementation of Plenum \cite{yosiphon_2009}. To compare this potential cost between CajeteCMA and DGGML, we have run a simple treadmilling model and tuned the parameters to one another. Effectively, all rules are turned off with the exception of stochastic growth and retraction, and deterministic growth and retraction, and all use the same associated parameters as in Appendix \ref{pcma_dgg}. 

Each simulation starts with 320 MTs uniformly distributed in a $10\mu\text{m} \times 10\mu\text{m}$ domain. For both CajeteCMA and DGGML we run with two subdivisions: $1\times1$ (equivalent to the exact algorithm) and $3\times3$ (equivalent to the approximate algorithm). The results can be found in table \ref{tab:compare}.

\begin{table}[ht]
\centering
\begin{tabular}{|c|c|c|}
\hline
\multicolumn{3}{|c|}{Runtimes} \\ \hline
Decomposition/Code & CajeteCMA & DGGML \\ \hline
$1 \times 1$ subdivision  &     6 min 37 sec     &   5 min 32 sec    \\ \hline
$3 \times 3$ subdivision &         17 sec  &    49 sec   \\ \hline  

\end{tabular}
\caption{Run-times for the simulations of the treadmilling model.}
\label{tab:compare}
\end{table}

On the $1\times1$ subdivision we see that DGGML performs slightly better than CajeteCMA. On the $3\times3$, CajeteCMA performs better than DGGML. Given the results for no subdivision (i.e. the the special case where the algorithm runs exactly), it may be that DGGML's true performance is equal to or better than CajeteCMA since it implements the incremental update. Currently, DGGML isn't fully optimized, and that may explain performance discrepancies when subdividing the domain. However, there still may be an unknown computational cost from the expressiveness in DGGML that would persist even if it were to outperform CajeteCMA on each of their respective implementations of the approximate algorithm. 

\section{Discussion and Conclusion}
In this work we introduced the Dynamical Graph Grammar (DGG) formalism and the connection it has to chemical kinetics. However, DGGs extend beyond chemical systems and are capable of simulating and modeling a broader class of complex dynamic systems, including spatially extended or structured objects such as cytoskeleton, by using graphs. An exact algorithm can be derived from a master equation to simulate a single stochastic trajectory of a DGG model, but it becomes slow for large systems, so work was done to improve the performance by introducing an approximate algorithm \cite{Medwedeff_2023} (Algorithm \ref{alg:original_approx}) to speed up the simulation at the cost of some errors and events occurring out of order. The algorithm itself was demonstrated by using a hard coded implementation of a cortical microtubule array (CMA) model, CajeteCMA. 

The approximate algorithm has been improved upon (Algorithm \ref{alg:approx}) in this work by introducing a match data structure, incremental update, and periodic rule recomputation. Additionally, the $\varphi$ rule-instances to domains mapping function, graph transformation, grammar analysis, and pattern matching, as well as the details on how they translate to implementation where necessary, have been detailed. These methods are demonstrated by the Dynamical Graph Grammar Modeling Library (DGGML), which is a generalized implementation and far more expressive than the model-specific and rigid implementation CajeteCMA. The improvement to the approximate algorithm (Algorithm \ref{alg:approx}) and an example of its implementation in the form of DGGML represents an example of foundational work for the simulation and modeling of complex systems in a generalized way. Future work on this algorithm would be able to bound the theoretical error incurred, by the commutator. Using the mathematical framework of the DGG formalism, automated theorem provers \cite{lean} could take advantage of the operator algebra to derive new algorithms or mathematically bound errors in Algorithm \ref{alg:approx}.

 DGGML was used to implement a plant cell periclinal microtubule cortical array (PCMA) DGG, and we were able to find that different boundary conditions and the shape of the plant cell face may have an effect on local and global alignment of the array. For a rectangular face shape, different boundary conditions are capable of reorienting the array between the long and short axes, which affects cell division direction and cellular response to mechanical stress. The computational nature of the PCMA DGG outlined in this work also makes collecting more measurements of the system, which are often difficult to collect during lab experiments, more feasible and can effectively enable a more sophisticated analysis of the system that would not otherwise be available. For example, we were able to collect MT angular orientations by just defining a metric collection function. An added benefit of tools like DGGML is that they can be used to computationally screen biological or other experiments that are too costly or cumbersome. DGG models defined using and simulated by DGGML can also be used to search for corresponding biological experiments that would distinguish between and test alternative hypotheses for emergent phenomena. There is also potential to reduce the model or learn governing equations for emergent behavior by using machine learning \cite{ernst, scott}.

The potential applications for Dynamical Graph Grammars are vast, because many structures and systems in biology, chemistry, and other fields can be represented as dynamic graphs. As the landscape for computing continuously changes and evolves, new implementations scaling with and accelerated by new hardware will be needed, and new, more complex models will be unlocked or discovered. The clearest future path for implementations of the Dynamical Graph Grammar Modeling Library and DGG research in general is the evolution to a Dynamical Graph Grammar compiler since the DGG formalism is a declarative modeling language, which would enable sophisticated analysis, automatic parallelization, and other optimizations to greatly improve performance and remove the burden of the complexities from user and make them a problem for the tool oriented researcher to solve. Such a compiler could target DGGML on the back end. If DGGML were combined with a modeling language front end and with machine learning for multi-scale model reduction, even more powerful models with intelligently learned graph dynamics could emerge with a broad range of applications in biology, chemical physics, and beyond.
% If in two-column mode, this environment will change to single-column format so that long equations can be displayed. 
%\begin{widetext}
%$$\mbox{put long equation here}$$
%\end{widetext}

% Figures should be put into the text as floats. 
% Use the graphics or graphics packages (distributed with LaTeX2e).
% Put the label that you will use with \ref{} command in the braces of the \label{} command.
%
% \begin{figure}
% \end{figure}

% \begin{table}
% \end{tabular}
% \end{table}

\begin{acknowledgments}
  Finally, this work was funded in part by the Human Frontiers Science Program grant HFSP—RGP0023/2018, U.S. NIH NIDA Brain Initiative grant 1RF1DA055668-01, and U.S. NIH National Institute of Aging grant R56AG059602, and the UCI Donald Bren School of Information and Computer Sciences. This work was supported in part by the UC Southern California Hub, with funding from the UC National Laboratories division of the University of California Office of the President. We would like to acknowledge valuable conversations with Jacques Dumais, Olivier Hamant, Elliot Meyerowitz, and Hina Arora.
\end{acknowledgments}

\section*{Author Declarations}
\subsubsection*{Conflict of Interest}
The authors have no conflicts to disclose.

\subsubsection*{Author Contributions}
\textbf{Eric Medwedeff}: Conceptualization (supporting); Formal analysis (supporting); Investigation (equal); Methodology (equal); Software (lead); Visualization (lead); Writing – original draft (lead); Writing – review and editing (supporting). 

\textbf{Eric Mjolsness}: Conceptualization (lead); Formal analysis (lead); Funding acquisition; (lead); Investigation (equal); Methodology (equal); Project administration Resources (lead); Supervision (lead); Writing – review and editing (lead).

\section*{Data Availability Statement}
The updated algorithm presented has been implemented in the Dynamical Graph Grammar Modeling Library (DGGML\cite{dggml}) and is freely downloadable from GitHub, and also includes the implementation of Periclinal Cortical Microtuble Array (PCMA) DGG model discussed in the results section of this work. The supporting library Yet Another Graph Library (YAGL\cite{yagl}) is also freely downloadable from GitHub. Any other data to support the findings of this study are available from the corresponding author upon reasonable request.

\appendix
\section{New PCMA DGG Rules} \label{pcma_dgg}
The PCMA DGG builds upon a previous CMA model\cite{Medwedeff_2023}. We include a listing of the rules and associated dynamics. 

\subsection{Growing Rules}
\begin{flalign} \label{rule2:cont_growth}
\nonumber &
\left(
\begin{diagram}[size=1em]
\;
\Circle_1 & \rLine & \CIRCLE_2 \\
\end{diagram}
\right)
\llangle 
(\text{${ \boldsymbol x}$}_1,\text{${ \boldsymbol u}$}_1)
(\text{${ \boldsymbol x}$}_2,\text{${ \boldsymbol u}$}_2)
\rrangle 
& \\ \nonumber
& \longrightarrow 
\left(
\begin{diagram}[size=1em]
\Circle_1 & \rLine & \CIRCLE_2
\end{diagram}
\right)
\llangle
(\text{${ \boldsymbol x}$}_1,\text{${ \boldsymbol u}$}_1),
(\text{${ \boldsymbol x}$}_2 + \text{${ d \boldsymbol x}$}_2,\text{${ \boldsymbol u}$}_2)
\rrangle 
& \\
& \quad \quad \text{\boldmath $\mathbf{solving}$} \quad
d\text{${\boldsymbol x}$}_2/dt = v_{plus} \; \text{${ \boldsymbol u}$}_2 & 
\end{flalign}

\begin{flalign}\label{rule2:disc_growth}
\nonumber & 
\left(
\begin{diagram}[size=1em]
\;
\Circle_1 & \rLine & \CIRCLE_2 \\
\end{diagram}
\right)
\llangle
(\text{${ \boldsymbol x}$}_1, \text{${ \boldsymbol u}$}_1), 
(\text{${ \boldsymbol x}$}_2, \text{${ \boldsymbol u}$}_2)
\rrangle 
& \\ 
\nonumber & \longrightarrow 
\left(
\begin{diagram}[size=1em]
\;
\Circle_1 &  \rLine & \Circle_3 & \rLine & \CIRCLE_2 \\
\end{diagram}
\right)
\llangle
(\text{${ \boldsymbol x}$}_1, \text{${ \boldsymbol u}$}_1), 
(\text{${ \boldsymbol x}$}_2, \text{${ \boldsymbol u}$}_2), 
(\text{${ \boldsymbol x}$}_3, \text{${ \boldsymbol u}$}_3)
\rrangle
& \\ 
\nonumber & \quad \quad \text{\boldmath $\mathbf{with}$} \ \
\hat{\rho}_{\text{grow}} \; H(\|\text{${ \boldsymbol x}$}_2 - \text{${ \boldsymbol x}$}_1\|; L_{\text{div}})
& \\ 
& \quad \quad \text{\boldmath $\mathbf{where}$} \ \ 
\begin{cases}
\text{${ \boldsymbol x}$}_3 = \text{${ \boldsymbol x}$}_2 - (\text{${ \boldsymbol x}$}_2-\text{${ \boldsymbol x}$}_1)\gamma & \\
\text{${ \boldsymbol u}$}_3 = \frac{\text{${ \boldsymbol x}$}_3-\text{${ \boldsymbol x}$}_2}{\|\text{${ \boldsymbol x}$}_3-\text{${ \boldsymbol x}$}_2\|}
\end{cases} & 
\end{flalign} \\ Note: in rule \ref{rule2:disc_growth} and any other occurrence, $H$ is short for Heaviside, as in table \ref{tab2:function}.

\subsection{Retraction Rules}
\begin{flalign}\label{rule2:cont_retract}
\nonumber & 
\left(
\begin{diagram}[size=1em]
\;
\blacksquare_1 & \rLine & \Circle_2
\end{diagram}
\right)
\llangle
(\text{${ \boldsymbol x}$}_{1}, \text{${ \boldsymbol u}$}_{1}), 
(\text{${ \boldsymbol x}$}_2, \text{${ \boldsymbol u}$}_2)
\rrangle
& \\ 
\nonumber & 
\longrightarrow
\left(
\begin{diagram}[size=1em]
\;
\blacksquare_1 & \rLine & \Circle_2
\end{diagram}
\right)
\llangle
(\text{${ \boldsymbol x}$}_{1}+d\text{${ \boldsymbol x}$}_{1}, \text{${ \boldsymbol u}$}_{1}), 
(\text{${ \boldsymbol x}$}_2, \text{${ \boldsymbol u}$}_2)
\rrangle
& \\ 
& \quad \quad \text{\boldmath $\mathbf{solving}$} \ \
d\text{${ \boldsymbol x}$}_{1}/dt = v_{{minus}}(L/L_{\text{min
}})\text{${ \boldsymbol u}$}_1 & 
\end{flalign}

\begin{flalign}\label{rule2:disc_retract}
\nonumber & 
\left(
\begin{diagram}[size=1em]
\;
\blacksquare_1 & \rLine & \Circle_2 & \rLine & \Circle_3
\end{diagram}
\right)
\llangle
(\text{${ \boldsymbol x}$}_{1}, \text{${ \boldsymbol u}$}_{1}), 
(\text{${ \boldsymbol x}$}_2, \text{${ \boldsymbol u}$}_2), 
(\text{${ \boldsymbol x}$}_3, \text{${ \boldsymbol u}$}_3)
\rrangle
& \\ 
\nonumber & \longrightarrow
\left(
\begin{diagram}[size=1em]
\;
\blacksquare_1 & \rLine & \Circle_3
\end{diagram}
\right)
\llangle
(\text{${ \boldsymbol x}$}_{1}, \text{${ \boldsymbol u}$}_{1}), 
\emptyset, 
(\text{${ \boldsymbol x}$}_3, \text{${ \boldsymbol u}$}_3)
\rrangle
& \\ 
& \quad \quad \text{\boldmath $\mathbf{with}$} \ \
\hat{\rho}_{retract} \; H(\|\text{${ \boldsymbol x}$}_2 - \text{${ \boldsymbol x}$}_1\|; L_{{min}}) & 
\end{flalign}

\subsection{Boundary Catastrophe Rules}
\begin{flalign}\label{rule2:bnd_cic_std}
\nonumber & 
\left(
\begin{diagram}[size=1em]
\;
\Circle_1 & \rLine & \CIRCLE_2 & \text{, } & \boxdot_3 & \rLine & \boxdot_4 
\end{diagram}
\right) 
\begin{array}{l}
\llangle
(\text{${ \boldsymbol x}$}_{1}, \text{${ \boldsymbol u}$}_{1}), 
(\text{${ \boldsymbol x}$}_2, \text{${ \boldsymbol u}$}_2), \\ 
(\text{${ \boldsymbol x}$}_3, \text{${ \boldsymbol u}$}_3)
(\text{${ \boldsymbol x}$}_4, \text{${ \boldsymbol u}$}_4)
\rrangle
\end{array}
& \\ 
\nonumber & \longrightarrow
\left(
\begin{diagram}[size=1em]
\;
\Circle_1 & \rLine & \blacksquare_2 & \text{, } & \boxdot_3 & \rLine & \boxdot_4 
\end{diagram}
\right)
\begin{array}{l}
\llangle
(\text{${ \boldsymbol x}$}_{1}, \text{${ \boldsymbol u}$}_{1}), 
(\text{${ \boldsymbol x}$}_2, \text{${ \boldsymbol u}$}_2),  \\ 
(\text{${ \boldsymbol x}$}_3, \text{${ \boldsymbol u}$}_3),
(\text{${ \boldsymbol x}$}_4, \text{${ \boldsymbol u}$}_4)
\rrangle
\end{array}
& \\ 
\nonumber & \quad \quad \text{\boldmath $\mathbf{with}$} \ \
\hat{\rho}_{bnd\_cic\_std} \; \Theta(M_d(x_2, x_3, x_4) \leq \sigma_{col}) & \\
& \quad \quad \text{\boldmath $\mathbf{where}$} \ \ 
\text{${ \boldsymbol u}$}_2 = -\text{${ \boldsymbol u}$}_2 & 
\end{flalign} \\ Note: in rule \ref{rule2:bnd_cic_std} and any other occurrence, $\Theta = \text{Heaviside}$ on a propositional True/False value as in table \ref{tab2:function}.

\begin{flalign}\label{rule2:bnd_cic_clasp}
\nonumber & 
\left(
\begin{diagram}[size=1em]
\;
\Circle_1 & \rLine & \CIRCLE_2 & \text{, } & \boxdot_3 & \rLine & \Circle_4 
\end{diagram}
\right)
\begin{array}{l}
\llangle
(\text{${ \boldsymbol x}$}_{1}, \text{${ \boldsymbol u}$}_{1}), 
(\text{${ \boldsymbol x}$}_2, \text{${ \boldsymbol u}$}_2),  \\
(\text{${ \boldsymbol x}$}_3, \text{${ \boldsymbol u}$}_3),  
(\text{${ \boldsymbol x}$}_4, \text{${ \boldsymbol u}$}_4)
\rrangle
\end{array}
& \\ 
\nonumber & \longrightarrow
\left(
\begin{diagram}[size=1em]
\;
\Circle_1 & \rLine & \blacksquare_2 & \text{, } & \boxdot_3 & \rLine & \Circle_4 
\end{diagram}
\right)
\begin{array}{l}
\llangle
(\text{${ \boldsymbol x}$}_{1}, \text{${ \boldsymbol u}$}_{1}), 
(\text{${ \boldsymbol x}$}_2, \text{${ \boldsymbol u}$}_2), \\
(\text{${ \boldsymbol x}$}_3, \text{${ \boldsymbol u}$}_3), 
(\text{${ \boldsymbol x}$}_4, \text{${ \boldsymbol u}$}_4)
\rrangle
\end{array}
& \\ 
\nonumber & \quad \quad \text{\boldmath $\mathbf{with}$} \ \
\hat{\rho}_{bnd\_cic\_clasp} \; \Theta(M_d(x_2, x_3, x_4) \leq \sigma_{col}) & \\
& \quad \quad \text{\boldmath $\mathbf{where}$} \ \ 
\text{${ \boldsymbol u}$}_2 = -\text{${ \boldsymbol u}$}_2 & 
\end{flalign}

\subsection{MT Collision Induced Catastrophe Rules}
\begin{flalign}\label{rule2:int_cic}
 & 
\left(
\begin{diagram}[size=1em]
\nonumber & \Circle_1 & \rLine & \CIRCLE_2 & \text{, } & \Circle_3 & \rLine & \Circle_4 & & 
\end{diagram}
\right)
\begin{array}{l}
\llangle
(\text{${ \boldsymbol x}$}_1, \text{${ \boldsymbol u}$}_1), 
(\text{${ \boldsymbol x}$}_2, \text{${ \boldsymbol u}$}_2),  \\
(\text{${ \boldsymbol x}$}_3, \text{${ \boldsymbol u}$}_3),
(\text{${ \boldsymbol x}$}_4, \text{${ \boldsymbol u}$}_4)
\rrangle
\end{array}
& \\ 
\nonumber & \longrightarrow
\left(
\begin{diagram}[size=1em]
\;
\Circle_1 & \rLine & \blacksquare_2 & \text{, } & \Circle_3 & \rLine & \Circle_4 
\end{diagram}
\right)
\begin{array}{l}
\llangle
(\text{${ \boldsymbol x}$}_{1}, \text{${ \boldsymbol u}$}_{1}), 
(\text{${ \boldsymbol x}$}_2, \text{${ \boldsymbol u}$}_2),  \\
(\text{${ \boldsymbol x}$}_3, \text{${ \boldsymbol u}$}_3),
(\text{${ \boldsymbol x}$}_4, \text{${ \boldsymbol u}$}_4)
\rrangle
\end{array}
& \\ 
\nonumber & \quad \quad \text{\boldmath $\mathbf{with}$} \ \
\begin{array}{l}
\hat{\rho}_{int\_cic} \; \Theta(M_d(x_2, x_3, x_4) \leq \sigma_{col}) \\ \times \Theta(\alpha_1 \geq 0) 
\; \Theta (M_\theta(u_2, x_2, u_3) \geq \theta_{cic}) 
\end{array} & \\
& \quad \quad \text{\boldmath $\mathbf{where}$} \ \ 
\text{${ \boldsymbol u}$}_2 = -\text{${ \boldsymbol u}$}_2 \text{, and } \alpha = I_p(x_2, u_2, x_3, x_4) & 
\end{flalign}

\begin{flalign}\label{rule2:grow_cic}
 & 
\left(
\begin{diagram}[size=1em]
\nonumber & \Circle_1 & \rLine & \CIRCLE_2 & \text{, } & \Circle_3 & \rLine & \CIRCLE_4 & & 
\end{diagram}
\right)
\begin{array}{l}
\llangle
(\text{${ \boldsymbol x}$}_1, \text{${ \boldsymbol u}$}_1), 
(\text{${ \boldsymbol x}$}_2, \text{${ \boldsymbol u}$}_2),  \\
(\text{${ \boldsymbol x}$}_3, \text{${ \boldsymbol u}$}_3),
(\text{${ \boldsymbol x}$}_4, \text{${ \boldsymbol u}$}_4)
\rrangle
\end{array}
& \\ 
\nonumber & \longrightarrow
\left(
\begin{diagram}[size=1em]
\;
\Circle_1 & \rLine & \blacksquare_2 & \text{, } & \Circle_3 & \rLine & \CIRCLE_4 
\end{diagram}
\right)
\begin{array}{l}
\llangle
(\text{${ \boldsymbol x}$}_{1}, \text{${ \boldsymbol u}$}_{1}), 
(\text{${ \boldsymbol x}$}_2, \text{${ \boldsymbol u}$}_2), \\
(\text{${ \boldsymbol x}$}_3, \text{${ \boldsymbol u}$}_3),
(\text{${ \boldsymbol x}$}_4, \text{${ \boldsymbol u}$}_4)
\rrangle
\end{array}
& \\ 
\nonumber & \quad \quad \text{\boldmath $\mathbf{with}$} \ \
\hat{\rho}_{grow\_cic} \; \Theta(M_d(x_2, x_3, x_4) \leq \sigma_{col}/2) & \\
& \quad \quad \text{\boldmath $\mathbf{where}$} \ \ 
\text{${ \boldsymbol u}$}_2 = -\text{${ \boldsymbol u}$}_2 & 
\end{flalign}

\begin{flalign}\label{rule2:retract_cic}
 & 
\left(
\begin{diagram}[size=1em]
\nonumber & \Circle_1 & \rLine & \CIRCLE_2 & \text{, } & \Circle_3 & \rLine & \blacksquare_4 & & 
\end{diagram}
\right)
\begin{array}{l}
\llangle
(\text{${ \boldsymbol x}$}_1, \text{${ \boldsymbol u}$}_1), 
(\text{${ \boldsymbol x}$}_2, \text{${ \boldsymbol u}$}_2),  \\
(\text{${ \boldsymbol x}$}_3, \text{${ \boldsymbol u}$}_3),
(\text{${ \boldsymbol x}$}_4, \text{${ \boldsymbol u}$}_4)
\rrangle
\end{array}
& \\ 
\nonumber & \longrightarrow
\left(
\begin{diagram}[size=1em]
\;
\Circle_1 & \rLine & \blacksquare_2 & \text{, } & \Circle_3 & \rLine & \blacksquare_4 
\end{diagram}
\right)
\begin{array}{l}
\llangle
(\text{${ \boldsymbol x}$}_{1}, \text{${ \boldsymbol u}$}_{1}), 
(\text{${ \boldsymbol x}$}_2, \text{${ \boldsymbol u}$}_2),  \\
(\text{${ \boldsymbol x}$}_3, \text{${ \boldsymbol u}$}_3),
(\text{${ \boldsymbol x}$}_4, \text{${ \boldsymbol u}$}_4)
\rrangle
\end{array}
& \\ 
\nonumber & \quad \quad \text{\boldmath $\mathbf{with}$} \ \
\hat{\rho}_{retract\_cic} \; \Theta(M_d(x_2, x_3, x_4) \leq \sigma_{col}/2) & \\
& \quad \quad \text{\boldmath $\mathbf{where}$} \ \ 
\text{${ \boldsymbol u}$}_2 = -\text{${ \boldsymbol u}$}_2 & 
\end{flalign}

\subsection{Crossover Rules}
\begin{flalign}\label{rule2:cross}
 & 
\left(
\begin{diagram}[size=1em]
\nonumber & \Circle_1 & \rLine & \CIRCLE_2 & \text{, } & \Circle_3 & \rLine & \Circle_4 & & 
\end{diagram}
\right)
\begin{array}{l}
\llangle
(\text{${ \boldsymbol x}$}_1, \text{${ \boldsymbol u}$}_1), 
(\text{${ \boldsymbol x}$}_2, \text{${ \boldsymbol u}$}_2), \\
(\text{${ \boldsymbol x}$}_3, \text{${ \boldsymbol u}$}_3),
(\text{${ \boldsymbol x}$}_4, \text{${ \boldsymbol u}$}_4)
\rrangle
\end{array}
& \\ 
\nonumber & \longrightarrow \ \
\left(
\begin{diagram}[size=1em]
& & & & & & \CIRCLE_6 \\
& & & & & \ldLine & \\
& &  & &\Circle_5 & & \\
&  & & \ldLine & & & \\
\Circle_3 & \rLine & \blacklozenge_2 & \rLine & \Circle_4 & &\\
& \ldLine & & & & &\\
\Circle_1 & & & & & & 
\end{diagram}
\right) 
\begin{array}{l}
\llangle
(\text{${ \boldsymbol x}$}_1, \text{${ \boldsymbol u}$}_1), 
(\text{${ \boldsymbol x}$}_2, \text{${ \boldsymbol u}$}_2), \\
(\text{${ \boldsymbol x}$}_3, \text{${ \boldsymbol u}$}_3), 
(\text{${ \boldsymbol x}$}_4, \text{${ \boldsymbol u}$}_4), \\
(\text{${ \boldsymbol x}$}_5, \text{${ \boldsymbol u}$}_2),
(\text{${ \boldsymbol x}$}_6, \text{${ \boldsymbol u}$}_2)
\rrangle
\end{array}
& \\ 
\nonumber & & \\
\nonumber & \quad \text{\boldmath $\mathbf{with}$} \ \
\begin{array}{l}
\hat{\rho}_{cross} \; \Theta(M_d(x_2, x_3, x_4) \leq \sigma_{col}) \\
\times \Theta (M_\theta(u_2, x_2, u_3) \geq \theta_{cross}) \\
\times \Theta(\alpha_1 \geq 0) \; \Theta(0 < \alpha_2 < 1)  
\end{array} & \\
&\quad \text{\boldmath $\mathbf{where}$} \ \
\begin{cases}
\alpha = I_p(x_2, u_2, x_3, x_4) \\
\text{${ \boldsymbol x}$}_2 = \text{${ \boldsymbol x}$}_4 + \alpha_2(\text{${ \boldsymbol x}$}_3 - \text{${ \boldsymbol x}$}_4) \text{, } \\
\text{${ \boldsymbol x}$}_5 = \text{${ \boldsymbol x}$}_2 + 0.01\;\text{${ \boldsymbol x}$}_2 \text{, } \\
\text{${ \boldsymbol x}$}_6 = \text{${ \boldsymbol x}$}_2 + 0.011\;\text{${ \boldsymbol x}$}_2 
\end{cases} & 
\end{flalign}

\begin{flalign}\label{rule2:uncross}
\nonumber & 
\left(
\begin{diagram}[size=1em]
& &  & &\Circle_3 & & &\\
&  & &  & \uLine & & &\\
\blacksquare_1 & \rLine & \Circle_2 & \rLine & \blacklozenge_5 & & &\\
& & & & \dLine & & & \\
& & & & \Circle_4 & & &
\end{diagram}
\right) 
\begin{array}{l}
\llangle
(\text{${ \boldsymbol x}$}_1, \text{${ \boldsymbol u}$}_1), 
(\text{${ \boldsymbol x}$}_2, \text{${ \boldsymbol u}$}_2), \\
(\text{${ \boldsymbol x}$}_3, \text{${ \boldsymbol u}$}_3), 
(\text{${ \boldsymbol x}$}_4, \text{${ \boldsymbol u}$}_4), \\
(\text{${ \boldsymbol x}$}_5, \text{${ \boldsymbol u}$}_5)
\rrangle
\end{array}
& \\ 
\nonumber & \longrightarrow \ \
\left(
\begin{diagram}[size=1em]
& &  & \Circle_3 & \rLine & \Circle_4 &  \\
\blacksquare_1 & \rLine & \Circle_2 & \rLine & \Circle_5 & & \\
\end{diagram}
\right) 
\begin{array}{l}
\llangle
(\text{${ \boldsymbol x}$}_1, \text{${ \boldsymbol u}$}_1), 
(\text{${ \boldsymbol x}$}_2, \text{${ \boldsymbol u}$}_2), \\
(\text{${ \boldsymbol x}$}_3, \text{${ \boldsymbol u}$}_3), 
(\text{${ \boldsymbol x}$}_4, \text{${ \boldsymbol u}$}_4), \\
(\text{${ \boldsymbol x}$}_5, \text{${ \boldsymbol u}$}_2)
\rrangle
\end{array}
& \\ 
\nonumber & & \\
 & \quad \text{\boldmath $\mathbf{with}$} \ \
\begin{array}{l}
\hat{\rho}_{\text{\text{uncross}}} \\ \times 
\Theta(|\text{${ \boldsymbol u}$}_2 \cdot \text{${ \boldsymbol u}$}_4| \leq 0.95 \text{ and } 
\|\text{${ \boldsymbol x}$}_4-\text{${ \boldsymbol x}$}_3\| \leq L_{div}) 
\end{array} & 
\end{flalign}

\subsection{Zippering Rules}
\begin{flalign}\label{rule2:zip_hit}
 & 
\left(
\begin{diagram}[size=1em]
\nonumber & \Circle_1 & \rLine & \CIRCLE_2 & \text{, } & \Circle_3 & \rLine & \Circle_4 & & 
\end{diagram}
\right)
\begin{array}{l}
\llangle
(\text{${ \boldsymbol x}$}_1, \text{${ \boldsymbol u}$}_1), 
(\text{${ \boldsymbol x}$}_2, \text{${ \boldsymbol u}$}_2), \\ 
(\text{${ \boldsymbol x}$}_3, \text{${ \boldsymbol u}$}_3),
(\text{${ \boldsymbol x}$}_4, \text{${ \boldsymbol u}$}_4)
\rrangle
\end{array}
& \\ 
\nonumber & \longrightarrow \ \
\left(
\begin{diagram}[size=1em]
&  & \Circle_3 & \rLine & & & \Circle_4 \\
& & \blacktriangle_2 & \rLine & \Circle_6 & \rLine & \CIRCLE_5  \\
& \ldLine & & & & & \\
\Circle_1 & & & & & & \\
\end{diagram}
\right) \begin{array}{l}
\llangle
(\text{${ \boldsymbol x}$}_1, \text{${ \boldsymbol u}$}_1), 
(\text{${ \boldsymbol x}$}_2, \text{${ \boldsymbol u}$}_2), \\
(\text{${ \boldsymbol x}$}_3, \text{${ \boldsymbol u}$}_3), 
\;\; (\text{${ \boldsymbol x}$}_4, \text{${ \boldsymbol u}$}_4), \\ 
 (\text{${ \boldsymbol x}$}_5, \text{${ \boldsymbol u}$}_5),
 (\text{${ \boldsymbol x}$}_6, \text{${ \boldsymbol u}$}_6)
\rrangle
\end{array} & \\ 
\nonumber & \quad \text{\boldmath $\mathbf{with}$} \ \
\begin{array}{l}
\hat{\rho}_{\text{\text{zip\_hit}}} \; \Theta(M_d(x_2, x_3, x_4) \leq \sigma_{sep}) \\
\times \Theta(M_\theta(u_2, x_2, u_3) \leq \theta_{crit})
\; \Theta(\alpha_1 \geq 0)
\end{array} & \\
&\quad \text{\boldmath $\mathbf{where}$} \ \
\begin{cases}
\alpha = I_p(x_2, u_2, x_3, x_4) \\
\text{${ \boldsymbol u}$}_2 = \text{${ \boldsymbol u}$}_3 \text{, } \\
\text{${ \boldsymbol u}$}_6 = \text{${ \boldsymbol u}$}_3 \text{, } \\
\text{${ \boldsymbol u}$}_5 = \text{${ \boldsymbol u}$}_3 \text{, } \\
\text{${ \boldsymbol x}$}_6 = \text{${ \boldsymbol u}$}_2 + 0.005 \; \text{${ \boldsymbol u}$}_3 \text{, } \\
\text{${ \boldsymbol x}$}_5 = \text{${ \boldsymbol u}$}_2 + 0.01 \; \text{${ \boldsymbol u}$}_3 
\end{cases} & 
\end{flalign}

\begin{flalign}\label{rule2:zip_guard}
 & 
\left(
\begin{diagram}[size=1em]
\nonumber & \Circle_1 & \rLine & \CIRCLE_2 &  \\ & \Circle_4 & \rLine & \blacktriangle_3 & \rLine & \Circle_5 & 
\end{diagram}
\right)
\begin{array}{l}
\llangle
(\text{${ \boldsymbol x}$}_1, \text{${ \boldsymbol u}$}_1), 
(\text{${ \boldsymbol x}$}_2, \text{${ \boldsymbol u}$}_2), \\
(\text{${ \boldsymbol x}$}_3, \text{${ \boldsymbol u}$}_3),
(\text{${ \boldsymbol x}$}_4, \text{${ \boldsymbol u}$}_4), \\
(\text{${ \boldsymbol x}$}_5, \text{${ \boldsymbol u}$}_5)
\rrangle
\end{array}
& \\ 
\nonumber & \longrightarrow \ \
\left(
\begin{diagram}[size=1em]
\nonumber & \Circle_1 & \rLine & \blacksquare_2 &  \\ & \Circle_4 & \rLine & \blacktriangle_3 & \rLine & \Circle_5 & 
\end{diagram}
\right) \begin{array}{l}
\llangle
(\text{${ \boldsymbol x}$}_1, \text{${ \boldsymbol u}$}_1), 
(\text{${ \boldsymbol x}$}_2, \text{${ \boldsymbol u}$}_2), \\
(\text{${ \boldsymbol x}$}_3, \text{${ \boldsymbol u}$}_3),
(\text{${ \boldsymbol x}$}_4, \text{${ \boldsymbol u}$}_4), \\
(\text{${ \boldsymbol x}$}_5, \text{${ \boldsymbol u}$}_5)
\rrangle
\end{array} & \\ 
\nonumber & \quad \text{\boldmath $\mathbf{with}$} \ \
\begin{array}{l}
\hat{\rho}_{\text{\text{zip\_guard}}} \; \Theta(\|\text{${ \boldsymbol x}$}_3-\text{${ \boldsymbol x}$}_2\| \leq \sigma_{sep}) \\
\end{array} & \\
&\quad \text{\boldmath $\mathbf{where}$} \ \
\begin{cases}
\text{${ \boldsymbol u}$}_2 = -\text{${ \boldsymbol u}$}_2
\end{cases} & 
\end{flalign}

\begin{flalign}\label{rule2:unzipper}
\nonumber & 
\left(
\begin{diagram}[size=1em]
\;
\blacksquare_1 & \rLine & \Circle_2 & \rLine & \blacktriangle_3
\end{diagram}
\right)
\llangle
(\text{${ \boldsymbol x}$}_{1}, \text{${ \boldsymbol u}$}_{1}), 
(\text{${ \boldsymbol x}$}_2, \text{${ \boldsymbol u}$}_2), 
(\text{${ \boldsymbol x}$}_3, \text{${ \boldsymbol u}$}_3)
\rrangle
& \\ 
\nonumber & \longrightarrow
\left(
\begin{diagram}[size=1em]
\;
\blacksquare_1 & \rLine & \Circle_3
\end{diagram}
\right)
\llangle
(\text{${ \boldsymbol x}$}_{1}, \text{${ \boldsymbol u}$}_{1}), 
\emptyset, 
(\text{${ \boldsymbol x}$}_3, \text{${ \boldsymbol u}$}_3)
\rrangle
& \\ 
& \quad \quad \text{\boldmath $\mathbf{with}$} \ \
\hat{\rho}_{retract} \; H(\|\text{${ \boldsymbol x}$}_2 - \text{${ \boldsymbol x}$}_1\|; L_{{min}}) & 
\end{flalign}

\subsection{CLASP Rules}
\begin{flalign}\label{rule2:clasp_entry}
 & 
\left(
\begin{diagram}[size=1em]
\nonumber & \boxdot_1 & \rLine & & & \boxdot_2  
\end{diagram}
\right)
\llangle
(\text{${ \boldsymbol x}$}_1, \text{${ \boldsymbol u}$}_1), 
(\text{${ \boldsymbol x}$}_2, \text{${ \boldsymbol u}$}_2)
\rrangle
& \\ 
\nonumber & \longrightarrow \ \
\left(
\begin{diagram}[size=1em]
\nonumber & \boxdot_1 & \rLine & \Circle_3 & \rLine & \boxdot_2 \\
& & & \uLine & & \\
& & & \CIRCLE_4 & & 
\end{diagram}
\right) \begin{array}{l}
\llangle
(\text{${ \boldsymbol x}$}_1, \text{${ \boldsymbol u}$}_1), 
(\text{${ \boldsymbol x}$}_2, \text{${ \boldsymbol u}$}_2), \\
(\text{${ \boldsymbol x}$}_3, \text{${ \boldsymbol u}$}_3), 
(\text{${ \boldsymbol x}$}_4, \text{${ \boldsymbol u}$}_4)
\rrangle
\end{array} & \\ 
\nonumber & \quad \text{\boldmath $\mathbf{with}$} \ \
\begin{array}{l}
\hat{\rho}_{\text{\text{claps\_entry}}}
\end{array} & \\
&\quad \text{\boldmath $\mathbf{where}$} \ \
\begin{cases}
\text{${ \boldsymbol x}$}_3 = (\text{${ \boldsymbol x}$}_1 + \text{${ \boldsymbol x}$}_2)/2 \text{, }
\text{${ \boldsymbol u}$}_4 = \text{${ \boldsymbol u}$}_3  \\
\theta_{rot} \sim \mathcal{U}(-\theta_{entry}, \theta_{entry}) \text{, } \\
\text{${ \boldsymbol x}$}_4 = rotate(\text{${ \boldsymbol x}$}_3+\sigma_{col} \; \text{${ \boldsymbol u}$}_3, \theta_{rot}) 
\end{cases} & \\ \nonumber
\end{flalign}

\begin{flalign}\label{rule2:clasp_exit}
 & 
\left(
\begin{diagram}[size=1em]
\nonumber & \boxdot_1 & \rLine & & & \boxdot_2 \\
& \Circle_3 & \rLine & \CIRCLE_4 & & 
\end{diagram}
\right)
\begin{array}{l}
\llangle
(\text{${ \boldsymbol x}$}_1, \text{${ \boldsymbol u}$}_1), 
(\text{${ \boldsymbol x}$}_2, \text{${ \boldsymbol u}$}_2), \\
(\text{${ \boldsymbol x}$}_3, \text{${ \boldsymbol u}$}_3), 
(\text{${ \boldsymbol x}$}_4, \text{${ \boldsymbol u}$}_4)
\rrangle
\end{array}
& \\ 
\nonumber & \longrightarrow \ \
\left(
\begin{diagram}[size=1em]
\nonumber & \boxdot_1 & \rLine & \Circle_3 & \rLine & \boxdot_2 \\
& & & \uLine & & \\
& & & \Circle_4 & & 
\end{diagram}
\right) \begin{array}{l}
\llangle
(\text{${ \boldsymbol x}$}_1, \text{${ \boldsymbol u}$}_1), 
(\text{${ \boldsymbol x}$}_2, \text{${ \boldsymbol u}$}_2), \\
(\text{${ \boldsymbol x}$}_3, \text{${ \boldsymbol u}$}_3), 
(\text{${ \boldsymbol x}$}_4, \text{${ \boldsymbol u}$}_4)
\rrangle
\end{array} & \\ 
\nonumber & \quad \text{\boldmath $\mathbf{with}$} \ \
\begin{array}{l}
\hat{\rho}_{\text{\text{clasp\_exit}}} \; \Theta(M_d(x_4, x_1, x_2) \leq \sigma_{col}) \\
\times \Theta(M_\theta(x_4, u_4, u_1) \leq \theta_{exit}) \\
\times \Theta(\alpha_1 \geq 0) \; \Theta(0 < \alpha_2 < 1)  
\end{array} & \\
&\quad \text{\boldmath $\mathbf{where}$} \ \
\begin{cases}
\alpha = I_p(x_4, u_4, x_1, x_2) \text{, } \\
\text{${ \boldsymbol x}$}_3 = \text{${ \boldsymbol x}$}_1 + \alpha_2(\text{${ \boldsymbol x}$}_1 - \text{${ \boldsymbol x}$}_2) 
\end{cases} & \\ \nonumber
\end{flalign}

\begin{flalign}\label{rule2:clasp_detach}
 & 
\left(
\begin{diagram}[size=1em]
\nonumber & \boxdot_1 & \rLine & \Circle_3 & \rLine & \boxdot_2 \\
& & & \uLine & & \\
& & & \Circle_4 & & 
\end{diagram}
\right)
\begin{array}{l}
\llangle
(\text{${ \boldsymbol x}$}_1, \text{${ \boldsymbol u}$}_1), 
(\text{${ \boldsymbol x}$}_2, \text{${ \boldsymbol u}$}_2), \\
(\text{${ \boldsymbol x}$}_3, \text{${ \boldsymbol u}$}_3), 
(\text{${ \boldsymbol x}$}_4, \text{${ \boldsymbol u}$}_4)
\rrangle
\end{array}
& \\ 
\nonumber & \longrightarrow \ \
\left(
\begin{diagram}[size=1em]
\nonumber & \boxdot_1 & \rLine & & & \boxdot_2 \\
& \blacksquare_3 & \rLine & \Circle_4 & & 
\end{diagram}
\right) \begin{array}{l}
\llangle
(\text{${ \boldsymbol x}$}_1, \text{${ \boldsymbol u}$}_1), 
(\text{${ \boldsymbol x}$}_2, \text{${ \boldsymbol u}$}_2), \\
(\text{${ \boldsymbol x}$}_3, \text{${ \boldsymbol u}$}_3), 
(\text{${ \boldsymbol x}$}_4, \text{${ \boldsymbol u}$}_4)
\rrangle
\end{array} & \\ 
 & \quad \text{\boldmath $\mathbf{with}$} \ \
\begin{array}{l}
\hat{\rho}_{\text{\text{clasp\_detach}}}
\end{array} & \\ \nonumber
\end{flalign}

\begin{flalign}\label{rule2:clasp_cat}
 &
\left(
\begin{diagram}[size=1em]
\nonumber & \boxdot_1 & \rLine & \Circle_3 & \rLine & \boxdot_2 \\
& & & \uLine & & \\
& & & \blacksquare_4 & & 
\end{diagram}
\right)
\begin{array}{l}
\llangle
(\text{${ \boldsymbol x}$}_1, \text{${ \boldsymbol u}$}_1), 
(\text{${ \boldsymbol x}$}_2, \text{${ \boldsymbol u}$}_2), \\
(\text{${ \boldsymbol x}$}_3, \text{${ \boldsymbol u}$}_3), 
(\text{${ \boldsymbol x}$}_4, \text{${ \boldsymbol u}$}_4)
\rrangle
\end{array}
& \\ 
\nonumber & \longrightarrow \ \
\left(
\begin{diagram}[size=1em]
\nonumber & \boxdot_1 & \rLine & & & \boxdot_2  
\end{diagram}
\right) \begin{array}{l}
\llangle
(\text{${ \boldsymbol x}$}_1, \text{${ \boldsymbol u}$}_1), 
(\text{${ \boldsymbol x}$}_2, \text{${ \boldsymbol u}$}_2), \\
(\text{${ \boldsymbol x}$}_3, \text{${ \boldsymbol u}$}_3), 
(\text{${ \boldsymbol x}$}_4, \text{${ \boldsymbol u}$}_4)
\rrangle
\end{array} & \\ 
 & \quad \text{\boldmath $\mathbf{with}$} \ \
\begin{array}{l}
\hat{\rho}_{\text{\text{clasp\_cat}}} \; \Theta(\|\text{${ \boldsymbol x}$}_4-\text{${ \boldsymbol x}$}_3\| \leq \epsilon)
\end{array} & \\ \nonumber
\end{flalign}

\subsection{Destruction and Creation Rules}
\begin{flalign}\label{rule2:destruct_1}
 &
\left(
\begin{diagram}[size=1em]
\nonumber & \blacksquare_1 & \rLine & \Circle_2 & \rLine & \blacksquare_3 \\
\end{diagram}
\right)
\llangle
(\text{${ \boldsymbol x}$}_1, \text{${ \boldsymbol u}$}_1), 
(\text{${ \boldsymbol x}$}_2, \text{${ \boldsymbol u}$}_2), 
(\text{${ \boldsymbol x}$}_3, \text{${ \boldsymbol u}$}_3)
\rrangle
\\ & \quad \quad \nonumber \longrightarrow 
\emptyset & &
& \\ 
& \quad \text{\boldmath $\mathbf{with}$} \ \
\begin{array}{l}
\hat{\rho}_{\text{\text{destruct}}} \; \Theta(\|\text{${ \boldsymbol x}$}_1-\text{${ \boldsymbol x}$}_2\| \leq s_{min}) \\ 
\times \Theta(\|\text{${ \boldsymbol x}$}_3-\text{${ \boldsymbol x}$}_2\| \leq s_{min})
\end{array} & \\ \nonumber
\end{flalign}

\begin{flalign}\label{rule2:create}
 &
\left(
\begin{diagram}[size=1em]
\nonumber  \boxslash_1 \\
\end{diagram}
\right)
\llangle
(\text{${ \boldsymbol x}$}_1)
\rrangle
\\ & \nonumber \longrightarrow 
\left(\begin{diagram}[size=1em]
\nonumber \boxslash_1 & \text{, } & \blacksquare_2 & \rLine & \Circle_3 & \rLine & \CIRCLE_ 4 \\
\end{diagram} \right) 
\begin{array}{l}
\llangle
(\text{${ \boldsymbol x}$}_1), 
(\text{${ \boldsymbol x}$}_2, \text{${ \boldsymbol u}$}_2), \\
(\text{${ \boldsymbol x}$}_3, \text{${ \boldsymbol u}$}_3),
(\text{${ \boldsymbol x}$}_4, \text{${ \boldsymbol u}$}_4)
\rrangle
\end{array} & &
& \\ 
 \nonumber & \quad \text{\boldmath $\mathbf{with}$} \ \
\begin{array}{l}
\hat{\rho}_{\text{\text{create}}} 
\end{array} & \\ 
\nonumber &\quad \text{\boldmath $\mathbf{where}$} \ \
\begin{cases}
\text{${ \boldsymbol x}$}_3 \sim \mathcal{U}([\text{${ \boldsymbol x}$}_{1x} - 0.5\epsilon\text{, } \text{${ \boldsymbol x}$}_{1x} + 0.5\epsilon] \\ \quad \times [\text{${ \boldsymbol x}$}_{1y} - 0.5\epsilon\text{, } \text{${ \boldsymbol x}$}_{1y} + 0.5\epsilon]) \\
\theta_{rot} \sim \mathcal{U}(0, 2\pi) \text{, and } d \sim \mathcal{U}(s_{min}, s_{max}) \\ \quad \text{ so }
\text{${ \boldsymbol x}$}_4 = rotate(\text{${ \boldsymbol x}$}_3 + d, \theta_{rot}) \\
\text{${ \boldsymbol x}$}_2 = \text{${ \boldsymbol x}$}_4 - 2d \text{, } \\
\text{${ \boldsymbol u}$}_4 = \frac{\text{${ \boldsymbol x}$}_3-\text{${ \boldsymbol x}$}_4}{\|\text{${ \boldsymbol x}$}_3-\text{${ \boldsymbol x}$}_4\|} \text{, } \\
\text{${ \boldsymbol u}$}_3 = \text{${ \boldsymbol u}$}_4 \text{, } \\ 
\text{${ \boldsymbol u}$}_2 = \text{${ \boldsymbol u}$}_4 
\end{cases} & \\
\end{flalign}

\begin{flalign}\label{rule2:destruct_2}
 &
\left(
\begin{diagram}[size=1em]
\nonumber & \blacksquare_1 & \rLine & \Circle_2 & \rLine & \CIRCLE_3 & \text{, } & \Circle_ 4 \\
\end{diagram}
\right)
\begin{array}{l}
\llangle
(\text{${ \boldsymbol x}$}_1, \text{${ \boldsymbol u}$}_1), 
(\text{${ \boldsymbol x}$}_2, \text{${ \boldsymbol u}$}_2), \\
(\text{${ \boldsymbol x}$}_3, \text{${ \boldsymbol u}$}_3),
(\text{${ \boldsymbol x}$}_4, \text{${ \boldsymbol u}$}_4)
\rrangle
\end{array}
\nonumber \longrightarrow 
\emptyset & &
& \\ 
 & \quad \text{\boldmath $\mathbf{with}$} \ \
\begin{array}{l}
\hat{\rho}_{\text{\text{destruct}}} \; \Theta(\|\text{${ \boldsymbol x}$}_4-\text{${ \boldsymbol x}$}_3\| \leq \sigma_{col}) 
\end{array} & \\ \nonumber
\end{flalign}

\subsection{State Changes}
\begin{flalign}\label{rule2:disc_state}
\nonumber &
\left(\begin{diagram}[size=1em]
\CIRCLE_1
\end{diagram}\right)
\llangle
\text{${ \boldsymbol x}$}_1, \text{${ \boldsymbol u}$}_1)
\rrangle 
\quad \longleftrightarrow \quad 
\left(\begin{diagram}[size=1.5em]
\blacksquare_1
\end{diagram}\right)
\llangle 
\text{${ \boldsymbol x}$}_1, \text{${ \boldsymbol u}$}_1)
\rrangle
 & \\ & \nonumber \quad \text{\boldmath $\mathbf{with}$} \ \
(\hat{\rho}_{\text{retract} \leftarrow \text{growth}}, \hat{\rho}_{\text{growth} \leftarrow \text{retract}}) & \\
&\quad \text{\boldmath $\mathbf{where}$} \ \
\begin{cases}
\text{${ \boldsymbol u}$}_1 = -\text{${ \boldsymbol u}$}_1
\end{cases} & \\ \nonumber
\end{flalign}
\FloatBarrier

\subsection{Parameters and Symbols} \label{sec:tables}
\begin{table}[!ht]
    \centering
    \begin{tabular}{|c|c|}
        \hline
        Rule Parameter & Description\\ 
        \hline
        $x_n$  &  a point in $\mathbb{R}^2$ \\
        \hline
        $u_n$  & a unit vector in $\mathbb{R}^2$ \\
        \hline
        $L$ & current length of the MT segment \\
        \hline
    \end{tabular}
    \caption{Rule Parameter Definitions}
    \label{tab2:rule_parameter}
\end{table}

\begin{table}[!ht]
    \centering
    \begin{tabular}{|c |c|} 
        \hline
        Graph Symbol & Type Name\\ 
        \hline
        $\CIRCLE$ & growing  \\ 
        \hline
        $\Circle$  & intermediate  \\
        \hline
        $\blacksquare$ & retraction  \\
        \hline
        $\blacktriangle$ & zipper  \\
        \hline
        $\blacklozenge$ & junction \\ 
        \hline
        $\boxslash$ & nucleator \\
        \hline
        $\boxdot$ & boundary \\
        \hline
    \end{tabular}
    \caption{Graph node type symbols for the periclinal cortical microtubule array (PCMA) grammar.}
    \label{tab2:graph_symbols}
\end{table}
\FloatBarrier
\begin{table*}[!ht]
    \centering
    \footnotesize
    \begin{tabular}{|c|c|c|c|}
        \hline
        Model Parameter & Description & Value & Source \\
        \hline 
        $L_{div}$ &  the maximal dividing length of a segment & 0.075 $\mu$m & Estimated \cite{mt_instability, update_cma} \\
        \hline
        $v_{plus}$ & growth rate & 0.0615 $\mu\text{m}/\text{sec}$ & \cite{dynamic_cmt_ordering, allard_sim} \\
        \hline
        $L_{min}$ &  the minimal length of a segment & 0.0025 $\mu$m & Estimated \cite{mt_instability, update_cma} \\
        \hline 
        $v_{minus}$ & retraction rate & 0.00883 $\mu\text{m}/\text{sec}$ & \cite{dynamic_cmt_ordering, allard_sim} \\
        \hline
        $\theta_{cic}$ & CIC angle threshold & $40^\circ$ & Estimated \cite{dynamic_cmt_ordering, allard_sim, CHAKRABORTTY20183031} \\
        \hline
        $\epsilon$& maximal reaction radius & 0.1 $\mu$m &  Estimated \cite{mt_instability, update_cma} \\
        \hline
        $\theta_{crit}$ & critical zippering angle threshold &  $40^\circ$ & \cite{dynamic_cmt_ordering, ambrose2008clasp} \\
        \hline
        $\sigma_{sep}$ & MT separation distance & 0.025 $\mu$m & \cite{sep_dist} \\
        \hline
        $\theta_{cross}$ & crossover angle threshold & $40^\circ$ & Estimated \cite{dynamic_cmt_ordering, allard_sim, CHAKRABORTTY20183031} \\
        \hline
        $\theta_{exit}$ & CLASP entry angle threshold & $15^\circ$ & Estimated \cite{picket_fence} \\
        \hline
        $\theta_{angle}$ & CLASP exit angle threshold & $15^\circ$ & Estimated \cite{picket_fence} \\
        \hline
        $\sigma_{col}$ & MT collision distance & 0.025 $\mu$m & Estimated \cite{CHAKRABORTTY20183031, update_cma} \\
        \hline
        $s_{min}$ & Minimum MT segment initialization length & 0.005 $\mu$m & Estimated \cite{update_cma} \\
        \hline
        $s_{max}$ & Minimum MT segment initialization length & 0.01 $\mu$m & Estimated \cite{update_cma} \\
        \hline
    \end{tabular}
    \caption{Model parameter definitions with default values.}
    \label{tab2:model_parameter}
\end{table*}
\begin{table*}[!ht]
    \centering
    \footnotesize
    \begin{tabular}{|c|c|c|c|}
        \hline
        Model Parameter & Description & Propensity Rate & Source \\
        \hline 
        $\hat{\rho}_{{grow}}$ &  growth rate factor & 100.0 & Estimated \\
        \hline 
        $\hat{\rho}_{{retract}}$ &  retraction rate factor & 10.0 & Estimated  \\
        \hline 
        $\hat{\rho}_{{bnd\_cic\_std}}$ & boundary CIC standard rate factor & 40,000 & Estimated \\
        \hline 
        $\hat{\rho}_{{bnd\_cic\_clasp}}$ & boundary CIC for CLASP rate factor & 40,000 & Estimated  \\
        \hline 
        $\hat{\rho}_{{int\_cic}}$ & intermediate CIC rate factor & 12,000 & Estimated \cite{dynamic_cmt_ordering} \\
        \hline 
        $\hat{\rho}_{{grow\_cic}}$ & growing end CIC rate factor & 12,000 & Estimated \cite{dynamic_cmt_ordering} \\
        \hline 
        $\hat{\rho}_{{retract\_cic}}$ & retracting end CIC rate factor & 12,000 & Estimated \cite{dynamic_cmt_ordering} \\
        \hline 
        $\hat{\rho}_{{zip\_hit}}$ & zippering entrainment rate factor & 4,000 & Estimated \cite{dynamic_cmt_ordering} \\
        \hline 
        $\hat{\rho}_{{zip\_guard}}$ & zippering guard rate factor & 12,000& Estimated  \cite{dynamic_cmt_ordering} \\
        \hline 
        $\hat{\rho}_{{cross}}$ & crossover rate factor & 200 & Estimated \cite{dynamic_cmt_ordering} \\
        \hline 
        $\hat{\rho}_{{uncross}}$ & uncrossover rate factor & 0.01 & Estimated \\
        \hline 
        $\hat{\rho}_{{clasp\_entry}}$ & clasp entry rate factor & 0.001 & Estimated \cite{picket_fence, light_reorg} \\
        \hline 
        $\hat{\rho}_{{clasp\_exit}}$ & clasp exit rate factor & 40,000 & Estimated \cite{picket_fence, light_reorg} \\
        \hline 
        $\hat{\rho}_{{clasp\_cat}}$ & clasp catastrophe rate factor & 1,000 & Estimated \\
        \hline 
        $\hat{\rho}_{{clasp\_detach}}$ & clasp detachment rate factor & 0.016 & Estimated \\
        \hline 
        $\hat{\rho}_{{destruct}}$ & destruction rate factor & 0.0026 & Estimated \\
        \hline 
        $\hat{\rho}_{{create}}$ & creation rate factor & 0.0026 & Estimated \cite{allard_sim} \\
        \hline 
        $\hat{\rho}_{{retract} \leftarrow \text{growth}}$ & retraction to growth conversion rate & 0.016 & Estimated  \cite{wightman} \\
        \hline
        $\hat{\rho}_{{growth} \leftarrow \text{retract}}$ & growth to retraction conversion rate & 0.016 & Estimated \cite{wightman} \\
        \hline
    \end{tabular}
    \caption{A table of model rate factors with default values.}
    \label{tab2:model_rates}
\end{table*}

\begin{table*}[!ht]
    \centering
    \begin{tabular}{|c|p{1.35in}|}
        \hline
        Function & Description\\ 
        \hline
        $M_d(x_1, x_2, x_3)$ & Minimum distance \newline from $x_1$ to the line \newline through $x_2$ and $x_3$ \\
        \hline
        $M_\theta(u_1, x_1, u_2)$ & Given a point $x_1$ \newline in the direction $u_1$ \newline find the angle \newline made with $u_2$ \\
        \hline
        $H(x; a) = \begin{cases} 1 & x \geq a \\ 0 & x < a \end{cases}$ & Heaviside function\\
        \hline
        $rotate(\Vec{x}, \theta) =  \begin{bmatrix} 
        \cos(\theta) & \sin(\theta)\\
    -\sin(\theta) & \cos(\theta)
    \end{bmatrix} \;  \cdot \; \Vec{x}$ & rotation function \\
        \hline 
        $\Theta(\text{stmt}) = \begin{cases} 1 & \text{if stmt is true} \\ 0 & \text{if stmt is false} \end{cases}$ & indication function \\
        \hline
        $
I_p(p_1, u_1, p_2, p_3) = 
\left[\left(
    \begin{array}{cc}
 u_{1x} & -({p}_{3x}-{p}_{2x}) \\
 u_{1y} & -({p}_{3y}-{p}_{2y}) \\
\end{array}
\right)^{-1} \; \left(
\begin{array}{c}
 {p}_{2x}-{p}_{1x} \\
 {p}_{2y}-{p}_{1y} \\
\end{array}
\right)\right]
= \left[
\begin{array}{cc} \alpha_1 \\ \alpha_2 \end{array}
\right]$ & Line intersection \\
\hline
    \end{tabular}
    \caption{Function Descriptions}
    \label{tab2:function}
\end{table*}
\FloatBarrier
\bibliography{paper.bib}

\begin{thebibliography}{73}%
\makeatletter
\providecommand \@ifxundefined [1]{%
 \@ifx{#1\undefined}
}%
\providecommand \@ifnum [1]{%
 \ifnum #1\expandafter \@firstoftwo
 \else \expandafter \@secondoftwo
 \fi
}%
\providecommand \@ifx [1]{%
 \ifx #1\expandafter \@firstoftwo
 \else \expandafter \@secondoftwo
 \fi
}%
\providecommand \natexlab [1]{#1}%
\providecommand \enquote  [1]{``#1''}%
\providecommand \bibnamefont  [1]{#1}%
\providecommand \bibfnamefont [1]{#1}%
\providecommand \citenamefont [1]{#1}%
\providecommand \href@noop [0]{\@secondoftwo}%
\providecommand \href [0]{\begingroup \@sanitize@url \@href}%
\providecommand \@href[1]{\@@startlink{#1}\@@href}%
\providecommand \@@href[1]{\endgroup#1\@@endlink}%
\providecommand \@sanitize@url [0]{\catcode `\\12\catcode `\$12\catcode `\&12\catcode `\#12\catcode `\^12\catcode `\_12\catcode `\%12\relax}%
\providecommand \@@startlink[1]{}%
\providecommand \@@endlink[0]{}%
\providecommand \url  [0]{\begingroup\@sanitize@url \@url }%
\providecommand \@url [1]{\endgroup\@href {#1}{\urlprefix }}%
\providecommand \urlprefix  [0]{URL }%
\providecommand \Eprint [0]{\href }%
\providecommand \doibase [0]{http://dx.doi.org/}%
\providecommand \selectlanguage [0]{\@gobble}%
\providecommand \bibinfo  [0]{\@secondoftwo}%
\providecommand \bibfield  [0]{\@secondoftwo}%
\providecommand \translation [1]{[#1]}%
\providecommand \BibitemOpen [0]{}%
\providecommand \bibitemStop [0]{}%
\providecommand \bibitemNoStop [0]{.\EOS\space}%
\providecommand \EOS [0]{\spacefactor3000\relax}%
\providecommand \BibitemShut  [1]{\csname bibitem#1\endcsname}%
\let\auto@bib@innerbib\@empty
%</preamble>
\bibitem [{\citenamefont {Diestel}(2017)}]{graph_theory}%
  \BibitemOpen
  \bibfield  {author} {\bibinfo {author} {\bibfnamefont {R.}~\bibnamefont {Diestel}},\ }\href {\doibase 0.1007/978-3-662-53622-3} {\emph {\bibinfo {title} {Graph Theory}}},\ \bibinfo {edition} {5th}\ ed.\ (\bibinfo  {publisher} {Springer},\ \bibinfo {year} {2017})\ Chap.~\bibinfo {chapter} {1}, pp.\ \bibinfo {pages} {1--31}\BibitemShut {NoStop}%
\bibitem [{\citenamefont {Mjolsness}(2013)}]{Mjolsness_2013}%
  \BibitemOpen
  \bibfield  {author} {\bibinfo {author} {\bibfnamefont {E.}~\bibnamefont {Mjolsness}},\ }\bibfield  {title} {\enquote {\bibinfo {title} {Time-ordered product expansions for computational stochastic system biology},}\ }\href {\doibase 10.1088/1478-3975/10/3/035009} {\bibfield  {journal} {\bibinfo  {journal} {Physical Biology}\ }\textbf {\bibinfo {volume} {10}},\ \bibinfo {pages} {035009} (\bibinfo {year} {2013})}\BibitemShut {NoStop}%
\bibitem [{\citenamefont {Mjolsness}(2022)}]{Mjolsness2022}%
  \BibitemOpen
  \bibfield  {author} {\bibinfo {author} {\bibfnamefont {E.}~\bibnamefont {Mjolsness}},\ }\bibfield  {title} {\enquote {\bibinfo {title} {Explicit calculation of structural commutation relations for stochastic and dynamical graph grammar rule operators in biological morphodynamics},}\ }\href {\doibase 10.3389/fsysb.2022.898858} {\bibfield  {journal} {\bibinfo  {journal} {Frontiers in Systems Biology}\ }\textbf {\bibinfo {volume} {2}} (\bibinfo {year} {2022}),\ 10.3389/fsysb.2022.898858}\BibitemShut {NoStop}%
\bibitem [{\citenamefont {Rozenberg}(1997)}]{graph_grammars_rozen}%
  \BibitemOpen
  \bibfield  {author} {\bibinfo {author} {\bibfnamefont {G.}~\bibnamefont {Rozenberg}},\ }\href {\doibase 10.1142/3303} {\emph {\bibinfo {title} {Handbook of Graph Grammars and Computing by Graph Transformation}}}\ (\bibinfo  {publisher} {WORLD SCIENTIFIC},\ \bibinfo {year} {1997})\BibitemShut {NoStop}%
\bibitem [{\citenamefont {Mjolsness}(2019)}]{Mjolsness2019}%
  \BibitemOpen
  \bibfield  {author} {\bibinfo {author} {\bibfnamefont {E.}~\bibnamefont {Mjolsness}},\ }\bibfield  {title} {\enquote {\bibinfo {title} {Prospects for declarative mathematical modeling of complex biological systems},}\ }\href {\doibase 10.1007/s11538-019-00628-7} {\bibfield  {journal} {\bibinfo  {journal} {Bulletin of Mathematical Biology}\ }\textbf {\bibinfo {volume} {81}},\ \bibinfo {pages} {3385--3420} (\bibinfo {year} {2019})}\BibitemShut {NoStop}%
\bibitem [{\citenamefont {Gillespie}(1977)}]{GILLESPIE1977_OG}%
  \BibitemOpen
  \bibfield  {author} {\bibinfo {author} {\bibfnamefont {D.~T.}\ \bibnamefont {Gillespie}},\ }\bibfield  {title} {\enquote {\bibinfo {title} {Exact stochastic simulation of coupled chemical reactions},}\ }\href {\doibase 10.1021/j100540a008} {\bibfield  {journal} {\bibinfo  {journal} {The Journal of Physical Chemistry}\ }\textbf {\bibinfo {volume} {81}},\ \bibinfo {pages} {2340--2361} (\bibinfo {year} {1977})}\BibitemShut {NoStop}%
\bibitem [{\citenamefont {Young}\ and\ \citenamefont {Elcock}(1966)}]{Kinetic_WMYoung1966}%
  \BibitemOpen
  \bibfield  {author} {\bibinfo {author} {\bibfnamefont {W.~M.}\ \bibnamefont {Young}}\ and\ \bibinfo {author} {\bibfnamefont {E.~W.}\ \bibnamefont {Elcock}},\ }\bibfield  {title} {\enquote {\bibinfo {title} {Monte carlo studies of vacancy migration in binary ordered alloys: I},}\ }\href {\doibase 10.1088/0370-1328/89/3/329} {\bibfield  {journal} {\bibinfo  {journal} {Proceedings of the Physical Society}\ }\textbf {\bibinfo {volume} {89}},\ \bibinfo {pages} {735--746} (\bibinfo {year} {1966})}\BibitemShut {NoStop}%
\bibitem [{\citenamefont {Medwedeff}\ and\ \citenamefont {Mjolsness}(2023)}]{Medwedeff_2023}%
  \BibitemOpen
  \bibfield  {author} {\bibinfo {author} {\bibfnamefont {E.}~\bibnamefont {Medwedeff}}\ and\ \bibinfo {author} {\bibfnamefont {E.}~\bibnamefont {Mjolsness}},\ }\bibfield  {title} {\enquote {\bibinfo {title} {Approximate simulation of cortical microtubule models using dynamical graph grammars},}\ }\href {\doibase 10.1088/1478-3975/acdbfb} {\bibfield  {journal} {\bibinfo  {journal} {Physical Biology}\ }\textbf {\bibinfo {volume} {20}},\ \bibinfo {pages} {055002} (\bibinfo {year} {2023})}\BibitemShut {NoStop}%
\bibitem [{\citenamefont {Metropolis}\ \emph {et~al.}(1953)\citenamefont {Metropolis}, \citenamefont {Rosenbluth}, \citenamefont {Rosenbluth}, \citenamefont {Teller},\ and\ \citenamefont {Teller}}]{metropolis}%
  \BibitemOpen
  \bibfield  {author} {\bibinfo {author} {\bibfnamefont {N.}~\bibnamefont {Metropolis}}, \bibinfo {author} {\bibfnamefont {A.~W.}\ \bibnamefont {Rosenbluth}}, \bibinfo {author} {\bibfnamefont {M.~N.}\ \bibnamefont {Rosenbluth}}, \bibinfo {author} {\bibfnamefont {A.~H.}\ \bibnamefont {Teller}}, \ and\ \bibinfo {author} {\bibfnamefont {E.}~\bibnamefont {Teller}},\ }\bibfield  {title} {\enquote {\bibinfo {title} {{Equation of State Calculations by Fast Computing Machines}},}\ }\href {\doibase 10.1063/1.1699114} {\bibfield  {journal} {\bibinfo  {journal} {The Journal of Chemical Physics}\ }\textbf {\bibinfo {volume} {21}},\ \bibinfo {pages} {1087--1092} (\bibinfo {year} {1953})}\BibitemShut {NoStop}%
\bibitem [{\citenamefont {Lecca}, \citenamefont {Laurenzi},\ and\ \citenamefont {Jordan}(2013{\natexlab{a}})}]{modeling}%
  \BibitemOpen
  \bibfield  {author} {\bibinfo {author} {\bibfnamefont {P.}~\bibnamefont {Lecca}}, \bibinfo {author} {\bibfnamefont {I.}~\bibnamefont {Laurenzi}}, \ and\ \bibinfo {author} {\bibfnamefont {F.}~\bibnamefont {Jordan}},\ }\bibfield  {title} {\enquote {\bibinfo {title} {4 - modelling in systems biology},}\ }in\ \href {\doibase https://doi.org/10.1533/9781908818218.117} {\emph {\bibinfo {booktitle} {Deterministic Versus Stochastic Modelling in Biochemistry and Systems Biology}}},\ \bibinfo {series and number} {Woodhead Publishing Series in Biomedicine},\ \bibinfo {editor} {edited by\ \bibinfo {editor} {\bibfnamefont {P.}~\bibnamefont {Lecca}}, \bibinfo {editor} {\bibfnamefont {I.}~\bibnamefont {Laurenzi}}, \ and\ \bibinfo {editor} {\bibfnamefont {F.}~\bibnamefont {Jordan}}}\ (\bibinfo  {publisher} {Woodhead Publishing},\ \bibinfo {year} {2013})\ pp.\ \bibinfo {pages} {117--180}\BibitemShut {NoStop}%
\bibitem [{\citenamefont {Lecca}, \citenamefont {Laurenzi},\ and\ \citenamefont {Jordan}(2013{\natexlab{b}})}]{reaction_rate}%
  \BibitemOpen
  \bibfield  {author} {\bibinfo {author} {\bibfnamefont {P.}~\bibnamefont {Lecca}}, \bibinfo {author} {\bibfnamefont {I.}~\bibnamefont {Laurenzi}}, \ and\ \bibinfo {author} {\bibfnamefont {F.}~\bibnamefont {Jordan}},\ }\bibfield  {title} {\enquote {\bibinfo {title} {1 - deterministic chemical kinetics},}\ }in\ \href {\doibase 10.1533/9781908818218.1} {\emph {\bibinfo {booktitle} {Deterministic Versus Stochastic Modelling in Biochemistry and Systems Biology}}},\ \bibinfo {series and number} {Woodhead Publishing Series in Biomedicine},\ \bibinfo {editor} {edited by\ \bibinfo {editor} {\bibfnamefont {P.}~\bibnamefont {Lecca}}, \bibinfo {editor} {\bibfnamefont {I.}~\bibnamefont {Laurenzi}}, \ and\ \bibinfo {editor} {\bibfnamefont {F.}~\bibnamefont {Jordan}}}\ (\bibinfo  {publisher} {Woodhead Publishing},\ \bibinfo {year} {2013})\ pp.\ \bibinfo {pages} {1--34}\BibitemShut {NoStop}%
\bibitem [{\citenamefont {Bortz}, \citenamefont {Kalos},\ and\ \citenamefont {Lebowitz}(1975)}]{kinetic_monte_carlo}%
  \BibitemOpen
  \bibfield  {author} {\bibinfo {author} {\bibfnamefont {A.}~\bibnamefont {Bortz}}, \bibinfo {author} {\bibfnamefont {M.}~\bibnamefont {Kalos}}, \ and\ \bibinfo {author} {\bibfnamefont {J.}~\bibnamefont {Lebowitz}},\ }\bibfield  {title} {\enquote {\bibinfo {title} {A new algorithm for monte carlo simulation of ising spin systems},}\ }\href {\doibase https://doi.org/10.1016/0021-9991(75)90060-1} {\bibfield  {journal} {\bibinfo  {journal} {Journal of Computational Physics}\ }\textbf {\bibinfo {volume} {17}},\ \bibinfo {pages} {10--18} (\bibinfo {year} {1975})}\BibitemShut {NoStop}%
\bibitem [{\citenamefont {Gillespie}(1992)}]{GILLESPIE1992_RIG}%
  \BibitemOpen
  \bibfield  {author} {\bibinfo {author} {\bibfnamefont {D.~T.}\ \bibnamefont {Gillespie}},\ }\bibfield  {title} {\enquote {\bibinfo {title} {A rigorous derivation of the chemical master equation},}\ }\href {\doibase 10.1016/0378-4371(92)90283-V} {\bibfield  {journal} {\bibinfo  {journal} {Physica A: Statistical Mechanics and its Applications}\ }\textbf {\bibinfo {volume} {188}},\ \bibinfo {pages} {404--425} (\bibinfo {year} {1992})}\BibitemShut {NoStop}%
\bibitem [{\citenamefont {Gillespie}(2001)}]{GILLESPIE2001_TAU}%
  \BibitemOpen
  \bibfield  {author} {\bibinfo {author} {\bibfnamefont {D.~T.}\ \bibnamefont {Gillespie}},\ }\bibfield  {title} {\enquote {\bibinfo {title} {Approximate accelerated stochastic simulation of chemically reacting systems},}\ }\href {\doibase 10.1063/1.1378322} {\bibfield  {journal} {\bibinfo  {journal} {The Journal of Chemical Physics}\ }\textbf {\bibinfo {volume} {115}},\ \bibinfo {pages} {1716--1733} (\bibinfo {year} {2001})}\BibitemShut {NoStop}%
\bibitem [{\citenamefont {Cao}, \citenamefont {Gillespie},\ and\ \citenamefont {Petzold}(2006)}]{TauLeapEfficient}%
  \BibitemOpen
  \bibfield  {author} {\bibinfo {author} {\bibfnamefont {Y.}~\bibnamefont {Cao}}, \bibinfo {author} {\bibfnamefont {D.~T.}\ \bibnamefont {Gillespie}}, \ and\ \bibinfo {author} {\bibfnamefont {L.~R.}\ \bibnamefont {Petzold}},\ }\bibfield  {title} {\enquote {\bibinfo {title} {Efficient step size selection for the tau-leaping simulation method},}\ }\href {\doibase 10.1063/1.2159468} {\bibfield  {journal} {\bibinfo  {journal} {The Journal of Chemical Physics}\ }\textbf {\bibinfo {volume} {124}},\ \bibinfo {pages} {044109} (\bibinfo {year} {2006})}\BibitemShut {NoStop}%
\bibitem [{\citenamefont {Auger}, \citenamefont {Chatelain},\ and\ \citenamefont {Koumoutsakos}(2006)}]{RLeap}%
  \BibitemOpen
  \bibfield  {author} {\bibinfo {author} {\bibfnamefont {A.}~\bibnamefont {Auger}}, \bibinfo {author} {\bibfnamefont {P.}~\bibnamefont {Chatelain}}, \ and\ \bibinfo {author} {\bibfnamefont {P.}~\bibnamefont {Koumoutsakos}},\ }\bibfield  {title} {\enquote {\bibinfo {title} {R-leaping: Accelerating the stochastic simulation algorithm by reaction leaps},}\ }\href {\doibase 10.1063/1.2218339} {\bibfield  {journal} {\bibinfo  {journal} {The Journal of Chemical Physics}\ }\textbf {\bibinfo {volume} {125}},\ \bibinfo {pages} {084103} (\bibinfo {year} {2006})}\BibitemShut {NoStop}%
\bibitem [{\citenamefont {Mjolsness}\ \emph {et~al.}(2009)\citenamefont {Mjolsness}, \citenamefont {Orendorff}, \citenamefont {Chatelain},\ and\ \citenamefont {Koumoutsakos}}]{ER_Leap}%
  \BibitemOpen
  \bibfield  {author} {\bibinfo {author} {\bibfnamefont {E.}~\bibnamefont {Mjolsness}}, \bibinfo {author} {\bibfnamefont {D.}~\bibnamefont {Orendorff}}, \bibinfo {author} {\bibfnamefont {P.}~\bibnamefont {Chatelain}}, \ and\ \bibinfo {author} {\bibfnamefont {P.}~\bibnamefont {Koumoutsakos}},\ }\bibfield  {title} {\enquote {\bibinfo {title} {An exact accelerated stochastic simulation algorithm},}\ }\href {\doibase 10.1063/1.3078490} {\bibfield  {journal} {\bibinfo  {journal} {The Journal of Chemical Physics}\ }\textbf {\bibinfo {volume} {130}},\ \bibinfo {pages} {144110} (\bibinfo {year} {2009})}\BibitemShut {NoStop}%
\bibitem [{\citenamefont {Orendorff}\ and\ \citenamefont {Mjolsness}(2012)}]{HiER_Leap}%
  \BibitemOpen
  \bibfield  {author} {\bibinfo {author} {\bibfnamefont {D.}~\bibnamefont {Orendorff}}\ and\ \bibinfo {author} {\bibfnamefont {E.}~\bibnamefont {Mjolsness}},\ }\bibfield  {title} {\enquote {\bibinfo {title} {A hierarchical exact accelerated stochastic simulation algorithm},}\ }\href {\doibase 10.1063/1.4766353} {\bibfield  {journal} {\bibinfo  {journal} {The Journal of Chemical Physics}\ }\textbf {\bibinfo {volume} {137}},\ \bibinfo {pages} {214104} (\bibinfo {year} {2012})}\BibitemShut {NoStop}%
\bibitem [{\citenamefont {Lipkov{\'a}}\ \emph {et~al.}(2019)\citenamefont {Lipkov{\'a}}, \citenamefont {Arampatzis}, \citenamefont {Chatelain}, \citenamefont {Menze},\ and\ \citenamefont {Koumoutsakos}}]{SLeap}%
  \BibitemOpen
  \bibfield  {author} {\bibinfo {author} {\bibfnamefont {J.}~\bibnamefont {Lipkov{\'a}}}, \bibinfo {author} {\bibfnamefont {G.}~\bibnamefont {Arampatzis}}, \bibinfo {author} {\bibfnamefont {P.}~\bibnamefont {Chatelain}}, \bibinfo {author} {\bibfnamefont {B.}~\bibnamefont {Menze}}, \ and\ \bibinfo {author} {\bibfnamefont {P.}~\bibnamefont {Koumoutsakos}},\ }\bibfield  {title} {\enquote {\bibinfo {title} {S-leaping: An adaptive, accelerated stochastic simulation algorithm, bridging $\tau$-leaping and r-leaping},}\ }\href {\doibase 10.1007/s11538-018-0464-9} {\bibfield  {journal} {\bibinfo  {journal} {Bulletin of Mathematical Biology}\ }\textbf {\bibinfo {volume} {81}},\ \bibinfo {pages} {3074--3096} (\bibinfo {year} {2019})}\BibitemShut {NoStop}%
\bibitem [{\citenamefont {Mjolsness}\ and\ \citenamefont {Yosiphon}(2007)}]{Mjolsness2006}%
  \BibitemOpen
  \bibfield  {author} {\bibinfo {author} {\bibfnamefont {E.}~\bibnamefont {Mjolsness}}\ and\ \bibinfo {author} {\bibfnamefont {G.}~\bibnamefont {Yosiphon}},\ }\bibfield  {title} {\enquote {\bibinfo {title} {Stochastic process semantics for dynamical grammars},}\ }\href {\doibase 10.1007/s10472-006-9034-1} {\bibfield  {journal} {\bibinfo  {journal} {Annals of Mathematics and Artificial Intelligence}\ }\textbf {\bibinfo {volume} {47}},\ \bibinfo {pages} {329--395} (\bibinfo {year} {2007})}\BibitemShut {NoStop}%
\bibitem [{\citenamefont {Mjolsness}(2010)}]{Mjolsness2010}%
  \BibitemOpen
  \bibfield  {author} {\bibinfo {author} {\bibfnamefont {E.}~\bibnamefont {Mjolsness}},\ }\bibfield  {title} {\enquote {\bibinfo {title} {Towards measurable types for dynamical process modeling languages},}\ }\href {\doibase 10.1016/j.entcs.2010.08.008} {\bibfield  {journal} {\bibinfo  {journal} {Electronic notes in theoretical computer science}\ }\textbf {\bibinfo {volume} {265}},\ \bibinfo {pages} {123--144} (\bibinfo {year} {2010})}\BibitemShut {NoStop}%
\bibitem [{\citenamefont {Yosiphon}\ and\ \citenamefont {Mjolsness}(2009)}]{Mjolsness2009}%
  \BibitemOpen
  \bibfield  {author} {\bibinfo {author} {\bibfnamefont {G.}~\bibnamefont {Yosiphon}}\ and\ \bibinfo {author} {\bibfnamefont {E.}~\bibnamefont {Mjolsness}},\ }\href@noop {} {\emph {\bibinfo {title} {Learning and Inference in Computational Systems Biology}}}\ (\bibinfo  {publisher} {MIT Press},\ \bibinfo {year} {2009})\ pp.\ \bibinfo {pages} {297--314}\BibitemShut {NoStop}%
\bibitem [{\citenamefont {Yosiphon}(2009)}]{yosiphon_2009}%
  \BibitemOpen
  \bibfield  {author} {\bibinfo {author} {\bibfnamefont {G.}~\bibnamefont {Yosiphon}},\ }\emph {\bibinfo {title} {Stochastic Parameterized Grammars : Formalization, Inference and Modeling Applications}},\ \href {http://computableplant.ics.uci.edu/theses/guy/downloads/papers/thesis.pdf} {Ph.D. thesis},\ \bibinfo  {school} {UC Irvine} (\bibinfo {year} {2009})\BibitemShut {NoStop}%
\bibitem [{\citenamefont {{Wolfram Research, Inc.}}(2021)}]{Mathematica}%
  \BibitemOpen
  \bibfield  {author} {\bibinfo {author} {\bibnamefont {{Wolfram Research, Inc.}}},\ }\href {https://www.wolfram.com/mathematica/} {\enquote {\bibinfo {title} {Mathematica},}\ } (\bibinfo {year} {2021})\BibitemShut {NoStop}%
\bibitem [{\citenamefont {Dechter}(1997)}]{bucket_elimination}%
  \BibitemOpen
  \bibfield  {author} {\bibinfo {author} {\bibfnamefont {R.}~\bibnamefont {Dechter}},\ }\bibfield  {title} {\enquote {\bibinfo {title} {Bucket elimination: a unifying framework for processing hard and soft constraints},}\ }\href {\doibase 10.1023/A:1009796922698} {\bibfield  {journal} {\bibinfo  {journal} {Constraints}\ }\textbf {\bibinfo {volume} {2}},\ \bibinfo {pages} {51--55} (\bibinfo {year} {1997})}\BibitemShut {NoStop}%
\bibitem [{\citenamefont {Dechter}(2003)}]{constraint_processing}%
  \BibitemOpen
  \bibfield  {author} {\bibinfo {author} {\bibfnamefont {R.}~\bibnamefont {Dechter}},\ }\href@noop {} {\emph {\bibinfo {title} {Constraint Processing}}},\ The Morgan Kaufmann Series in Artificial Intelligence\ (\bibinfo  {publisher} {Morgan Kaufmann},\ \bibinfo {address} {San Francisco},\ \bibinfo {year} {2003})\BibitemShut {NoStop}%
\bibitem [{\citenamefont {Giavitto}\ and\ \citenamefont {Michel}(2001)}]{MGS_long}%
  \BibitemOpen
  \bibfield  {author} {\bibinfo {author} {\bibfnamefont {J.-L.}\ \bibnamefont {Giavitto}}\ and\ \bibinfo {author} {\bibfnamefont {O.}~\bibnamefont {Michel}},\ }\bibfield  {title} {\enquote {\bibinfo {title} {Mgs: A rule-based programming language for complex objects and collections},}\ }\href {\doibase https://doi.org/10.1016/S1571-0661(04)00293-2} {\bibfield  {journal} {\bibinfo  {journal} {Electronic Notes in Theoretical Computer Science}\ }\textbf {\bibinfo {volume} {59}},\ \bibinfo {pages} {286--304} (\bibinfo {year} {2001})},\ \bibinfo {note} {rULE 2001, Second International Workshop on Rule-Based Programming (Satellite Event of PLI 2001)}\BibitemShut {NoStop}%
\bibitem [{\citenamefont {Stiles}, \citenamefont {Bartol}\ \emph {et~al.}(2001)\citenamefont {Stiles}, \citenamefont {Bartol} \emph {et~al.}}]{mcell}%
  \BibitemOpen
  \bibfield  {author} {\bibinfo {author} {\bibfnamefont {J.~R.}\ \bibnamefont {Stiles}}, \bibinfo {author} {\bibfnamefont {T.~M.}\ \bibnamefont {Bartol}},  \emph {et~al.},\ }\bibfield  {title} {\enquote {\bibinfo {title} {Monte carlo methods for simulating realistic synaptic microphysiology using mcell},}\ }\href@noop {} {\bibfield  {journal} {\bibinfo  {journal} {Computational neuroscience: Realistic modeling for experimentalists}\ ,\ \bibinfo {pages} {87--127}} (\bibinfo {year} {2001})}\BibitemShut {NoStop}%
\bibitem [{\citenamefont {Shapiro}\ and\ \citenamefont {Mjolsness}(2015)}]{pycellerator}%
  \BibitemOpen
  \bibfield  {author} {\bibinfo {author} {\bibfnamefont {B.~E.}\ \bibnamefont {Shapiro}}\ and\ \bibinfo {author} {\bibfnamefont {E.}~\bibnamefont {Mjolsness}},\ }\bibfield  {title} {\enquote {\bibinfo {title} {Pycellerator: an arrow-based reaction-like modelling language for biological simulations},}\ }\href {\doibase 10.1093/bioinformatics/btv596} {\bibfield  {journal} {\bibinfo  {journal} {Bioinformatics}\ }\textbf {\bibinfo {volume} {32}},\ \bibinfo {pages} {629--631} (\bibinfo {year} {2015})}\BibitemShut {NoStop}%
\bibitem [{\citenamefont {Blinov}\ \emph {et~al.}(2004)\citenamefont {Blinov}, \citenamefont {Faeder}, \citenamefont {Goldstein},\ and\ \citenamefont {Hlavacek}}]{bionetgen2}%
  \BibitemOpen
  \bibfield  {author} {\bibinfo {author} {\bibfnamefont {M.~L.}\ \bibnamefont {Blinov}}, \bibinfo {author} {\bibfnamefont {J.~R.}\ \bibnamefont {Faeder}}, \bibinfo {author} {\bibfnamefont {B.}~\bibnamefont {Goldstein}}, \ and\ \bibinfo {author} {\bibfnamefont {W.~S.}\ \bibnamefont {Hlavacek}},\ }\bibfield  {title} {\enquote {\bibinfo {title} {{BioNetGen: software for rule-based modeling of signal transduction based on the interactions of molecular domains}},}\ }\href {\doibase 10.1093/bioinformatics/bth378} {\bibfield  {journal} {\bibinfo  {journal} {Bioinformatics}\ }\textbf {\bibinfo {volume} {20}},\ \bibinfo {pages} {3289--3291} (\bibinfo {year} {2004})}\BibitemShut {NoStop}%
\bibitem [{\citenamefont {Harris}\ \emph {et~al.}(2016)\citenamefont {Harris}, \citenamefont {Hogg}, \citenamefont {Tapia}, \citenamefont {Sekar}, \citenamefont {Gupta}, \citenamefont {Korsunsky}, \citenamefont {Arora}, \citenamefont {Barua}, \citenamefont {Sheehan},\ and\ \citenamefont {Faeder}}]{bionetgen}%
  \BibitemOpen
  \bibfield  {author} {\bibinfo {author} {\bibfnamefont {L.~A.}\ \bibnamefont {Harris}}, \bibinfo {author} {\bibfnamefont {J.~S.}\ \bibnamefont {Hogg}}, \bibinfo {author} {\bibfnamefont {J.-J.}\ \bibnamefont {Tapia}}, \bibinfo {author} {\bibfnamefont {J.~A.~P.}\ \bibnamefont {Sekar}}, \bibinfo {author} {\bibfnamefont {S.}~\bibnamefont {Gupta}}, \bibinfo {author} {\bibfnamefont {I.}~\bibnamefont {Korsunsky}}, \bibinfo {author} {\bibfnamefont {A.}~\bibnamefont {Arora}}, \bibinfo {author} {\bibfnamefont {D.}~\bibnamefont {Barua}}, \bibinfo {author} {\bibfnamefont {R.~P.}\ \bibnamefont {Sheehan}}, \ and\ \bibinfo {author} {\bibfnamefont {J.~R.}\ \bibnamefont {Faeder}},\ }\bibfield  {title} {\enquote {\bibinfo {title} {{BioNetGen 2.2: advances in rule-based modeling}},}\ }\href {\doibase 10.1093/bioinformatics/btw469} {\bibfield  {journal} {\bibinfo  {journal} {Bioinformatics}\ }\textbf {\bibinfo {volume} {32}},\ \bibinfo {pages} {3366--3368} (\bibinfo {year} {2016})}\BibitemShut {NoStop}%
\bibitem [{\citenamefont {Danos}\ and\ \citenamefont {Laneve}(2004)}]{kappa}%
  \BibitemOpen
  \bibfield  {author} {\bibinfo {author} {\bibfnamefont {V.}~\bibnamefont {Danos}}\ and\ \bibinfo {author} {\bibfnamefont {C.}~\bibnamefont {Laneve}},\ }\bibfield  {title} {\enquote {\bibinfo {title} {Formal molecular biology},}\ }\href {\doibase https://doi.org/10.1016/j.tcs.2004.03.065} {\bibfield  {journal} {\bibinfo  {journal} {Theoretical Computer Science}\ }\textbf {\bibinfo {volume} {325}},\ \bibinfo {pages} {69--110} (\bibinfo {year} {2004})},\ \bibinfo {note} {computational Systems Biology}\BibitemShut {NoStop}%
\bibitem [{\citenamefont {Spicher}\ and\ \citenamefont {Michel}(2007)}]{MGS_short}%
  \BibitemOpen
  \bibfield  {author} {\bibinfo {author} {\bibfnamefont {A.}~\bibnamefont {Spicher}}\ and\ \bibinfo {author} {\bibfnamefont {O.}~\bibnamefont {Michel}},\ }\bibfield  {title} {\enquote {\bibinfo {title} {Declarative modeling of a neurulation-like process},}\ }\href {\doibase https://doi.org/10.1016/j.biosystems.2006.09.024} {\bibfield  {journal} {\bibinfo  {journal} {Biosystems}\ }\textbf {\bibinfo {volume} {87}},\ \bibinfo {pages} {281--288} (\bibinfo {year} {2007})},\ \bibinfo {note} {papers presented at the Sixth International Workshop on Information Processing in Cells and Tissues, York, UK, 2005}\BibitemShut {NoStop}%
\bibitem [{\citenamefont {Lane}(2015)}]{Lane2015CellCT}%
  \BibitemOpen
  \bibfield  {author} {\bibinfo {author} {\bibfnamefont {B.}~\bibnamefont {Lane}},\ }\emph {\bibinfo {title} {Cell Complexes: The Structure of Space and the Mathematics of Modularity}},\ \href {http://algorithmicbotany.org/papers/laneb.th2015.html} {Ph.D. thesis},\ \bibinfo  {school} {University of Calgary Computer Science Department} (\bibinfo {year} {2015})\BibitemShut {NoStop}%
\bibitem [{\citenamefont {Shapiro}\ \emph {et~al.}(2003)\citenamefont {Shapiro}, \citenamefont {Levchenko}, \citenamefont {Meyerowitz}, \citenamefont {Wold},\ and\ \citenamefont {Mjolsness}}]{cellerator}%
  \BibitemOpen
  \bibfield  {author} {\bibinfo {author} {\bibfnamefont {B.~E.}\ \bibnamefont {Shapiro}}, \bibinfo {author} {\bibfnamefont {A.}~\bibnamefont {Levchenko}}, \bibinfo {author} {\bibfnamefont {E.~M.}\ \bibnamefont {Meyerowitz}}, \bibinfo {author} {\bibfnamefont {B.~J.}\ \bibnamefont {Wold}}, \ and\ \bibinfo {author} {\bibfnamefont {E.}~\bibnamefont {Mjolsness}},\ }\bibfield  {title} {\enquote {\bibinfo {title} {Cellerator: extending a computer algebra system to include biochemical arrows for signal transduction simulations},}\ }\href {\doibase 10.1093/BIOINFORMATICS/BTG042} {\bibfield  {journal} {\bibinfo  {journal} {Bioinform.}\ }\textbf {\bibinfo {volume} {19}},\ \bibinfo {pages} {677--678} (\bibinfo {year} {2003})}\BibitemShut {NoStop}%
\bibitem [{\citenamefont {Deeds}\ \emph {et~al.}(2012)\citenamefont {Deeds}, \citenamefont {Krivine}, \citenamefont {Feret}, \citenamefont {Danos},\ and\ \citenamefont {Fontana}}]{deeds}%
  \BibitemOpen
  \bibfield  {author} {\bibinfo {author} {\bibfnamefont {E.~J.}\ \bibnamefont {Deeds}}, \bibinfo {author} {\bibfnamefont {J.}~\bibnamefont {Krivine}}, \bibinfo {author} {\bibfnamefont {J.}~\bibnamefont {Feret}}, \bibinfo {author} {\bibfnamefont {V.}~\bibnamefont {Danos}}, \ and\ \bibinfo {author} {\bibfnamefont {W.}~\bibnamefont {Fontana}},\ }\bibfield  {title} {\enquote {\bibinfo {title} {Combinatorial complexity and compositional drift in protein interaction networks},}\ }\href {\doibase 10.1371/journal.pone.0032032} {\bibfield  {journal} {\bibinfo  {journal} {PLOS ONE}\ }\textbf {\bibinfo {volume} {7}},\ \bibinfo {pages} {1--14} (\bibinfo {year} {2012})}\BibitemShut {NoStop}%
\bibitem [{\citenamefont {Rand}\ and\ \citenamefont {Walkington}(2009)}]{collars}%
  \BibitemOpen
  \bibfield  {author} {\bibinfo {author} {\bibfnamefont {A.}~\bibnamefont {Rand}}\ and\ \bibinfo {author} {\bibfnamefont {N.}~\bibnamefont {Walkington}},\ }\bibfield  {title} {\enquote {\bibinfo {title} {Collars and intestines: Practical conforming {D}elaunay refinement},}\ }in\ \href@noop {} {\emph {\bibinfo {booktitle} {Proceedings of the 18th International Meshing Roundtable}}},\ \bibinfo {editor} {edited by\ \bibinfo {editor} {\bibfnamefont {B.~W.}\ \bibnamefont {Clark}}}\ (\bibinfo  {publisher} {Springer},\ \bibinfo {address} {Berlin},\ \bibinfo {year} {2009})\ pp.\ \bibinfo {pages} {481--497}\BibitemShut {NoStop}%
\bibitem [{\citenamefont {Siek}, \citenamefont {Lee},\ and\ \citenamefont {Lumsdaine}(2002)}]{BGL}%
  \BibitemOpen
  \bibfield  {author} {\bibinfo {author} {\bibfnamefont {J.}~\bibnamefont {Siek}}, \bibinfo {author} {\bibfnamefont {L.-Q.}\ \bibnamefont {Lee}}, \ and\ \bibinfo {author} {\bibfnamefont {A.}~\bibnamefont {Lumsdaine}},\ }\enquote {\bibinfo {title} {The boost graph library: User guide and reference manual},}\ \ (\bibinfo {year} {2002})\BibitemShut {NoStop}%
\bibitem [{\citenamefont {Allen}\ and\ \citenamefont {Tildesley}(2017)}]{cell_list}%
  \BibitemOpen
  \bibfield  {author} {\bibinfo {author} {\bibfnamefont {M.~P.}\ \bibnamefont {Allen}}\ and\ \bibinfo {author} {\bibfnamefont {D.~J.}\ \bibnamefont {Tildesley}},\ }\href {\doibase 10.1093/oso/9780198803195.001.0001} {\emph {\bibinfo {title} {{Computer Simulation of Liquids}}}}\ (\bibinfo  {publisher} {Oxford University Press},\ \bibinfo {year} {2017})\BibitemShut {NoStop}%
\bibitem [{\citenamefont {Löwe}(1993)}]{single_pushout}%
  \BibitemOpen
  \bibfield  {author} {\bibinfo {author} {\bibfnamefont {M.}~\bibnamefont {Löwe}},\ }\bibfield  {title} {\enquote {\bibinfo {title} {Algebraic approach to single-pushout graph transformation},}\ }\href {\doibase https://doi.org/10.1016/0304-3975(93)90068-5} {\bibfield  {journal} {\bibinfo  {journal} {Theoretical Computer Science}\ }\textbf {\bibinfo {volume} {109}},\ \bibinfo {pages} {181--224} (\bibinfo {year} {1993})}\BibitemShut {NoStop}%
\bibitem [{\citenamefont {Ehrig}, \citenamefont {Pfender},\ and\ \citenamefont {Schneider}(1973)}]{double_pushout}%
  \BibitemOpen
  \bibfield  {author} {\bibinfo {author} {\bibfnamefont {H.}~\bibnamefont {Ehrig}}, \bibinfo {author} {\bibfnamefont {M.}~\bibnamefont {Pfender}}, \ and\ \bibinfo {author} {\bibfnamefont {H.~J.}\ \bibnamefont {Schneider}},\ }\bibfield  {title} {\enquote {\bibinfo {title} {Graph-grammars: An algebraic approach},}\ }in\ \href {\doibase 10.1109/SWAT.1973.11} {\emph {\bibinfo {booktitle} {14th Annual Symposium on Switching and Automata Theory (swat 1973)}}}\ (\bibinfo {year} {1973})\ pp.\ \bibinfo {pages} {167--180}\BibitemShut {NoStop}%
\bibitem [{\citenamefont {Valiente}(2021)}]{Valiente2021}%
  \BibitemOpen
  \bibfield  {author} {\bibinfo {author} {\bibfnamefont {G.}~\bibnamefont {Valiente}},\ }\enquote {\bibinfo {title} {Graph isomorphism},}\ in\ \href {\doibase 10.1007/978-3-030-81885-2_7} {\emph {\bibinfo {booktitle} {Algorithms on Trees and Graphs: With Python Code}}}\ (\bibinfo  {publisher} {Springer International Publishing},\ \bibinfo {address} {Cham},\ \bibinfo {year} {2021})\ pp.\ \bibinfo {pages} {255--285}\BibitemShut {NoStop}%
\bibitem [{\citenamefont {Coen}\ and\ \citenamefont {Cosgrove}(2023)}]{plant_mechanics}%
  \BibitemOpen
  \bibfield  {author} {\bibinfo {author} {\bibfnamefont {E.}~\bibnamefont {Coen}}\ and\ \bibinfo {author} {\bibfnamefont {D.~J.}\ \bibnamefont {Cosgrove}},\ }\bibfield  {title} {\enquote {\bibinfo {title} {The mechanics of plant morphogenesis},}\ }\href {\doibase 10.1126/science.ade8055} {\bibfield  {journal} {\bibinfo  {journal} {Science}\ }\textbf {\bibinfo {volume} {379}},\ \bibinfo {pages} {eade8055} (\bibinfo {year} {2023})}\BibitemShut {NoStop}%
\bibitem [{\citenamefont {Shaw}(2013)}]{shaw_reorg}%
  \BibitemOpen
  \bibfield  {author} {\bibinfo {author} {\bibfnamefont {S.~L.}\ \bibnamefont {Shaw}},\ }\bibfield  {title} {\enquote {\bibinfo {title} {Reorganization of the plant cortical microtubule array},}\ }\href {\doibase https://doi.org/10.1016/j.pbi.2013.09.006} {\bibfield  {journal} {\bibinfo  {journal} {Current Opinion in Plant Biology}\ }\textbf {\bibinfo {volume} {16}},\ \bibinfo {pages} {693--697} (\bibinfo {year} {2013})},\ \bibinfo {note} {cell biology}\BibitemShut {NoStop}%
\bibitem [{\citenamefont {Elliott}\ and\ \citenamefont {Shaw}(2017)}]{update_cma}%
  \BibitemOpen
  \bibfield  {author} {\bibinfo {author} {\bibfnamefont {A.}~\bibnamefont {Elliott}}\ and\ \bibinfo {author} {\bibfnamefont {S.~L.}\ \bibnamefont {Shaw}},\ }\bibfield  {title} {\enquote {\bibinfo {title} {{Update: Plant Cortical Microtubule Arrays}},}\ }\href {\doibase 10.1104/pp.17.01329} {\bibfield  {journal} {\bibinfo  {journal} {Plant Physiology}\ }\textbf {\bibinfo {volume} {176}},\ \bibinfo {pages} {94--105} (\bibinfo {year} {2017})}\BibitemShut {NoStop}%
\bibitem [{\citenamefont {Abrash}\ and\ \citenamefont {Bergmann}(2009)}]{cell_div}%
  \BibitemOpen
  \bibfield  {author} {\bibinfo {author} {\bibfnamefont {E.~B.}\ \bibnamefont {Abrash}}\ and\ \bibinfo {author} {\bibfnamefont {D.~C.}\ \bibnamefont {Bergmann}},\ }\bibfield  {title} {\enquote {\bibinfo {title} {Asymmetric cell divisions: A view from plant development},}\ }\href {\doibase https://doi.org/10.1016/j.devcel.2009.05.014} {\bibfield  {journal} {\bibinfo  {journal} {Developmental Cell}\ }\textbf {\bibinfo {volume} {16}},\ \bibinfo {pages} {783--796} (\bibinfo {year} {2009})}\BibitemShut {NoStop}%
\bibitem [{\citenamefont {Allard}, \citenamefont {Wasteneys},\ and\ \citenamefont {Cytrynbaum}(2010)}]{allard_sim}%
  \BibitemOpen
  \bibfield  {author} {\bibinfo {author} {\bibfnamefont {J.~F.}\ \bibnamefont {Allard}}, \bibinfo {author} {\bibfnamefont {G.~O.}\ \bibnamefont {Wasteneys}}, \ and\ \bibinfo {author} {\bibfnamefont {E.~N.}\ \bibnamefont {Cytrynbaum}},\ }\bibfield  {title} {\enquote {\bibinfo {title} {Mechanisms of self-organization of cortical microtubules in plants revealed by computational simulations},}\ }\href {\doibase 10.1091/mbc.e09-07-0579} {\bibfield  {journal} {\bibinfo  {journal} {Molecular Biology of the Cell}\ }\textbf {\bibinfo {volume} {21}},\ \bibinfo {pages} {278--286} (\bibinfo {year} {2010})},\ \bibinfo {note} {pMID: 19910489}\BibitemShut {NoStop}%
\bibitem [{\citenamefont {Tindemans}\ \emph {et~al.}(2014)\citenamefont {Tindemans}, \citenamefont {Deinum}, \citenamefont {Lindeboom},\ and\ \citenamefont {Mulder}}]{cylinder_sim}%
  \BibitemOpen
  \bibfield  {author} {\bibinfo {author} {\bibfnamefont {S.}~\bibnamefont {Tindemans}}, \bibinfo {author} {\bibfnamefont {E.}~\bibnamefont {Deinum}}, \bibinfo {author} {\bibfnamefont {J.}~\bibnamefont {Lindeboom}}, \ and\ \bibinfo {author} {\bibfnamefont {B.}~\bibnamefont {Mulder}},\ }\bibfield  {title} {\enquote {\bibinfo {title} {Efficient event-driven simulations shed new light on microtubule organization in the plant cortical array},}\ }\href {\doibase 10.3389/fphy.2014.00019} {\bibfield  {journal} {\bibinfo  {journal} {Frontiers in Physics}\ }\textbf {\bibinfo {volume} {2}} (\bibinfo {year} {2014}),\ 10.3389/fphy.2014.00019}\BibitemShut {NoStop}%
\bibitem [{\citenamefont {Eren}, \citenamefont {Dixit},\ and\ \citenamefont {Gautam}(2010)}]{cylinder_sim2}%
  \BibitemOpen
  \bibfield  {author} {\bibinfo {author} {\bibfnamefont {E.~C.}\ \bibnamefont {Eren}}, \bibinfo {author} {\bibfnamefont {R.}~\bibnamefont {Dixit}}, \ and\ \bibinfo {author} {\bibfnamefont {N.}~\bibnamefont {Gautam}},\ }\bibfield  {title} {\enquote {\bibinfo {title} {A three-dimensional computer simulation model reveals the mechanisms for self-organization of plant cortical microtubules into oblique arrays},}\ }\href {\doibase 10.1091/mbc.e10-02-0136} {\bibfield  {journal} {\bibinfo  {journal} {Molecular Biology of the Cell}\ }\textbf {\bibinfo {volume} {21}},\ \bibinfo {pages} {2674--2684} (\bibinfo {year} {2010})},\ \bibinfo {note} {pMID: 20519434}\BibitemShut {NoStop}%
\bibitem [{\citenamefont {Thoms}\ \emph {et~al.}(2018)\citenamefont {Thoms}, \citenamefont {Vineyard}, \citenamefont {Elliott},\ and\ \citenamefont {Shaw}}]{CLASP_facilitates}%
  \BibitemOpen
  \bibfield  {author} {\bibinfo {author} {\bibfnamefont {D.}~\bibnamefont {Thoms}}, \bibinfo {author} {\bibfnamefont {L.}~\bibnamefont {Vineyard}}, \bibinfo {author} {\bibfnamefont {A.}~\bibnamefont {Elliott}}, \ and\ \bibinfo {author} {\bibfnamefont {S.~L.}\ \bibnamefont {Shaw}},\ }\bibfield  {title} {\enquote {\bibinfo {title} {{CLASP Facilitates Transitions between Cortical Microtubule Array Patterns}},}\ }\href {\doibase 10.1104/pp.18.00961} {\bibfield  {journal} {\bibinfo  {journal} {Plant Physiology}\ }\textbf {\bibinfo {volume} {178}},\ \bibinfo {pages} {1551--1567} (\bibinfo {year} {2018})}\BibitemShut {NoStop}%
\bibitem [{\citenamefont {Ambrose}\ \emph {et~al.}(2011)\citenamefont {Ambrose}, \citenamefont {Allard}, \citenamefont {Cytrynbaum},\ and\ \citenamefont {Wasteneys}}]{Ambrose2011}%
  \BibitemOpen
  \bibfield  {author} {\bibinfo {author} {\bibfnamefont {C.}~\bibnamefont {Ambrose}}, \bibinfo {author} {\bibfnamefont {J.~F.}\ \bibnamefont {Allard}}, \bibinfo {author} {\bibfnamefont {E.~N.}\ \bibnamefont {Cytrynbaum}}, \ and\ \bibinfo {author} {\bibfnamefont {G.~O.}\ \bibnamefont {Wasteneys}},\ }\bibfield  {title} {\enquote {\bibinfo {title} {A clasp-modulated cell edge barrier mechanism drives cell-wide cortical microtubule organization in arabidopsis},}\ }\href {\doibase 10.1038/ncomms1444} {\bibfield  {journal} {\bibinfo  {journal} {Nature Communications}\ }\textbf {\bibinfo {volume} {2}},\ \bibinfo {pages} {430} (\bibinfo {year} {2011})}\BibitemShut {NoStop}%
\bibitem [{\citenamefont {Tindemans}, \citenamefont {Hawkins},\ and\ \citenamefont {Mulder}(2010)}]{aligned_survival}%
  \BibitemOpen
  \bibfield  {author} {\bibinfo {author} {\bibfnamefont {S.}~\bibnamefont {Tindemans}}, \bibinfo {author} {\bibfnamefont {R.}~\bibnamefont {Hawkins}}, \ and\ \bibinfo {author} {\bibfnamefont {B.}~\bibnamefont {Mulder}},\ }\bibfield  {title} {\enquote {\bibinfo {title} {Survival of the aligned: Ordering of the plant cortical microtubule array},}\ }\href {\doibase 10.1103/PhysRevLett.104.058103} {\bibfield  {journal} {\bibinfo  {journal} {Physical review letters}\ }\textbf {\bibinfo {volume} {104}},\ \bibinfo {pages} {058103} (\bibinfo {year} {2010})}\BibitemShut {NoStop}%
\bibitem [{\citenamefont {Schneider}\ \emph {et~al.}(2021)\citenamefont {Schneider}, \citenamefont {Klooster}, \citenamefont {Picard}, \citenamefont {van~der Gucht}, \citenamefont {Demura}, \citenamefont {Janson}, \citenamefont {Sampathkumar}, \citenamefont {Deinum}, \citenamefont {Ketelaar},\ and\ \citenamefont {Persson}}]{band_fromation}%
  \BibitemOpen
  \bibfield  {author} {\bibinfo {author} {\bibfnamefont {R.}~\bibnamefont {Schneider}}, \bibinfo {author} {\bibfnamefont {K.~v.}\ \bibnamefont {Klooster}}, \bibinfo {author} {\bibfnamefont {K.~L.}\ \bibnamefont {Picard}}, \bibinfo {author} {\bibfnamefont {J.}~\bibnamefont {van~der Gucht}}, \bibinfo {author} {\bibfnamefont {T.}~\bibnamefont {Demura}}, \bibinfo {author} {\bibfnamefont {M.}~\bibnamefont {Janson}}, \bibinfo {author} {\bibfnamefont {A.}~\bibnamefont {Sampathkumar}}, \bibinfo {author} {\bibfnamefont {E.~E.}\ \bibnamefont {Deinum}}, \bibinfo {author} {\bibfnamefont {T.}~\bibnamefont {Ketelaar}}, \ and\ \bibinfo {author} {\bibfnamefont {S.}~\bibnamefont {Persson}},\ }\bibfield  {title} {\enquote {\bibinfo {title} {Long-term single-cell imaging and simulations of microtubules reveal principles behind wall patterning during proto-xylem development},}\ }\href {\doibase 10.1038/s41467-021-20894-1} {\bibfield  {journal} {\bibinfo  {journal} {Nature Communications}\ }\textbf {\bibinfo {volume} {12}},\
  \bibinfo {pages} {669} (\bibinfo {year} {2021})}\BibitemShut {NoStop}%
\bibitem [{\citenamefont {Vineyard}\ \emph {et~al.}(2013)\citenamefont {Vineyard}, \citenamefont {Elliott}, \citenamefont {Dhingra}, \citenamefont {Lucas},\ and\ \citenamefont {Shaw}}]{picket_fence}%
  \BibitemOpen
  \bibfield  {author} {\bibinfo {author} {\bibfnamefont {L.}~\bibnamefont {Vineyard}}, \bibinfo {author} {\bibfnamefont {A.}~\bibnamefont {Elliott}}, \bibinfo {author} {\bibfnamefont {S.}~\bibnamefont {Dhingra}}, \bibinfo {author} {\bibfnamefont {J.~R.}\ \bibnamefont {Lucas}}, \ and\ \bibinfo {author} {\bibfnamefont {S.~L.}\ \bibnamefont {Shaw}},\ }\bibfield  {title} {\enquote {\bibinfo {title} {{Progressive Transverse Microtubule Array Organization in Hormone-Induced Arabidopsis Hypocotyl Cells}},}\ }\href {\doibase 10.1105/tpc.112.107326} {\bibfield  {journal} {\bibinfo  {journal} {The Plant Cell}\ }\textbf {\bibinfo {volume} {25}},\ \bibinfo {pages} {662--676} (\bibinfo {year} {2013})}\BibitemShut {NoStop}%
\bibitem [{\citenamefont {Sambade}\ \emph {et~al.}(2012)\citenamefont {Sambade}, \citenamefont {Pratap}, \citenamefont {Buschmann}, \citenamefont {Morris},\ and\ \citenamefont {Lloyd}}]{light_reorg}%
  \BibitemOpen
  \bibfield  {author} {\bibinfo {author} {\bibfnamefont {A.}~\bibnamefont {Sambade}}, \bibinfo {author} {\bibfnamefont {A.}~\bibnamefont {Pratap}}, \bibinfo {author} {\bibfnamefont {H.}~\bibnamefont {Buschmann}}, \bibinfo {author} {\bibfnamefont {R.~J.}\ \bibnamefont {Morris}}, \ and\ \bibinfo {author} {\bibfnamefont {C.}~\bibnamefont {Lloyd}},\ }\bibfield  {title} {\enquote {\bibinfo {title} {{The Influence of Light on Microtubule Dynamics and Alignment in the Arabidopsis Hypocotyl}},}\ }\href@noop {} {\bibfield  {journal} {\bibinfo  {journal} {The Plant Cell}\ }\textbf {\bibinfo {volume} {24}},\ \bibinfo {pages} {192--201} (\bibinfo {year} {2012})}\BibitemShut {NoStop}%
\bibitem [{\citenamefont {Chan}\ \emph {et~al.}(1999)\citenamefont {Chan}, \citenamefont {Jensen}, \citenamefont {Jensen}, \citenamefont {Bush},\ and\ \citenamefont {Lloyd}}]{sep_dist}%
  \BibitemOpen
  \bibfield  {author} {\bibinfo {author} {\bibfnamefont {J.}~\bibnamefont {Chan}}, \bibinfo {author} {\bibfnamefont {C.~G.}\ \bibnamefont {Jensen}}, \bibinfo {author} {\bibfnamefont {L.~C.}\ \bibnamefont {Jensen}}, \bibinfo {author} {\bibfnamefont {M.}~\bibnamefont {Bush}}, \ and\ \bibinfo {author} {\bibfnamefont {C.~W.}\ \bibnamefont {Lloyd}},\ }\bibfield  {title} {\enquote {\bibinfo {title} {The 65-kda carrot microtubule-associated protein forms regularly arranged filamentous cross-bridges between microtubules},}\ }\href@noop {} {\bibfield  {journal} {\bibinfo  {journal} {Proceedings of the National Academy of Sciences}\ }\textbf {\bibinfo {volume} {96}},\ \bibinfo {pages} {14931--14936} (\bibinfo {year} {1999})}\BibitemShut {NoStop}%
\bibitem [{\citenamefont {Murata}\ \emph {et~al.}(2005)\citenamefont {Murata}, \citenamefont {Sonobe}, \citenamefont {Baskin}, \citenamefont {Hyodo}, \citenamefont {Hasezawa}, \citenamefont {Nagata}, \citenamefont {Horio},\ and\ \citenamefont {Hasebe}}]{NucleateMurata2005}%
  \BibitemOpen
  \bibfield  {author} {\bibinfo {author} {\bibfnamefont {T.}~\bibnamefont {Murata}}, \bibinfo {author} {\bibfnamefont {S.}~\bibnamefont {Sonobe}}, \bibinfo {author} {\bibfnamefont {T.~I.}\ \bibnamefont {Baskin}}, \bibinfo {author} {\bibfnamefont {S.}~\bibnamefont {Hyodo}}, \bibinfo {author} {\bibfnamefont {S.}~\bibnamefont {Hasezawa}}, \bibinfo {author} {\bibfnamefont {T.}~\bibnamefont {Nagata}}, \bibinfo {author} {\bibfnamefont {T.}~\bibnamefont {Horio}}, \ and\ \bibinfo {author} {\bibfnamefont {M.}~\bibnamefont {Hasebe}},\ }\bibfield  {title} {\enquote {\bibinfo {title} {Microtubule-dependent microtubule nucleation based on recruitment of $\gamma$-tubulin in higher plants},}\ }\href {\doibase 10.1038/ncb1306} {\bibfield  {journal} {\bibinfo  {journal} {Nature Cell Biology}\ }\textbf {\bibinfo {volume} {7}},\ \bibinfo {pages} {961--968} (\bibinfo {year} {2005})}\BibitemShut {NoStop}%
\bibitem [{\citenamefont {Nakamura}\ and\ \citenamefont {Hashimoto}(2009)}]{NucleateNakamure2009}%
  \BibitemOpen
  \bibfield  {author} {\bibinfo {author} {\bibfnamefont {M.}~\bibnamefont {Nakamura}}\ and\ \bibinfo {author} {\bibfnamefont {T.}~\bibnamefont {Hashimoto}},\ }\bibfield  {title} {\enquote {\bibinfo {title} {{A mutation in the Arabidopsis $\gamma$-tubulin-containing complex causes helical growth and abnormal microtubule branching}},}\ }\href {\doibase 10.1242/jcs.044131} {\bibfield  {journal} {\bibinfo  {journal} {Journal of Cell Science}\ }\textbf {\bibinfo {volume} {122}},\ \bibinfo {pages} {2208--2217} (\bibinfo {year} {2009})}\BibitemShut {NoStop}%
\bibitem [{\citenamefont {Wightman}\ \emph {et~al.}(2013)\citenamefont {Wightman}, \citenamefont {Chomicki}, \citenamefont {Kumar}, \citenamefont {Carr},\ and\ \citenamefont {Turner}}]{crossover_severing}%
  \BibitemOpen
  \bibfield  {author} {\bibinfo {author} {\bibfnamefont {R.}~\bibnamefont {Wightman}}, \bibinfo {author} {\bibfnamefont {G.}~\bibnamefont {Chomicki}}, \bibinfo {author} {\bibfnamefont {M.}~\bibnamefont {Kumar}}, \bibinfo {author} {\bibfnamefont {P.}~\bibnamefont {Carr}}, \ and\ \bibinfo {author} {\bibfnamefont {S.}~\bibnamefont {Turner}},\ }\bibfield  {title} {\enquote {\bibinfo {title} {Spiral2 determines plant microtubule organization by modulating microtubule severing},}\ }\href {\doibase https://doi.org/10.1016/j.cub.2013.07.061} {\bibfield  {journal} {\bibinfo  {journal} {Current Biology}\ }\textbf {\bibinfo {volume} {23}},\ \bibinfo {pages} {1902--1907} (\bibinfo {year} {2013})}\BibitemShut {NoStop}%
\bibitem [{\citenamefont {Himmelspach}, \citenamefont {Williamson},\ and\ \citenamefont {Wasteneys}(2003)}]{orientation_angle_measure}%
  \BibitemOpen
  \bibfield  {author} {\bibinfo {author} {\bibfnamefont {R.}~\bibnamefont {Himmelspach}}, \bibinfo {author} {\bibfnamefont {R.~E.}\ \bibnamefont {Williamson}}, \ and\ \bibinfo {author} {\bibfnamefont {G.~O.}\ \bibnamefont {Wasteneys}},\ }\bibfield  {title} {\enquote {\bibinfo {title} {Cellulose microfibril alignment recovers from dcb-induced disruption despite microtubule disorganization},}\ }\href {\doibase https://doi.org/10.1046/j.1365-313X.2003.01906.x} {\bibfield  {journal} {\bibinfo  {journal} {The Plant Journal}\ }\textbf {\bibinfo {volume} {36}},\ \bibinfo {pages} {565--575} (\bibinfo {year} {2003})}\BibitemShut {NoStop}%
\bibitem [{\citenamefont {Baulin}, \citenamefont {Marques},\ and\ \citenamefont {Thalmann}(2007)}]{angular_cost}%
  \BibitemOpen
  \bibfield  {author} {\bibinfo {author} {\bibfnamefont {V.~A.}\ \bibnamefont {Baulin}}, \bibinfo {author} {\bibfnamefont {C.~M.}\ \bibnamefont {Marques}}, \ and\ \bibinfo {author} {\bibfnamefont {F.}~\bibnamefont {Thalmann}},\ }\bibfield  {title} {\enquote {\bibinfo {title} {Collision induced spatial organization of microtubules},}\ }\href {\doibase https://doi.org/10.1016/j.bpc.2007.04.009} {\bibfield  {journal} {\bibinfo  {journal} {Biophysical Chemistry}\ }\textbf {\bibinfo {volume} {128}},\ \bibinfo {pages} {231--244} (\bibinfo {year} {2007})}\BibitemShut {NoStop}%
\bibitem [{\citenamefont {Deinum}\ \emph {et~al.}(2017)\citenamefont {Deinum}, \citenamefont {Tindemans}, \citenamefont {Lindeboom},\ and\ \citenamefont {Mulder}}]{katanin_mulder}%
  \BibitemOpen
  \bibfield  {author} {\bibinfo {author} {\bibfnamefont {E.~E.}\ \bibnamefont {Deinum}}, \bibinfo {author} {\bibfnamefont {S.~H.}\ \bibnamefont {Tindemans}}, \bibinfo {author} {\bibfnamefont {J.~J.}\ \bibnamefont {Lindeboom}}, \ and\ \bibinfo {author} {\bibfnamefont {B.~M.}\ \bibnamefont {Mulder}},\ }\bibfield  {title} {\enquote {\bibinfo {title} {How selective severing by katanin promotes order in the plant cortical microtubule array},}\ }\href {\doibase 10.1073/pnas.1702650114} {\bibfield  {journal} {\bibinfo  {journal} {Proceedings of the National Academy of Sciences}\ }\textbf {\bibinfo {volume} {114}},\ \bibinfo {pages} {6942--6947} (\bibinfo {year} {2017})}\BibitemShut {NoStop}%
\bibitem [{\citenamefont {Pampaloni}\ \emph {et~al.}(2006)\citenamefont {Pampaloni}, \citenamefont {Lattanzi}, \citenamefont {Jonáš}, \citenamefont {Surrey}, \citenamefont {Frey},\ and\ \citenamefont {Florin}}]{mt_thermal}%
  \BibitemOpen
  \bibfield  {author} {\bibinfo {author} {\bibfnamefont {F.}~\bibnamefont {Pampaloni}}, \bibinfo {author} {\bibfnamefont {G.}~\bibnamefont {Lattanzi}}, \bibinfo {author} {\bibfnamefont {A.}~\bibnamefont {Jonáš}}, \bibinfo {author} {\bibfnamefont {T.}~\bibnamefont {Surrey}}, \bibinfo {author} {\bibfnamefont {E.}~\bibnamefont {Frey}}, \ and\ \bibinfo {author} {\bibfnamefont {E.-L.}\ \bibnamefont {Florin}},\ }\bibfield  {title} {\enquote {\bibinfo {title} {Thermal fluctuations of grafted microtubules provide evidence of a length-dependent persistence length},}\ }\href {\doibase 10.1073/pnas.0603931103} {\bibfield  {journal} {\bibinfo  {journal} {Proceedings of the National Academy of Sciences}\ }\textbf {\bibinfo {volume} {103}},\ \bibinfo {pages} {10248--10253} (\bibinfo {year} {2006})}\BibitemShut {NoStop}%
\bibitem [{\citenamefont {de~Moura}\ \emph {et~al.}(2015)\citenamefont {de~Moura}, \citenamefont {Kong}, \citenamefont {Avigad}, \citenamefont {van Doorn},\ and\ \citenamefont {von Raumer}}]{lean}%
  \BibitemOpen
  \bibfield  {author} {\bibinfo {author} {\bibfnamefont {L.}~\bibnamefont {de~Moura}}, \bibinfo {author} {\bibfnamefont {S.}~\bibnamefont {Kong}}, \bibinfo {author} {\bibfnamefont {J.}~\bibnamefont {Avigad}}, \bibinfo {author} {\bibfnamefont {F.}~\bibnamefont {van Doorn}}, \ and\ \bibinfo {author} {\bibfnamefont {J.}~\bibnamefont {von Raumer}},\ }\bibfield  {title} {\enquote {\bibinfo {title} {The lean theorem prover (system description)},}\ }in\ \href@noop {} {\emph {\bibinfo {booktitle} {Automated Deduction - CADE-25}}},\ \bibinfo {editor} {edited by\ \bibinfo {editor} {\bibfnamefont {A.~P.}\ \bibnamefont {Felty}}\ and\ \bibinfo {editor} {\bibfnamefont {A.}~\bibnamefont {Middeldorp}}}\ (\bibinfo  {publisher} {Springer International Publishing},\ \bibinfo {address} {Cham},\ \bibinfo {year} {2015})\ pp.\ \bibinfo {pages} {378--388}\BibitemShut {NoStop}%
\bibitem [{\citenamefont {Ernst}\ \emph {et~al.}(2018)\citenamefont {Ernst}, \citenamefont {Bartol}, \citenamefont {Sejnowski},\ and\ \citenamefont {Mjolsness}}]{ernst}%
  \BibitemOpen
  \bibfield  {author} {\bibinfo {author} {\bibfnamefont {O.~K.}\ \bibnamefont {Ernst}}, \bibinfo {author} {\bibfnamefont {T.}~\bibnamefont {Bartol}}, \bibinfo {author} {\bibfnamefont {T.}~\bibnamefont {Sejnowski}}, \ and\ \bibinfo {author} {\bibfnamefont {E.}~\bibnamefont {Mjolsness}},\ }\bibfield  {title} {\enquote {\bibinfo {title} {Learning dynamic boltzmann distributions as reduced models of spatial chemical kinetics},}\ }\href {\doibase 10.1063/1.5026403} {\bibfield  {journal} {\bibinfo  {journal} {The Journal of Chemical Physics}\ }\textbf {\bibinfo {volume} {149}},\ \bibinfo {pages} {034107} (\bibinfo {year} {2018})}\BibitemShut {NoStop}%
\bibitem [{\citenamefont {Scott}\ and\ \citenamefont {Mjolsness}(2019)}]{scott}%
  \BibitemOpen
  \bibfield  {author} {\bibinfo {author} {\bibfnamefont {C.~B.}\ \bibnamefont {Scott}}\ and\ \bibinfo {author} {\bibfnamefont {E.}~\bibnamefont {Mjolsness}},\ }\bibfield  {title} {\enquote {\bibinfo {title} {Multilevel artificial neural network training for spatially correlated learning},}\ }\href {\doibase 10.1137/18M1191506} {\bibfield  {journal} {\bibinfo  {journal} {SIAM Journal on Scientific Computing}\ }\textbf {\bibinfo {volume} {41}},\ \bibinfo {pages} {S297--S320} (\bibinfo {year} {2019})}\BibitemShut {NoStop}%
\bibitem [{\citenamefont {Eric~Medwedeff}(2024)}]{dggml}%
  \BibitemOpen
  \bibfield  {author} {\bibinfo {author} {\bibfnamefont {E.~M.}\ \bibnamefont {Eric~Medwedeff}},\ }\href@noop {} {\enquote {\bibinfo {title} {The dynamical graph grammar modeling library},}\ }\bibinfo {howpublished} {\url{https://github.com/emedwede/DGGML}} (\bibinfo {year} {2024})\BibitemShut {NoStop}%
\bibitem [{\citenamefont {Medwedeff}(2023)}]{yagl}%
  \BibitemOpen
  \bibfield  {author} {\bibinfo {author} {\bibfnamefont {E.}~\bibnamefont {Medwedeff}},\ }\href@noop {} {\enquote {\bibinfo {title} {Yet another graph library},}\ }\bibinfo {howpublished} {\url{https://github.com/emedwede/YAGL}} (\bibinfo {year} {2023})\BibitemShut {NoStop}%
\bibitem [{\citenamefont {Burbank}\ and\ \citenamefont {Mitchison}(2006)}]{mt_instability}%
  \BibitemOpen
  \bibfield  {author} {\bibinfo {author} {\bibfnamefont {K.~S.}\ \bibnamefont {Burbank}}\ and\ \bibinfo {author} {\bibfnamefont {T.~J.}\ \bibnamefont {Mitchison}},\ }\bibfield  {title} {\enquote {\bibinfo {title} {Microtubule dynamic instability},}\ }\href {\doibase 10.1016/j.cub.2006.06.044} {\bibfield  {journal} {\bibinfo  {journal} {Current Biology}\ }\textbf {\bibinfo {volume} {16}},\ \bibinfo {pages} {R516--R517} (\bibinfo {year} {2006})}\BibitemShut {NoStop}%
\bibitem [{\citenamefont {Dixit}\ and\ \citenamefont {Cyr}(2004)}]{dynamic_cmt_ordering}%
  \BibitemOpen
  \bibfield  {author} {\bibinfo {author} {\bibfnamefont {R.}~\bibnamefont {Dixit}}\ and\ \bibinfo {author} {\bibfnamefont {R.}~\bibnamefont {Cyr}},\ }\bibfield  {title} {\enquote {\bibinfo {title} {{Encounters between Dynamic Cortical Microtubules Promote Ordering of the Cortical Array through Angle-Dependent Modifications of Microtubule Behavior}},}\ }\href {\doibase 10.1105/tpc.104.026930} {\bibfield  {journal} {\bibinfo  {journal} {The Plant Cell}\ }\textbf {\bibinfo {volume} {16}},\ \bibinfo {pages} {3274--3284} (\bibinfo {year} {2004})}\BibitemShut {NoStop}%
\bibitem [{\citenamefont {Chakrabortty}\ \emph {et~al.}(2018)\citenamefont {Chakrabortty}, \citenamefont {Willemsen}, \citenamefont {{de Zeeuw}}, \citenamefont {Liao}, \citenamefont {Weijers}, \citenamefont {Mulder},\ and\ \citenamefont {Scheres}}]{CHAKRABORTTY20183031}%
  \BibitemOpen
  \bibfield  {author} {\bibinfo {author} {\bibfnamefont {B.}~\bibnamefont {Chakrabortty}}, \bibinfo {author} {\bibfnamefont {V.}~\bibnamefont {Willemsen}}, \bibinfo {author} {\bibfnamefont {T.}~\bibnamefont {{de Zeeuw}}}, \bibinfo {author} {\bibfnamefont {C.-Y.}\ \bibnamefont {Liao}}, \bibinfo {author} {\bibfnamefont {D.}~\bibnamefont {Weijers}}, \bibinfo {author} {\bibfnamefont {B.}~\bibnamefont {Mulder}}, \ and\ \bibinfo {author} {\bibfnamefont {B.}~\bibnamefont {Scheres}},\ }\bibfield  {title} {\enquote {\bibinfo {title} {A plausible microtubule-based mechanism for cell division orientation in plant embryogenesis},}\ }\href {\doibase 10.1016/j.cub.2018.07.025} {\bibfield  {journal} {\bibinfo  {journal} {Current Biology}\ }\textbf {\bibinfo {volume} {28}},\ \bibinfo {pages} {3031--3043.e2} (\bibinfo {year} {2018})}\BibitemShut {NoStop}%
\bibitem [{\citenamefont {Ambrose}\ and\ \citenamefont {Wasteneys}(2008)}]{ambrose2008clasp}%
  \BibitemOpen
  \bibfield  {author} {\bibinfo {author} {\bibfnamefont {J.~C.}\ \bibnamefont {Ambrose}}\ and\ \bibinfo {author} {\bibfnamefont {G.~O.}\ \bibnamefont {Wasteneys}},\ }\bibfield  {title} {\enquote {\bibinfo {title} {Clasp modulates microtubule-cortex interaction during self-organization of acentrosomal microtubules},}\ }\href@noop {} {\bibfield  {journal} {\bibinfo  {journal} {Molecular biology of the cell}\ }\textbf {\bibinfo {volume} {19}},\ \bibinfo {pages} {4730--4737} (\bibinfo {year} {2008})}\BibitemShut {NoStop}%
\bibitem [{\citenamefont {Wightman}\ and\ \citenamefont {Turner}(2007)}]{wightman}%
  \BibitemOpen
  \bibfield  {author} {\bibinfo {author} {\bibfnamefont {R.}~\bibnamefont {Wightman}}\ and\ \bibinfo {author} {\bibfnamefont {S.~R.}\ \bibnamefont {Turner}},\ }\bibfield  {title} {\enquote {\bibinfo {title} {Severing at sites of microtubule crossover contributes to microtubule alignment in cortical arrays},}\ }\href {\doibase 10.1111/j.1365-313X.2007.03271.x} {\bibfield  {journal} {\bibinfo  {journal} {The Plant Journal}\ }\textbf {\bibinfo {volume} {52}},\ \bibinfo {pages} {742--751} (\bibinfo {year} {2007})}\BibitemShut {NoStop}%
\end{thebibliography}%

\end{document}